\documentclass[final,3p,times]{elsarticle}

\biboptions{numbers,sort&compress}

\usepackage{amssymb}

\usepackage{placeins}

\usepackage[table]{xcolor} 
\usepackage{multirow}      
\usepackage{array}         
\usepackage{caption}
\usepackage{subcaption}
\usepackage{longtable}
\usepackage{booktabs}
 
\usepackage{hyperref}
\usepackage{pdflscape}
\usepackage{mathtools}
\usepackage{array}
\usepackage{float}
\usepackage{algorithm}
\usepackage{algorithmic}
\usepackage{pgfplots}
\pgfplotsset{compat=1.18}
\usepackage{graphicx}

\begin{document}
\begin{frontmatter}

\title{Core-Periphery Dynamics in Market-Conditioned Financial Networks: A Conditional P-Threshold Mutual Information Approach}

\author[inst1]{Kundan Mukhia}
\ead{kundanmukhia07@gmail.com} 

\author[inst2]{Imran Ansari\corref{cor1}}
\ead{imranansari@iisc.ac.in} 

\author[inst1]{S R Luwang}
\ead{salamrabindrajit@gmail.com} 

\author[inst1]{Md Nurujjaman\corref{cor1}}
\ead{md.nurujjaman@nitsikkim.ac.in} 

\affiliation[inst1]{organization={Department of Physics, National Institute of Technology},
            postcode={737139}, 
            state={Sikkim},
            country={India}}

\affiliation[inst2]{organization={Department of Management Studies, Indian Institute of Science},
            postcode={560012},
            state={Bengaluru},
            country={India}}

\cortext[cor1]{Corresponding Author}

\begin{abstract}
This study investigates how financial market structure reorganizes during the COVID-19 crash using a conditional p-threshold mutual information (MI) based Minimum Spanning Tree (MST) network framework. We analyze nonlinear dependencies among the largest stocks from four geographically and economically diverse QUAD countries: the United States, Japan, Australia, and India. The crash period is identified using the Hellinger distance and further characterized by the Hilbert spectrum. A crash is defined when the Hellinger distance exceeds the threshold $H_D = \mu_H + 2\sigma_H$, enabling the segmentation of the data into pre-crash, crash, and post-crash periods. To isolate direct stock-level dependencies, the conditional p-threshold MI approach filters out common market effects and applies permutation-based significance testing. The resulting statistically validated dependencies are used to construct MST networks, allowing consistent comparison across market periods. The network analysis reveals common crisis-related dynamics across all markets. During the crash, networks become more integrated, with shorter path lengths and higher centrality, while algebraic connectivity declines, indicating increased structural fragility. A clear reorganization of the core–periphery structure is observed, with declining core concentration and increasing periphery fragility, supported by disassortative mixing that facilitates shock transmission. In the post-crash period, network topology shows only partial recovery, suggesting persistent structural effects. This interpretation is further supported by an aftershock analysis based on the Gutenberg–Richter law, which indicates a higher relative frequency of large volatility events following the crash. The consistency of these findings across all four markets highlights the effectiveness of the conditional p-threshold MI framework for capturing nonlinear interdependencies and systemic vulnerability in financial markets.
\end{abstract}

\end{frontmatter}


\section{Introduction}

Financial markets are complex adaptive systems where the collective behavior of numerous interacting assets gives rise to emergent phenomena, including periods of stability, bubbles, and systemic crashes. Understanding the structure and dynamics of these interactions is fundamental to financial stability analysis and risk management. Network theory has emerged as a powerful model for this purpose, transforming multivariate time series of stock returns into graphs that reveal the underlying structure of financial systems~\cite{mantegna1999hierarchical,Yang2008,mukhia2024complex,tumminello2007correlation,ansari2025comprehensive,Tse2010}. The most established approach involves constructing networks from correlation matrices, with filtering techniques like the Minimum Spanning Tree (MST), planar maximally filtered graph (PMFG) and threshold-based filtering method, used to distill meaningful, sparse topologies from dense correlation structures~\cite{mantegna1999hierarchical,tumminello2005tool,boginski2005statistical,Heiberger2014}. These methods have successfully mapped hierarchical organization, sectoral clustering and core--periphery and community structures in financial networks~\cite{Onnela2003,Boginski2006,pawanesh2025exploring,ansari2025novel}.

Periods of financial crisis expose latent vulnerabilities in market structure and often trigger abrupt reorganization of asset interdependencies~\cite{han2019network,pawanesh2025exploring,ansari2025novel,li2019portfolio}. The COVID-19 pandemic-induced crash of March 2020, like the global financial crisis of 2008, represents an extreme stress episode marked by heightened volatility, synchronized market movements, and large-scale capital reallocation across asset classes~\cite{pawanesh2025exploring,ansari2025novel,peng2024global}. During such periods, pairwise correlations across assets rise sharply and tend to converge, leading to dense and homogeneous correlation networks in which meaningful structural information is lost~\cite{kenett2010dominating,pozzi2013spread,longin1995correlation}. As a result, correlation-based network approaches struggle to distinguish genuine channels of contagion from spurious dependencies induced by common market-wide factors~\cite{gopikrishnan2001quantifying,pan2007collective}. Although noise-filtering techniques applied to correlation matrices can partially recover global organization and sectoral structure~\cite{macmahon2015community,jiang2014structure,ansari2025comprehensive}, many traditional crisis-analysis frameworks based on volatility measures, raw correlations, or standard econometric models remain limited in their ability to isolate intrinsic stock-to-stock interactions under extreme stress~\cite{pozzi2013spread,han2019network,tumminello2007correlation,pozzi2007dynamical}. To address these limitations, mutual information (MI) has been proposed as a more general, model-free measure of dependence that captures both linear and nonlinear relationships among financial assets~\cite{fiedor2014networks,sharma2019mutual}. However, MI estimates from finite samples are themselves sensitive to noise and are strongly influenced by market-wide co-movement during crises, highlighting the need for conditioning and statistically grounded filtering procedures to extract direct and meaningful dependencies.

Existing empirical studies applying network methods to finance have largely focused on economically homogeneous groups of markets. Examples include analyses of the G7 countries \cite{korkusuz2023complex, polat2020frequency}, regionally and institutionally integrated blocs such as BRICS \cite{vizgunov2013comparative,dong2020research}, as well as sector-wise network studies within individual national stock markets that examine industry-level or firm-level interactions under normal and crisis conditions \cite{pozzi2013spread,huang2009network,bielinskyi2022high,rabindrajit2024high}. Although these studies offer useful insights, their emphasis on relatively homogeneous economic groupings makes it difficult to assess whether observed crisis-induced network patterns, particularly changes in core–periphery organization, reflect universal properties of financial systems or are specific to economically integrated regions. Comparative evidence from geographically and economically diverse markets that are not bound by formal economic pacts remains limited.

In this context, the QUAD countries, namely the United States, Japan, Australia, and India, offer a distinctive and underexplored setting. Unlike economic alliances such as the G7 or BRICS, the QUAD is not an economic pact but a strategic grouping encompassing markets with markedly different levels of development, financial structures, and economic drivers~\cite{quadrilateral_security_dialogue}. It includes a dominant global financial market (USA), a mature Asian economy (Japan), a developed commodity-oriented market (Australia), and a large emerging economy (India). This heterogeneity provides a natural testbed for examining whether crisis-induced network reorganization, particularly in terms of core–periphery structure, exhibits universal characteristics across fundamentally different market environments.

In this study, we address these methodological and empirical gaps by proposing a conditional p-threshold MI framework to analyze stock market dynamics before, during, and after the COVID-19 crash across the QUAD countries. First, we identify the crash period using the rolling Hellinger distance(HD), which detects statistically significant shifts in the cross-sectional return distribution, and validate the intensity and persistence of the crash using Hilbert spectrum(HS) analysis. This allows us to divide the time series into pre-crash, crash, and post-crash periods. Second, we compute MI between stock returns after removing market-wide effects through regression on the corresponding market index. We then apply permutation-based significance testing to retain only statistically significant and direct nonlinear dependencies. Finally, these validated dependencies are mapped into MST networks to ensure sparse, comparable topologies across different periods.

Our results show a consistent pattern across all four markets. During the COVID-19 crash, financial networks experienced a clear core–periphery reorganization, with reduced core concentration and increased periphery fragility, indicating a more integrated but structurally fragile market state. Post-crash networks exhibit only partial recovery, suggesting persistent structural aftershocks, a finding further supported by the Gutenberg–Richter law. The similarity of these patterns across the diverse QUAD markets highlights the ability of the conditional p-threshold MI framework to capture universal features of systemic vulnerability.

The remainder of the paper is organized as follows. Section~\ref{Methodology} describes the methodology employed in this study, while Section~\ref{Data_description} presents the data used for the analysis. Section~\ref{Results} reports and discusses the empirical results. Finally, Section~\ref{conclusion} concludes the paper and outlines directions for future research.

\section{Method of Analysis}
\label{Methodology}

To study the evolution of dependency structures and network organization in QUAD countries' stock markets during the COVID-19 crash, we applied several methods. Initially, we detect the COVID-19 stock market crash using the HD method and further characterize it through the HS. Based on the identified crash regions, we divide the data into pre-crash, crash, and post-crash periods. Next, we remove the market effect from individual stock returns using a CAPM-based regression framework~\cite{kisman2015m}. After removing the market effect, we estimate the MI among stocks to capture nonlinear dependencies. A conditional p-threshold MI approach based on permutation testing is used to filter statistically significant dependencies. Using this conditional p-threshold MI, an MST method is applied to construct the stock market network. Finally, we perform a comprehensive topological analysis using network metrics, including core concentration, periphery fragility, centrality distributions, and modularity, to quantify structural reconfiguration and systemic vulnerability across the different market periods.

\subsection{Crash detection}

In this subsection, we discuss the methodology used to identify and characterize stock market crash periods. First, crash events are detected using the HD, which quantifies abrupt, systemic shifts in the cross-sectional distribution of stock returns, enabling the simultaneous identification of a market-wide crash. Second, the detected crash regions are further characterized using the HS, allowing for a time-frequency-energy analysis of market volatility. The combined use of these two methods ensures robust identification of crash periods by integrating both statistical distance measures and spectral characteristics. Detailed descriptions of each technique are provided in the following subsubsections.

\subsubsection{Hellinger-Distance}
\label{subsec:hellinger_method}

The Hellinger distance (HD) is a measure of dissimilarity between probability distributions~\cite{deza2009encyclopedia,gibbs2002choosing}. The HD has been successfully applied in change detection, concept drift analysis, and fault detection \cite{aggoune2016change,ditzler2011hellinger,tang2009sketch}. In this study, we used the HD as an indicator to detect the crash periods and structural stress in market data.

Let $P_{t,i}$ denote the closing price of stock $i$ on trading day $t$. We define the log-return of  stock $i$  as
\begin{equation}
    r_{t,i} = \log P_{t,i} - \log P_{t-1,i}, \qquad
    t = 2,\dots,T,\quad i = 1,\dots,N.
\end{equation}

At each time $t$, the cross-section of returns is denoted by ${r}_t = (r_{t,1},\dots,r_{t,N})$.

To characterise the normal market state at time $t$, we construct a reference distribution from a rolling window of length $W$:
\begin{equation}
    \mathcal{P}_0(t)
    =
    \big\{
        r_{\tau,i} : \tau = t-W,\dots,t-1,\;
        i = 1,\dots,N
    \big\}.
\end{equation} 

This set aggregates all returns observed in the $W$ days prior to $t$ across all assets and serves as a representation for the recent cross-sectional behaviour of the market. The corresponding cross-sectional state is represented by
\begin{equation}
    \mathcal{P}_1(t)
    =
    \big\{
        r_{t,i} : i = 1,\dots,N
    \big\}.
\end{equation}

Both $\mathcal{P}_0(t)$ and $\mathcal{P}_1(t)$ are transformed into discrete probability distributions by histogram-based density estimation. Let $[a,b]$ denotes the lower and upper bounds of the range used for estimation, which in practice is restricted to the central quantiles of the pooled return distribution to limit the influence of extreme outliers.This interval is partitioned into $B$ bins with edges $\{a = \xi_0 < \xi_1 < \dots < \xi_B = b\}$. The empirical probability assigned to bin $b$ for the reference window and the corresponding cross-section is
given by
\begin{equation}
    p_b(t)
    =
    \frac{
        \#\big\{ x \in \mathcal{P}_0(t) : x \in [\xi_{b-1},\xi_b) \big\}
    }{
        \sum_{j=1}^B 
        \#\big\{ x \in \mathcal{P}_0(t) : x \in [\xi_{j-1},\xi_j) \big\}
    }, \end{equation}

    \begin{equation}
     q_b(t)
    =
    \frac{
        \#\big\{ y \in \mathcal{P}_1(t) : y \in [\xi_{b-1},\xi_b) \big\}
    }{
        \sum_{j=1}^B 
        \#\big\{ y \in \mathcal{P}_1(t) : y \in [\xi_{j-1},\xi_j) \big\}
    },
\end{equation}
where $\#\{\cdot\}$ denotes the cardinality operator. By construction,
$\sum_{b=1}^B p_b(t) = \sum_{b=1}^B q_b(t) = 1$ for all $t$.

For two discrete probability distributions $p(t) = (p_1(t),\dots,p_B(t))$ and $q(t) = (q_1(t),\dots,q_B(t))$, the HD is given as,
\begin{equation}
    H\big(p(t),q(t)\big)
    =
    \sqrt{
        \frac{1}{2}
        \sum_{b=1}^B
        \Big(
            \sqrt{p_b(t)} - \sqrt{q_b(t)}
        \Big)^2
    }.
    \label{eq:hellinger_def}
\end{equation}

The HD lies between $0 \leq H(p,q) \leq 1$, with $H(p,q) = 0$ if and only if $p \equiv q$ and $H(p,q)$ approaches $1$ when the two distributions assign probability mass to almost disjoint regions of the support. These properties of the HD make it suitable for detecting distributional shifts.

Using Eq.~\eqref{eq:hellinger_def}, we compute a univariate time series of rolling Hellinger distances for $t = W+1,\dots,T$,
\begin{equation}
    H_t = H\big(p(t), q(t)\big).
\end{equation}

Under normal market conditions, $H_t$ fluctuates around a baseline level, while sudden changes in market behaviour, such as crashes or regime shifts, appear as pronounced spikes. To identify such periods, we define an empirical decision threshold based on the mean and standard deviation of the HD series,
\begin{equation}
    H_D = \mu_H + 2\sigma_H,
\end{equation}
where $\mu_H$ and $\sigma_H$ denote the sample mean and standard deviation of $\{H_t\}$. Days for which $H_t > H_D$ are classified as the crash region.

In this study, we interpret $H\big(p(t), q(t)\big)$ as a measure of how strongly the cross-sectional return distribution at time $t$ deviates from the reference distribution formed over the preceding $W$ trading days. Large values of the Hellinger distance indicate significant changes in the distributional shape. Accordingly, time periods for which $H_t$ exceeds the decision threshold $T_D$ are classified as crash regions.

\subsubsection{Hilbert-Huang Transform}
\label{Hilbert spectrum}

The Hilbert–Huang Transform (HHT) is a data-driven technique designed to analyze nonlinear and nonstationary time series. Its first stage, Empirical Mode Decomposition (EMD), adaptively decomposes a signal into a finite set of oscillatory components known as Intrinsic Mode Functions (IMFs), each representing a characteristic time scale of the underlying process~\cite{huang1998empirical,rai2023detection}.

A component extracted from the time series is identified as an IMF if it satisfies the following criteria,
\begin{itemize}
    \item The number of local extrema and the number of zero crossings are equal or differ by at most one.
    \item The local mean, defined as the average of the upper and lower envelopes formed by the local maxima and minima, is zero at every point.
\end{itemize}

The EMD procedure is implemented as follows. For a given input time series $D_t$, the local maxima and minima are first identified, and spline interpolation is used to construct the upper envelope ($UE_t$) and lower envelope ($LE_t$), respectively. The local mean of these envelopes is then computed as
\begin{equation}
    m_t = \frac{UE_t + LE_t}{2}.
\end{equation}
Removing this mean from the original signal produces an updated series,
\begin{equation}
    ND_t = D_t - m_t.
\end{equation}

This sifting process is iteratively applied to $ND_t$ until the resulting signal satisfies the IMF conditions. The extracted component is then designated as the first intrinsic mode function, $IMF_1$. The residual signal is obtained as
\begin{equation}
    N_t = D_t - IMF_1,
\end{equation}

The same procedure is then applied iteratively to extract additional IMFs. The decomposition process continues until the remaining residual becomes a monotonic function, representing the long-term trend of the original time series. As a result, the original signal can be reconstructed as

\begin{equation}
    D_t = \sum_{j=1}^{n} IMF_j + \text{residue},
\end{equation}
where $n$ denotes the total number of extracted IMFs.

In the second stage of the HHT, each IMF is analyzed using the Hilbert transform to obtain instantaneous frequency information. The Hilbert transform of an IMF is defined as
\begin{equation}
    H(t) = \frac{1}{\pi} \, \text{P.V.} \int_{-\infty}^{\infty} \frac{\text{IMF}(\tau)}{t - \tau} \, d\tau,
\end{equation}
Here, P.V. is the Cauchy principal value.  We defined the instantaneous phase as~\cite{huang1998empirical}
\begin{equation}
    \phi(t) = \tan^{-1}\left( \frac{H(t)}{\text{IMF}(t)} \right),
\end{equation}

and the instantaneous frequency is given by
\begin{equation}
    \omega(t) = \frac{d\phi(t)}{dt}.
\end{equation}

The Hilbert spectrum $H(t,\omega)$ provides a time–frequency representation of the signal and is expressed as
\begin{equation}
    H(t,\omega) = \Re \left\{ \sum_{i} K_i(t) \, e^{j \int \omega(t) \, dt} \right\},
\end{equation}
Here, $K_i(t)$ denotes the instantaneous amplitude and $\Re{\cdot}$ denotes the real part.

From the Hilbert spectrum, the instantaneous energy is defined by the following equation,
\begin{equation}
    IE(t) = \int_{\omega} H^2(t,\omega) \, d\omega.
\end{equation}
To facilitate comparison across time, the instantaneous energy is normalized as
\begin{equation}
    IE_{N}(t) = \frac{IE(t)}{\max[IE(t)]}.
\end{equation}

In our study, we apply this framework to pinpoint the crash for the QUAD countries' stock index.

\subsection{Methodology for Network Construction: Conditional P-Threshold Mutual Information}

In this subsection, we describe the methodology used to construct stock dependency networks. Market-adjusted abnormal returns are first obtained using the Capital Asset Pricing Model (CAPM). Mutual information(MI) is then used to measure linear and non-linear dependencies between stocks, followed by permutation testing to assess statistical significance. The resulting conditional p-thresholded MI matrices are used as adjacency matrices for network analysis.

\subsubsection{Returns and Abnormal Returns}

Let $p_i(t)$ denote the daily closing price of stock $i$ at time $t$ $(i = 1, 2, \ldots, N, t = 1, 2, \ldots, T)$. The logarithmic return $r_i(t)$ of stock $i$ over time interval $\Delta t$ is defined as:

\begin{equation}
r_i(t) = \ln[p_i(t)] - \ln[p_i(t - \Delta t)]
\label{eq:log_return}
\end{equation}

In this study, we set $\Delta t = 1$, making $r_i(t)$ the daily return of stock $i$ at time $t$.

However, computing MI directly from raw stock returns $r_i(t)$ may not accurately reflect true inter-stock dependencies. A high MI value for a given stock pair does not automatically indicate a strong direct dependence, as it may be influenced by common market movements, macroeconomic factors, or systemic events \cite{longin1995correlation,lintner1975valuation,xu2017topological}. In such cases, the observed dependence may arise mainly from shared exposure to market-wide fluctuations rather than from intrinsic interactions between the stocks. To address this issue, we follow the CAPM framework and remove the systematic market component from stock returns. The resulting abnormal returns capture stock-specific behavior and allow for a more reliable analysis of pure inter-stock dependencies, reducing the impact of common market effects.

\subsubsection{Removal of Market Influence via CAPM Regression}

To isolate stock-specific components from common market movements, we employ the CAPM framework\cite {xu2017topological}. For each QUAD country, we use the respective benchmark market index, S\&P 500 for the United States, NIKKEI 225 for Japan, ASX 200 for Australia, and NIFTY 50 for India, to model and remove systematic market effects, as overall stock sentiment in each market is predominantly driven by these indices. For each stock $i$, we specify the market model:

\begin{equation}
r_i(t) = \alpha_i + \beta_i r_M(t) + \varepsilon_i(t), \quad \varepsilon_i(t) \sim \mathcal{N}(0, \sigma_i^2)
\label{eq:capm_model}
\end{equation}

\noindent where, $r_i(t)$ is return of stock $i$ at time $t$, $r_M(t)$ is the return of market index at time $t$, $\alpha_i$ is the stock-specific intercept (Jensen's alpha), $\beta_i$ is systematic risk coefficient, and $\varepsilon_i(t)$ is idiosyncratic error term, which is assumed to be independent and identically distributed (i.i.d.) normal.

The parameters $\alpha_i$ and $\beta_i$ are estimated using ordinary least squares (OLS) regression as
\begin{equation}
(\hat{\alpha}_i, \hat{\beta}_i)
=
\arg\min_{\alpha_i,\,\beta_i}
\sum_{t=1}^{T}
\left[
r_i(t) - \alpha_i - \beta_i r_M(t)
\right]^2 ,
\label{eq:ols_estimation}
\end{equation}
where $T$ is the total number of observations.

The market-adjusted abnormal return is then defined as
\begin{equation}
ar_i(t)
=
r_i(t)
-
\hat{\alpha}_i
-
\hat{\beta}_i r_M(t) .
\label{eq:abnormal_return}
\end{equation}

\noindent where, $ar_i(t)$ is abnormal return of stock $i$ at time $t$ and $\hat{\alpha}_i, \hat{\beta}_i$ is the OLS estimates of CAPM parameters.

These abnormal returns represent the portion of stock returns unexplained by market movements, enabling analysis of pure stock interdependencies.  We then quantify dependencies between these abnormal returns using MI, which captures both linear and non-linear relationships between stocks, as described in the following section.

\subsubsection{Mutual Information}

Mutual information (MI) measures both linear and non-linear dependencies between random variables $X_i$ and $X_j$ representing abnormal returns of stocks $i$ and $j$, making it superior to traditional correlation measures for capturing complex relationships in financial markets~\cite{guo2018development,fiedor2014networks,lahmiri2020renyi}. By applying MI to abnormal returns rather than raw returns, we ensure that the measured dependencies reflect genuine inter-stock relationships rather than spurious correlations induced by common market factors.

For two continuous random variables $X$ and $Y$, the MI is defined as~\cite{kraskov2004estimating,hacine2012low,steuer2002mutual}
\begin{equation}
I(X;Y) = \iint_{\mathbb{R}^2} p(x,y) \log\left(\frac{p(x,y)}{p(x)p(y)}\right) dx dy
\label{eq:mi_continuous}
\end{equation}

\noindent where, $p(x,y)$ is the joint probability density function of $X$ and $Y$, $p(x), p(y)$ is marginal probability density functions and $\log(\cdot)$ is natural logarithm.

MI can be expressed in terms of entropy as
\begin{equation}
I(X;Y) = H(X) + H(Y) - H(X,Y)
\label{eq:mi_entropy}
\end{equation}

\noindent where $H(X)$ and $H(Y)$ are marginal entropies and $H(X,Y)$ is joint entropy.

In this study, we used a histogram-based estimator with $n_{\text{bins}} = 16$ bins~\cite{hacine2013new}
\begin{equation}
\hat{I}(X;Y)
=
\sum_{k=1}^{n_{\text{bins}}}
\sum_{l=1}^{n_{\text{bins}}}
\hat{p}_{kl}
\log\left(
\frac{\hat{p}_{kl}}{\hat{p}_k \hat{p}_l}
\right),
\label{eq:mi_estimator}
\end{equation}

\noindent where, $\hat{p}{kl} = \frac{n{kl}}{N}$ represent empirical joint probability, $\hat{p}k = \sum{l=1}^{n_{\text{bins}}} \hat{p}{kl}$ is the empirical marginal probability of $X$, $\hat{p}l = \sum{k=1}^{n{\text{bins}}} \hat{p}{kl}$ is empirical marginal probability of $Y$, $n{kl}$ represent the number of observations in bin $(k,l)$,and $N$ is the total number of observations.

The bin boundaries are determined using equidistant binning over the range of observed values. This non-parametric estimator is consistent and converges to the true MI as $N \to \infty$.

\subsubsection{Significance Testing: Permutation Test}

To distinguish statistically significant dependencies from random noise, we employ a non-parametric permutation test that makes minimal distributional assumptions~\cite{good2005permutation}.

The hypothesis framework is:
\begin{equation}
\begin{aligned}
H_0 &: I(ar_i; ar_j) = 0 \quad 
H_1 &: I(ar_i; ar_j) > 0 \quad 
\end{aligned}
\label{eq:mi_hypothesis}
\end{equation}

For each stock pair $(i,j)$, the testing procedure is:

\begin{enumerate}
\item Compute observed MI on abnormal returns:
\begin{equation}
I_{\text{obs}} = \hat{I}(ar_i; ar_j)
\label{eq:observed_mi}
\end{equation}

\item Generate $N_{\text{perm}} = 100$ permuted samples by randomly shuffling one time series:
\begin{equation}
{ar_i^{(1)}, ar_i^{(2)}, \ldots, ar_i^{(N_{\text{perm}})}} \sim \text{Permutation}(ar_i)
\label{eq:permutation}
\end{equation}

\item Compute MI for each permuted sample:
\begin{equation}
I_{\text{perm}}^{(k)} = \hat{I}(ar_i^{(k)}; ar_j), \quad k = 1, \ldots, N_{\text{perm}}
\label{eq:permuted_mi}
\end{equation}

\item Calculate the empirical p-value:
\begin{equation}
p = \frac{\left|{k : I_{\text{perm}}^{(k)} \geq I_{\text{obs}}}\right| + 1}{N_{\text{perm}} + 1}
\label{eq:permutation_pvalue}
\end{equation}
\end{enumerate}
We reject $H_0$ at significance level $\alpha = 0.05$ if $p \leq \alpha$.

\subsubsection{Conditional P-Thresholded MI Matrix}
\label{Conditional P-Thresholded MI Matrix}

The conditional p-threshold mutual information matrix 
$MI = [mi_{ij}]_{N \times N}$ is constructed as
\begin{equation}
mi_{ij}
=
\begin{cases}
\hat{I}(ar_i; ar_j), & \text{if } p_{ij} \leq \alpha, \\
0, & \text{otherwise},
\end{cases}
\label{eq:thresholded_mi}
\end{equation}

\noindent where $p_{ij}$ denotes the permutation test $p$-value for stocks $i$ and $j$ and $\alpha = 0.05$ is the significance level. In this study, the conditional p-thresholded MI matrix is used to construct the heatmap, which helps us to direct comparison between the unconditional MI and the conditional p-thresholded MI.

\subsubsection{Transformation to Distance Matrix}

For network construction, the MI matrix is transformed into a distance matrix using a monotonic mapping that preserves the dependency structure.

The distance matrix $D = [d_{ij}]_{N \times N}$ is defined as
\begin{equation}
d_{ij}
=
\begin{cases}
\dfrac{1}{mi_{ij} + \epsilon}, & \text{if } mi_{ij} \neq 0, \\
\infty, & \text{if } mi_{ij} = 0,
\end{cases}
\label{eq:distance_transform}
\end{equation}

\noindent where $\epsilon = 10^{-9}$ is a small constant introduced for numerical stability.

The distance metric satisfies the following properties:
\begin{equation}
\begin{aligned}
d_{ij} &\geq 0, \\
d_{ij} &= d_{ji}, \\
d_{ij} &= 0 \iff i = j .
\end{aligned}
\label{eq:distance_properties}
\end{equation}

This distance formulation provides a suitable basis for MST construction, where smaller distances correspond to stronger dependencies.

\subsubsection{Minimum Spanning Tree}

From the distance matrix $D$, we construct an undirected weighted graph $G = (V, E, W)$ representing the stock dependency network, where $V = {v_1, v_2, \ldots, v_N}$ represents the set of nodes, each corresponding to an individual stock. The edge set $E$ consists of all pairs $(v_i, v_j)$ for which the distance $d_{ij}$ is finite, indicating statistically significant dependencies between stocks. The weight set $W$ assigns to each edge $(v_i, v_j) \in E$ a weight $w_{ij} = d_{ij}$, representing the transformed distance between the corresponding stock pair.

The graph $G$ may contain multiple connected components. In this study, we focus on the largest connected component to ensure network coherence.
\begin{equation}
G_{\text{LCC}} = (V_{\text{LCC}}, E_{\text{LCC}}, W_{\text{LCC}}),
\quad \text{where } 
V_{\text{LCC}} = \arg\max_{C \subseteq V,\, C \in \mathcal{C}(G)} |C|.
\label{eq:largest_component}
\end{equation}

The Minimum Spanning Tree (MST)~\cite{mantegna1999hierarchical} is extracted from $G_{\text{LCC}}$ using Prim’s algorithm~\cite{dutta2014development,huda2023modified}, which iteratively selects the edge with the smallest weight that connects a new node to the growing tree. This procedure ensures that all nodes in the largest connected component are spanned while minimizing the total sum of edge weights. As a result, the MST provides a sparse representation of the stock dependency network that preserves the strongest inter-stock dependencies and removes redundant connections.

\begin{algorithm}
\caption{Prim's Algorithm for MST Construction}
\begin{algorithmic}[1]
\REQUIRE Connected graph $G = (V, E, W)$ with $n = |V|$ nodes
\ENSURE Minimum Spanning Tree $T = (V, E_T)$
\STATE Initialize $E_T \gets \emptyset$
\STATE Initialize $U \gets \{v_1\}$ (arbitrary starting node)
\STATE Initialize min-priority queue $Q$ with all edges from $v_1$
\FOR {$i = 2$ to $n$}
\STATE Extract edge $(u,v)$ with minimum $w_{uv}$ from $Q$ where $u \in U$, $v \notin U$
\STATE $E_T \gets E_T \cup \{(u,v)\}$
\STATE $U \gets U \cup \{v\}$
\STATE Add all edges from $v$ to nodes not in $U$ to $Q$
\ENDFOR
\RETURN $T = (V, E_T)$
\end{algorithmic}
\label{alg:prim}
\end{algorithm}

The optimization is subject to the constraints that $T$ forms a tree (connected and acyclic), spans all nodes in $V_{\text{LCC}}$, and has the minimum total edge weight among all possible spanning trees. This construction yields a hierarchical backbone that captures the essential dependency structure while filtering out weaker connections. Based on the resulting conditional p-threshold MST, we compute a comprehensive set of network topological metrics to quantify the structural properties and hierarchical organization of the stock dependency network across different market periods.

\subsection{Topological features of the network}
To characterize the network structure in each market period, we compute topological features on the MST. This sparse representation, extracted from the largest connected component of the conditional p-threshold MI network, preserves the strongest dependencies while ensuring a connected, acyclic structure for robust analysis.

\subsubsection{Average Closeness Centrality}
Closeness centrality~\cite{freeman1979centrality} quantifies a node's closeness to all other nodes in the network. It reflects how efficiently a node can access or disseminate information across the network, with higher values indicating a more central and well-connected position.

For a node $i$, closeness centrality is defined as the reciprocal of the average shortest-path distance to all other nodes:
\begin{equation}
C_C(i) = \frac{N-1}{\sum_{j \neq i} l_{ij}},
\end{equation}
where $l_{ij}$ denotes the shortest-path length (in terms of number of edges) between nodes $i$ and $j$ in the MST, and $N$ is the total number of nodes in the connected component. The network-level average closeness centrality is then computed as
\begin{equation}
\langle C_C \rangle = \frac{1}{N} \sum_{i=1}^{N} C_C(i).
\end{equation}

This measure provides insight into the overall integration and potential speed of information propagation within the network.

\subsubsection{Average eccentricity}
Eccentricity~\cite{caldarelli2007scale} measures the maximum distance from a node to any other node in the network, capturing how peripheral a node is. Higher eccentricity values indicate weaker integration within the network.

The eccentricity of node $i$ is defined as
\begin{equation}
\varepsilon(i) = \max_{j \in V} l_{ij},
\end{equation}
and the average eccentricity is
\begin{equation}
\langle \varepsilon \rangle = \frac{1}{N} \sum_{i=1}^{N} \varepsilon(i).
\end{equation}

\subsubsection{Average eigenvector centrality}
Eigenvector centrality measures~\cite{bonacich1991simultaneous} the influence of a node by accounting not only for its number of connections but also for the importance of its neighbors. Nodes connected to highly central nodes receive higher scores, capturing hierarchical influence patterns in the network.

Mathematically, the eigenvector centrality of node $i$ is defined as
\begin{equation}
C_E(i) = \frac{1}{\lambda} \sum_{j=1}^{N} A_{ij} C_E(j),
\end{equation}
where $A_{ij}$ is the (weighted) adjacency matrix and $\lambda$ is the largest eigenvalue of $A$. The network-level average eigenvector centrality is given by
\begin{equation}
\langle C_E \rangle = \frac{1}{N} \sum_{i=1}^{N} C_E(i).
\end{equation}

\subsubsection{Average weighted degree}
Weighted degree measures the total strength of connections of a node by accounting for both the number of links and their associated weights. Nodes with higher weighted degree are more strongly connected to the network, reflecting their overall interaction intensity.

Mathematically, the weighted degree of node $i$ is defined as
\begin{equation}
k_i^{(w)} = \sum_{j=1}^{N} w_{ij},
\end{equation}
where $w_{ij}$ denotes the weight of the edge between nodes $i$ and $j$. The network-level average weighted degree is given by
\begin{equation}
\langle k^{(w)} \rangle = \frac{1}{N} \sum_{i=1}^{N} k_i^{(w)}.
\end{equation}

\subsubsection{Average betweenness centrality}
Betweenness centrality~\cite{freeman1979centrality} quantifies the extent to which a node lies on the shortest paths between other node pairs, reflecting its role as an intermediary or bridge in information transmission. Nodes with high betweenness can significantly influence the propagation of shocks across the network.

For node $i$, betweenness centrality is defined as
\begin{equation}
C_B(i) = \sum_{m \neq n \neq i} \frac{\sigma_{mn}(i)}{\sigma_{mn}},
\end{equation}
where $\sigma_{mn}$ is the total number of shortest paths between nodes $m$ and $n$ and $\sigma_{mn}(i)$ is the number of those paths passing through node $i$. The average betweenness centrality is
\begin{equation}
\langle C_B \rangle = \frac{1}{N} \sum_{i=1}^{N} C_B(i).
\end{equation}

\subsubsection{Average path length}
The average path length~\cite{albert2002statistical} measures the typical separation between node pairs in the network. Shorter path lengths imply a more compact and integrated structure.

It is computed as
\begin{equation}
\bar{l} = \frac{2}{N(N-1)} \sum_{m<n} l_{mn}.
\end{equation}

\subsubsection{Global efficiency}
Global efficiency~\cite{latora2001efficient} quantifies the overall efficiency of information transfer in the network by accounting for inverse shortest path lengths. Higher values indicate faster diffusion of information across the system.

It is defined as
\begin{equation}
E_{\text{glob}} = \frac{1}{N(N-1)} \sum_{i \neq j} \frac{1}{l_{ij}}.
\end{equation}

\subsubsection{Assortativity}

Assortativity~\cite{newman2003mixing} measures the tendency of nodes to connect with other nodes of similar degree. In financial networks, negative assortativity typically indicates a core--periphery structure, where highly connected core nodes preferentially link to weakly connected peripheral nodes.

For an undirected network, the degree assortativity coefficient $r$ is defined as
\begin{equation}
r =
\frac{
\sum_{i,j} A_{ij} k_i k_j
-
\frac{1}{2m}
\left( \sum_i k_i^2 \right)^2
}{
\sum_i k_i^3
-
\frac{1}{2m}
\left( \sum_i k_i^2 \right)^2
},
\label{eq:assortativity}
\end{equation}
where $A_{ij}$ denotes the binary adjacency matrix of the minimum spanning tree (MST), with $A_{ij}=1$ if nodes $i$ and $j$ are directly connected and $A_{ij}=0$ otherwise; $k_i$ is the degree of node $i$; and $m$ is the total number of edges in the MST ($m=N-1$ for a tree with $N$ nodes).

The assortativity coefficient satisfies $-1 \le r \le 1$. Positive values indicate assortative mixing, where nodes preferentially connect to others with similar degree, while negative values indicate disassortative mixing, where highly connected nodes tend to link with low-degree nodes. Values close to zero imply weak or no degree of correlation. In MST-based stock networks, strongly negative $r$ values reflect a pronounced core--periphery topology, a characteristic feature of financial markets during periods of stress.

\subsubsection{Algebraic connectivity}

Algebraic connectivity~\cite{fiedler1973algebraic}, denoted by $\lambda_2$, is defined as the second smallest eigenvalue of the weighted graph Laplacian matrix $L_w = D_w - A_w$, where $A_w(i,j)=w_{ij}$ is the weighted adjacency matrix and $D_w(i,i)=\sum_j w_{ij}$ is the weighted degree matrix. It can be equivalently expressed through the Rayleigh quotient as
\begin{equation}
\lambda_2 = \min_{\substack{x \neq 0 \\ x \perp \mathbf{1}}} \frac{x^T L_w x}{x^T x}.
\end{equation}

For the MST constructed from conditional $p$-threshold mutual information, the edge weights are defined as distances $w_{ij}=d_{ij}=1/(mi_{ij}+\epsilon)$. Higher values of $\lambda_2$ indicate stronger network cohesion and robustness, whereas lower values reflect increased structural fragility.

\subsubsection{Tree length}
The tree length corresponds to the total distance of edges in the minimum spanning tree (MST)~\cite{mantegna1999hierarchical}, representing the backbone of the network. Shorter tree length implies stronger overall connectivity.

It is defined as
\begin{equation}
L_{\text{tree}} = \sum_{(i,j) \in T} d_{ij}.
\end{equation}

\subsection{Core--Periphery Analysis}

\label{core--pha}
Core--periphery analysis provides a compact representation of market organization by distinguishing a densely connected core of influential stocks from a sparsely connected periphery of weaker stocks~\cite{borgatti2000models,ansari2025uncovering,ansari2025comprehensive}. In financial networks, core stocks typically correspond to large, systemically important firms that exert broad influence on market dynamics. The following measures quantify the structural concentration and fragility of the network.

\subsubsection{Core Concentration (Eigenvector--Closeness Ratio)}
We introduce a ratio-based measure of core concentration defined as the ratio of the network-average eigenvector centrality to the network-average closeness centrality,
\begin{equation}
C_{\text{core}} = \frac{\langle C_E \rangle}{\langle C_C \rangle},
\label{eq:core_concentration_ratio}
\end{equation}
where $\langle C_E \rangle$ and $\langle C_C \rangle$ denote the average eigenvector and closeness centralities, respectively.

This ratio captures the relative strength of core influence compared to overall network integration. Eigenvector centrality reflects the concentration of influence through connections to highly influential nodes, thereby serving as a proxy for core dominance, while closeness centrality measures the efficiency of global connectivity and the degree of peripheral integration. Higher values of $C_{\text{core}}$ indicate that influence is more strongly concentrated within central nodes relative to the level of network-wide integration, consistent with a segmented core–periphery structure. Conversely, lower values suggest a more distributed influence pattern accompanied by higher integration, corresponding to a flatter and more homogeneous network hierarchy. Temporal increases in the ratio signal a strengthening of core dominance that outpaces gains in global integration, whereas decreases indicate a diffusion of core influence as the network becomes more uniformly connected.

\subsubsection{Periphery Fragility}
Periphery fragility quantifies the vulnerability of the network's peripheral structure. It is computed as the ratio of the absolute assortativity to algebraic connectivity:

\begin{equation}
F_{\text{peri}} = \frac{|r|}{\lambda_2}
\label{eq:periphery_fragility}
\end{equation}

where:
\begin{itemize}
\item $|r|$ is the absolute value of the assortativity coefficient \cite{newman2003mixing}, which measures the tendency of nodes to connect with similar nodes. In core-periphery structures, $r$ is typically negative (core nodes connect to peripheral nodes), so $|r|$ captures the strength of this disassortative mixing.
\item $\lambda_2$ is the algebraic connectivity (second smallest eigenvalue of the Laplacian), which measures the overall robustness of the network \cite{Fiedler1973}.
\end{itemize}

This ratio captures the trade-off between structural segregation (high $|r|$) and network robustness (high $\lambda_2$). Higher values indicate greater fragility, where strong core-periphery segregation coexists with weak overall connectivity. While the individual components assortativity and algebraic connectivity are well-established network metrics \cite{newman2003mixing, Fiedler1973}, their combination into a fragility index specifically for financial network peripheries represents a methodological contribution of this study for assessing systemic vulnerability during market stress.

\subsection{Gutenberg--Richter Power Law}
\label{method:GR}

The Gutenberg--Richter (GR) power law is a classical empirical relation in geophysics describing the statistical distribution of earthquake magnitudes~\cite{gutenberg1942earthquake}. It relates the cumulative number of events \(N(M)\) with magnitude greater than or equal to \(M\) to the magnitude itself and is expressed as
\begin{equation}
\log_{10} N(M) = a - b M ,
\label{Gr_eq}
\end{equation}
where \(a\) and \(b\) are positive constants and \(b\) represents the slope of the scaling relation.

In financial markets, the GR framework has been adapted to examine aftershock-like volatility dynamics following major market crashes~\cite{siokis2012stock,selccuk2004financial}. In this study, the market crash is treated as an exogenous mainshock and the analysis is restricted to the post-crash period. The magnitude \(M\) of an aftershock event is defined as the absolute logarithmic price difference between consecutive local peaks and troughs:
\begin{equation}
M_i = \left| \log_{10}(P_{\text{peak},i}) - \log_{10}(P_{\text{trough},i}) \right|,
\end{equation}
where \(P_{\text{peak},i}\) and \(P_{\text{trough},i}\) denote adjacent local maxima and minima in the closing price series. This extremum-based definition captures discrete post-crash volatility fluctuations without imposing arbitrary thresholds.

The parameter \(b\) characterizes the scaling behavior of post-crash volatility events. Larger values of \(b\) indicate a faster decay in the frequency of large-magnitude fluctuations, while smaller values suggest a relatively higher occurrence of large volatility events. The intercept \(a\) reflects the overall number of post-crash events and is independent of their magnitude distribution.

\section{Data Description}
\label{Data_description}

We analyze the top 200 stocks by market capitalization from each of the QUAD countries, the United States, Japan, Australia, and India, to examine the evolution of nonlinear dependencies during the COVID-19 market crash. Daily closing price data for the period spanning September 1, 2019, to July 15, 2020, were obtained via the yfinance Python library from Yahoo Finance~\cite{yahoofinance}. This interval encompasses the pre-crash, crash, and post-crash periods of the COVID-19-induced financial crisis, providing a natural stress environment to study how stock-specific dependencies reorganize under extreme market conditions after filtering out common market factors.

For each market, the constituent stocks of the respective benchmark index, S\&P 500 (USA), NIKKEI 225 (Japan), ASX 200 (Australia), and NIFTY 50 (India), served as the initial set. From these, the 200 largest companies by average market capitalization over the sample period were selected to ensure liquidity and market representativeness. Daily closing prices were chosen for their stability and wide adoption in financial network analysis, and were transformed into logarithmic returns for subsequent analysis. The selection of QUAD countries provides a strategically diverse sample, with diverse economies across distinct geographical regions. This heterogeneity allows us to test whether the topological signatures of crisis identified by our conditional p-threshold MI framework represent universal patterns of systemic stress.

\section{Results and Discussion}
\label{Results}
This section presents and discusses the empirical findings of our study, organized as follows. First, Section~\ref{Market Crash Detection and Characterization} identifies and characterizes the COVID-19 market crash using the HD and HS, respectively, dividing the data into pre-crash, crash, and post-crash periods. Next, Section~\ref{Condition P-threshold Mutual Information_heatmap} compares MI with the conditional p-threshold MI, demonstrating the importance of removing market-wide effects and applying statistical filtering. Section~\ref{Conditional P-Threshold Network Dynamics} then examines the topological evolution of the conditional p-threshold MI networks across the three periods, with particular focus on core--periphery reconfiguration. Section~\ref{After_shock} investigates post-crash aftershock dynamics using the GR law. Finally, Section~\ref{conclusion} presents the conclusions of the study and outlines directions for future research.
 
\subsection{Market Crash Detection and Characterization}
\label{Market Crash Detection and Characterization}

In this section, we employ two complementary methodologies to identify and characterize market crashes across the QUAD countries during the COVID-19 period. 
We first detect crash periods using a rolling HD. This HD method captures structural changes in the cross-sectional return distributions of groups of stocks. We then characterize the dynamic behavior of the market during these periods using the HS applied to the benchmark stock indices of the QUAD countries. Together, these approaches provide a robust method for detecting the COVID-19 market crash.

\subsubsection{Crash detection using Hellinger Distance}

We identify the timing of the COVID-19-induced stock market crash of the QUAD countries, namely, the United States, Japan, Australia, and India, using the rolling HD as described in the Section~\ref{subsec:hellinger_method}. The HD provides a distribution-based measure of market instability by quantifying the divergence between two probability distributions. Unlike pointwise measures of volatility or correlation, it captures changes in the entire cross-sectional return distribution, making it particularly suitable for detecting systemic structural breaks in financial markets. The HD at time $t$ measures the divergence between the cross-sectional distribution of stock returns constructed from a rolling historical window $(t-60,\dots,t-1)$ and the cross-sectional return distribution observed on day $t$. A sudden increase in HD reflects an abrupt redistribution of returns across stocks, indicating heightened dispersion and synchronized extreme movements. Such behavior is characteristic of market crash periods, during which the normal cross-sectional market structure breaks down due to systemic shocks.

\begin{figure}[!t]
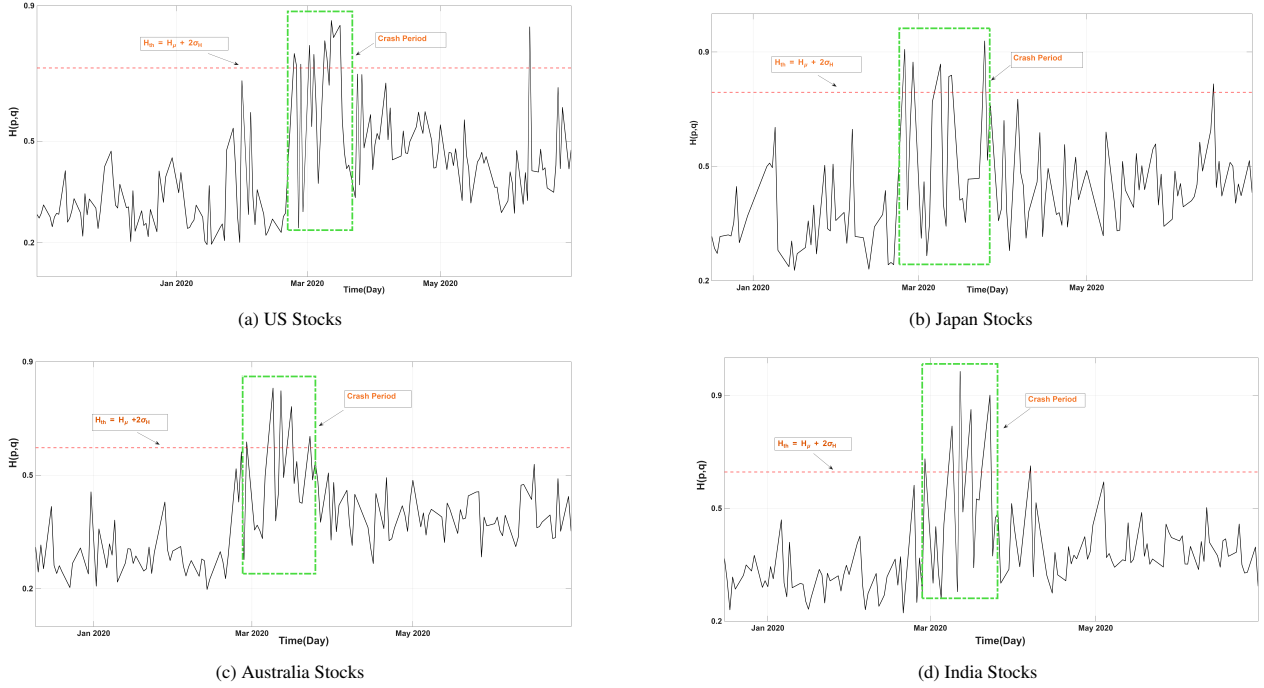

    \centering
    \begin{subfigure}[b]{0.45\textwidth}
        \centering
        \includegraphics[width=\textwidth]{US_Hill_group.png}
        \caption{US Stocks}
        \label{fig:US_Hel1}
    \end{subfigure}
    \hfill
    \begin{subfigure}[b]{0.45\textwidth}
        \centering
        \includegraphics[width=\textwidth]{Japan_Hill_group.png}
        \caption{ Japan Stocks}
        \label{fig:Japan_Hel2}
    \end{subfigure}
    \hfill

 \vspace{0.3cm}
    
    \begin{subfigure}[b]{0.45\textwidth}
        \centering
        \includegraphics[width=\textwidth]{Aus_hill_group.png}
        \caption{Australia Stocks}
        \label{fig:Aus_Hel3}
    \end{subfigure}
     \hfill
    \begin{subfigure}[b]{0.45\textwidth}
        \centering
        \includegraphics[width=\textwidth]{India_group_Hil.png}
        \caption{India Stocks}
        \label{fig:India_Hel4}
    \end{subfigure}
    \caption{
    Rolling Hellinger distance(HD) for the top 100 market-capitalization stocks from the QUAD countries. The x-axis represents time, while the y-axis shows the HD. The red dashed line indicates the $2\sigma$ threshold (\( H_D = \mu_H + 2\sigma_H \)), with the green rectangular box showing the period when HD exceeds this threshold. Plot (a) shows the HD for 100 U.S. stocks, Plot (b) for 100 Japanese stocks, Plot (c) for 100 Australian stocks, and Plot (d) for 100 Indian stocks. All four markets exhibit a pronounced and synchronized increase in HD during February--March 2020, indicating a systemic breakdown in cross-sectional market structure, which represents the COVID-19 market crash.
}
    \label{fig:Hel}
\end{figure}

Fig.~\ref{fig:Hel} shows the rolling HD for groups of stocks across the QUAD countries. The x-axis represents time, while the y-axis denotes the magnitude of the HD. 
Figs.~\ref{fig:US_Hel1},~\ref{fig:Japan_Hel2},~\ref{fig:Aus_Hel3} and~\ref{fig:India_Hel4} show the HD calculated from the top 100 market-capitalization stocks of the U.S., Japan, Australia, and Indian stock markets, respectively. To identify crash periods, we define a threshold as $ H_D = \mu_H + 2\sigma_H $, where $\mu_H$ and $\sigma_H$ denote the mean and standard deviation of the rolling HD, respectively. The time interval during which the HD exceeds this threshold is classified as a market crash period. From Fig.~\ref{fig:Hel}, we see a pronounced spike in the HD values crossing the threshold during February--March 2020. This spike region is highlighted by the green rectangular box.  Hence, we conclude that the HD method successfully detects the COVID-19 stock market crash across all QUAD countries. This synchronized threshold crossing indicates a sudden and statistically significant change in the cross-sectional return distributions of stocks across all markets. The peaks during this interval reflect a breakdown of normal market structure, characterized by increased return dispersion and collective extreme movements among stocks. Based on these threshold crossings, we divide the time series data into three periods: a pre-crash period, defined as the interval before HD exceeds the threshold; a crash period, defined as the interval during which HD remains above the threshold; and a post-crash period, defined as the interval after HD falls back below the threshold. This data-driven segmentation ensures a consistent event window for subsequent analysis and allows for a systematic comparison of market structure before, during, and after the crash

\subsubsection{Crash characterization using the Hilbert Spectrum}

The rolling HD provides a robust, data-driven identification of the timing and duration of systemic market disruptions by detecting abrupt changes in cross-sectional return distributions. However, while HD effectively signals when a market-wide structural break occurs, it does not describe the internal temporal dynamics of the market during the crash period. In particular, it does not capture how the intensity, frequency and persistence of market fluctuations evolve over time. Crucially, the HD cannot distinguish between a single day of extreme dislocation and a sustained period of chaotic, high-amplitude oscillations; both would manifest as a high HD value. To characterize this internal market dynamics and to further validate the crash period identified by the HD, we employ the HHT on the benchmark stock market indices of the QUAD countries, namely  S\&P~500, NIKKEI~225, ASX~200 and NIFTY~50 indices. The key advantage of the HS in this context is its ability to reveal the time-varying persistence and spectral composition of extreme movements. While the HD indicates a break in the cross-sectional structure, the HS quantifies whether that break corresponds to transient volatility or a prolonged regime of high-energy, chaotic market states.

\begin{figure}[!t]
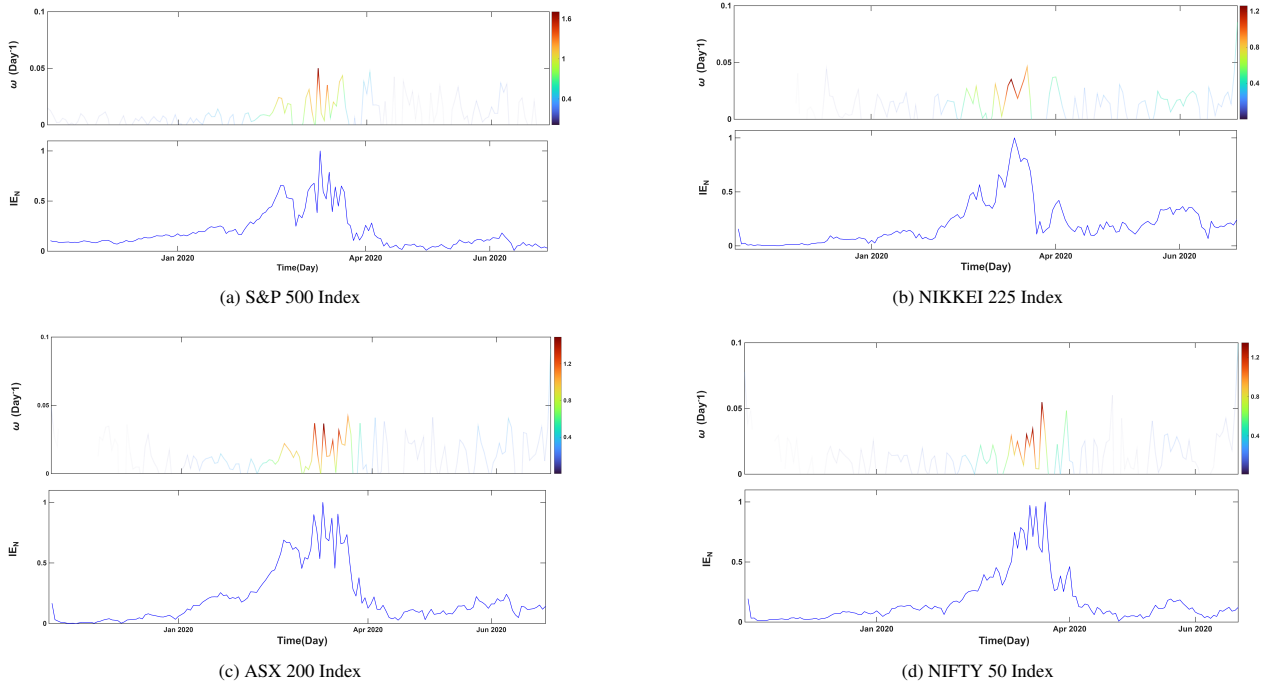

    \centering
    \begin{subfigure}[b]{0.45\textwidth}
        \centering
        \includegraphics[width=\textwidth,height= 8cm,keepaspectratio]{USA_index.png}
        \caption{S\&P 500 Index}
        \label{fig:US_Hil}
    \end{subfigure}
    \hfill
    \begin{subfigure}[b]{0.45\textwidth}
        \centering
        \includegraphics[width=\textwidth,height= 8cm,keepaspectratio]{Japan_index.png}
        \caption{NIKKEI 225 Index}
        \label{fig:Japan_Hil}
    \end{subfigure}

    \vspace{0.3cm}

    \begin{subfigure}[b]{0.45\textwidth}
        \centering
        \includegraphics[width=\textwidth,height= 8 cm,keepaspectratio]{Aus_indx.png}
        \caption{ASX 200 Index}
        \label{fig:Aus_Hil}
    \end{subfigure}
    \hfill
    \begin{subfigure}[b]{0.45\textwidth}
        \centering
        \includegraphics[width=\textwidth,height= 8 cm,keepaspectratio]{India_indx.png}
        \caption{NIFTY 50 Index}
        \label{fig:India_Hil}
    \end{subfigure}

\caption{
    Hilbert spectrum in the top panels and normalized instantaneous energy in the bottom panels of benchmark stock market indices for the QUAD countries. Plots (a), (b), (c), and (d)  represent the HS of the S\&P 500 Index (USA), NIKKEI 225 Index (Japan),  ASX 200 Index (Australia), and the NIFTY 50 Index (India), respectively. The x-axis represents time, while the y-axis in each subplot of the upper plot denotes the instantaneous frequency. In the lower plots, the y-axis represents the normalized instantaneous energy. A pronounced concentration of high-energy components is observed during Feb-Mar 2020 in all four indices, indicating heightened market volatility during the COVID-19 market crash.
}
   
    \label{fig:Hilbert}
\end{figure}

Fig.~\ref{fig:Hilbert} presents the HS and the corresponding normalized instantaneous energy for the S\&P~500, NIKKEI~225, ASX~200, and NIFTY~50 indices. In each plot, the 
x-axis denotes time, while the color intensity represents the magnitude of instantaneous energy. Across all four stock indices, a pronounced concentration of high instantaneous energy appears during the February–March 2020 period. This period coincides exactly with the crash period identified by the HD-based method. These high-energy regions, visible as intense red patches in the HS, indicate large-amplitude oscillations in the stocks. The normalized instantaneous energy further shows a sharp increase over this period, confirming the crash. The simultaneous energy surges across geographically distinct markets underscore the global and synchronized nature of the shock. Using the crash window identified from the HS and HD analyses, the data were divided into three periods: pre-crash (01-11-2019–14-02-2020), crash (15-02-2020–25-03-2020), and post-crash (26-03-2020–15-07-2020).

Using both HD and HS, we reliably identify the COVID-19 market crash. HD detects the onset and duration of the crash by capturing statistically significant shifts in the cross-sectional return distribution, indicating systemic structural change. HS then confirms this period by measuring the intensity and persistence of market fluctuations through instantaneous energy dynamics, confirming a sustained high-energy market state. Together, these methods provide a robust identification and characterization of the COVID-19 crash, forming a basis for subsequent network-based analysis.

\subsection{Comparison between the MI and Condition P-threshold MI}
\label{Condition P-threshold Mutual Information_heatmap}

\begin{figure}[!t]
    \centering

    \begin{subfigure}[b]{0.32\textwidth}
        \centering
        \includegraphics[width=\textwidth]{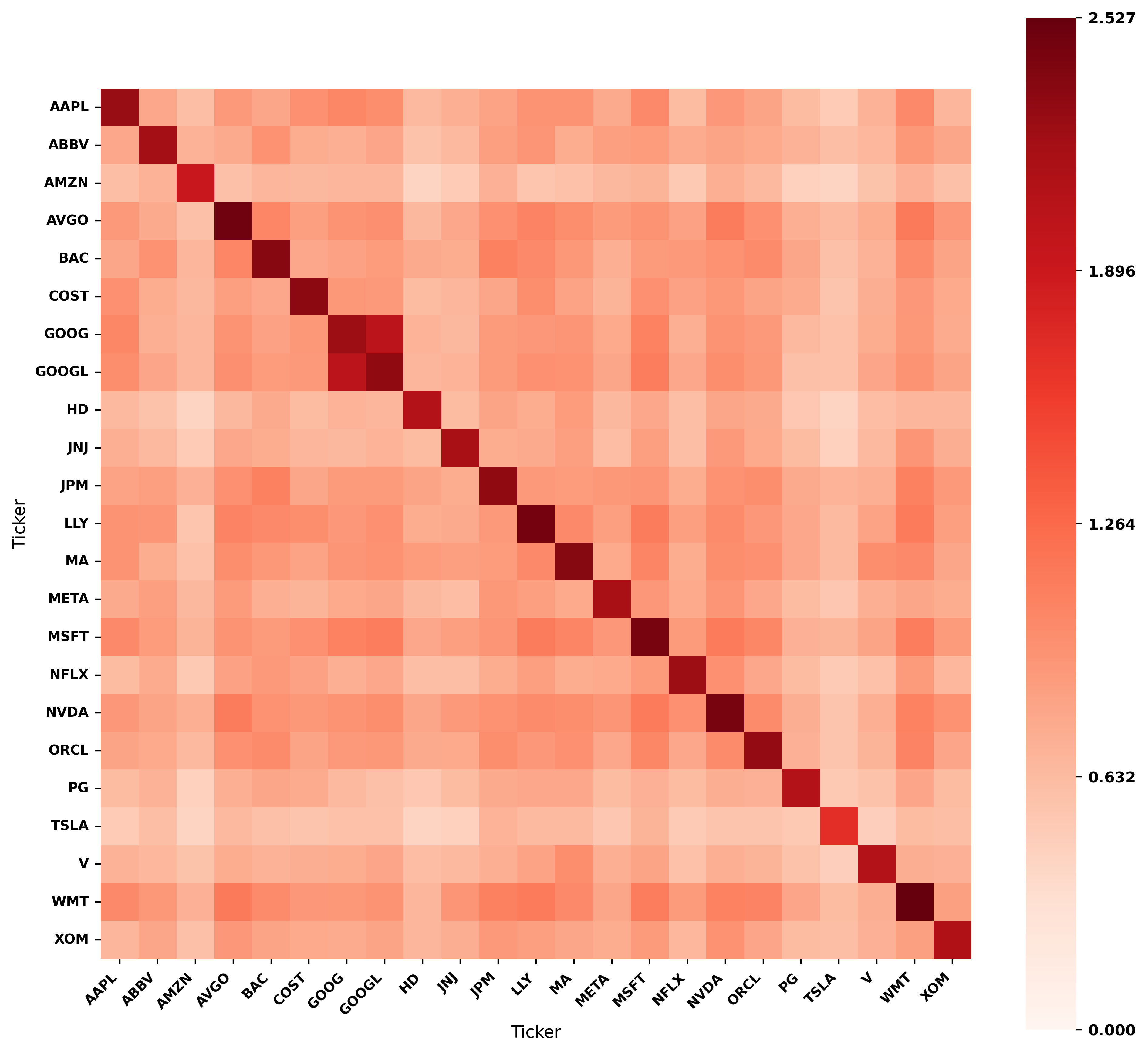}
        \caption{Pre-Crash (MI)}
        \label{1}
    \end{subfigure}
    \hfill
    \begin{subfigure}[b]{0.32\textwidth}
        \centering
        \includegraphics[width=\textwidth]{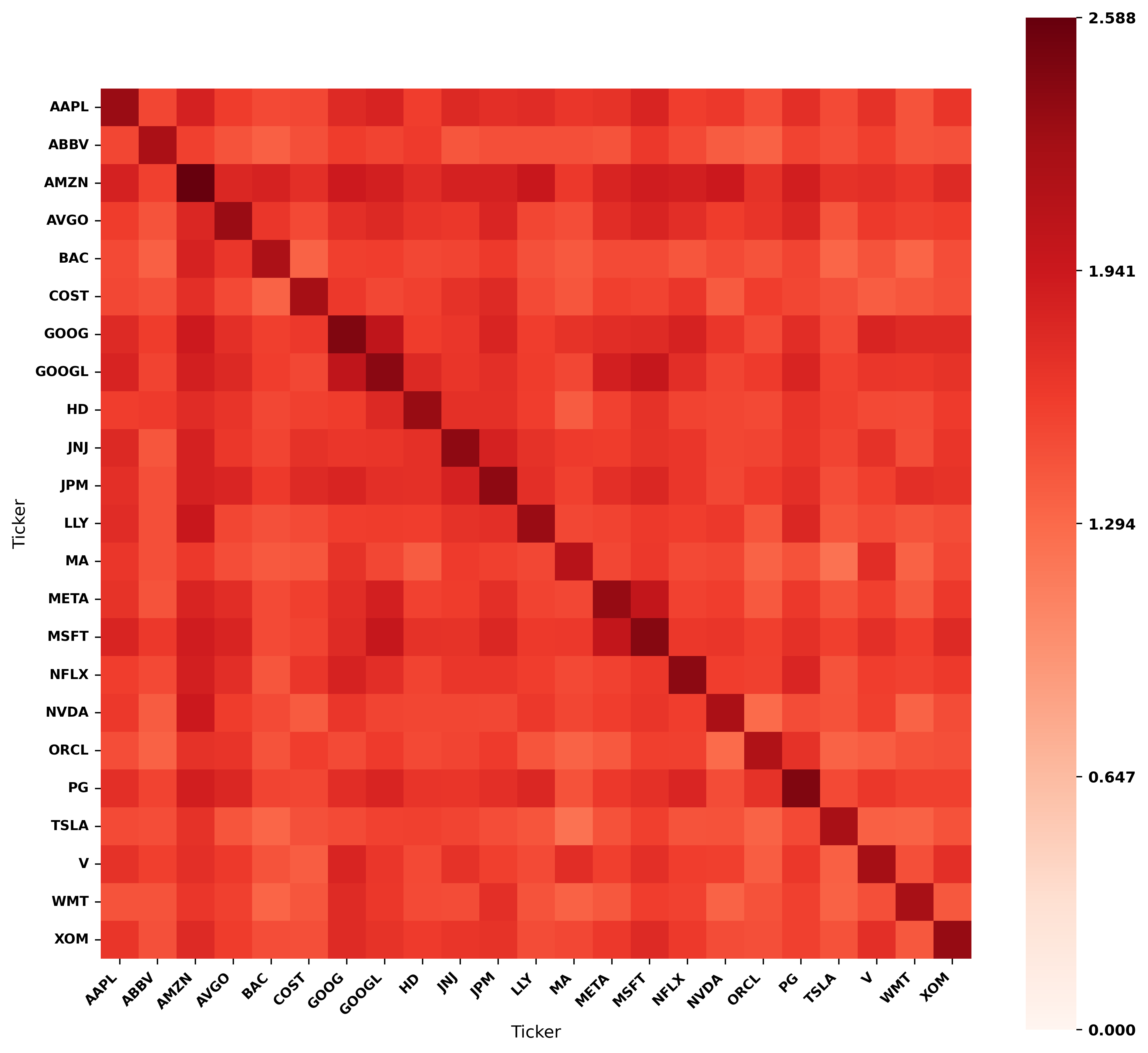}
        \caption{Crash (MI)}  \label{2}
    \end{subfigure}
    \hfill
    \begin{subfigure}[b]{0.32\textwidth}
        \centering
        \includegraphics[width=\textwidth]{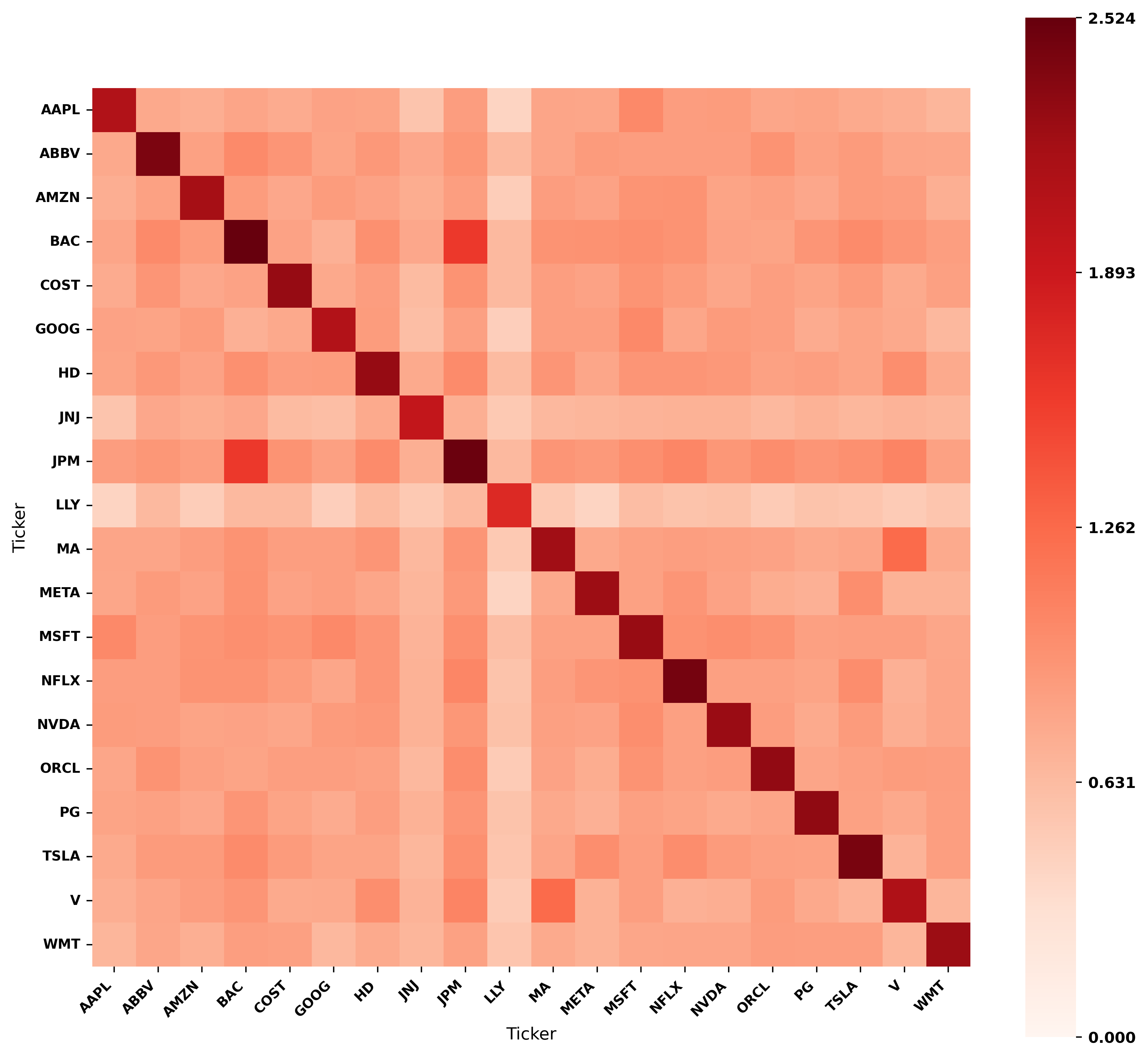}
        \caption{Post-Crash (MI)} \label{3}
    \end{subfigure}

    \vspace{0.3cm}

    \begin{subfigure}[b]{0.32\textwidth}
        \centering
        \includegraphics[width=\textwidth]{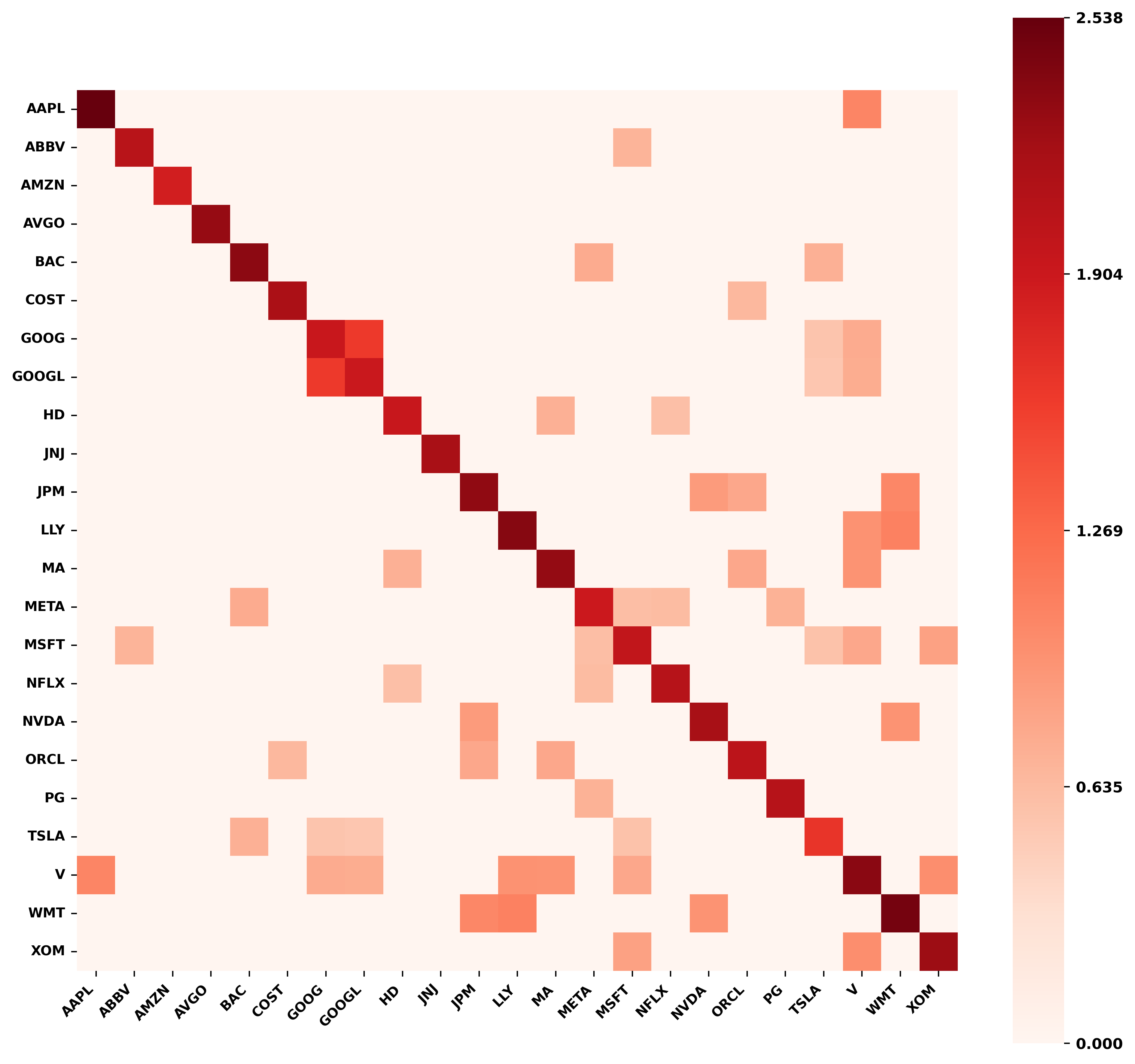}
        \caption{Pre-Crash (Cond.\ P-threshold MI)}  \label{4}
    \end{subfigure}
    \hfill
    \begin{subfigure}[b]{0.32\textwidth}
        \centering
        \includegraphics[width=\textwidth]{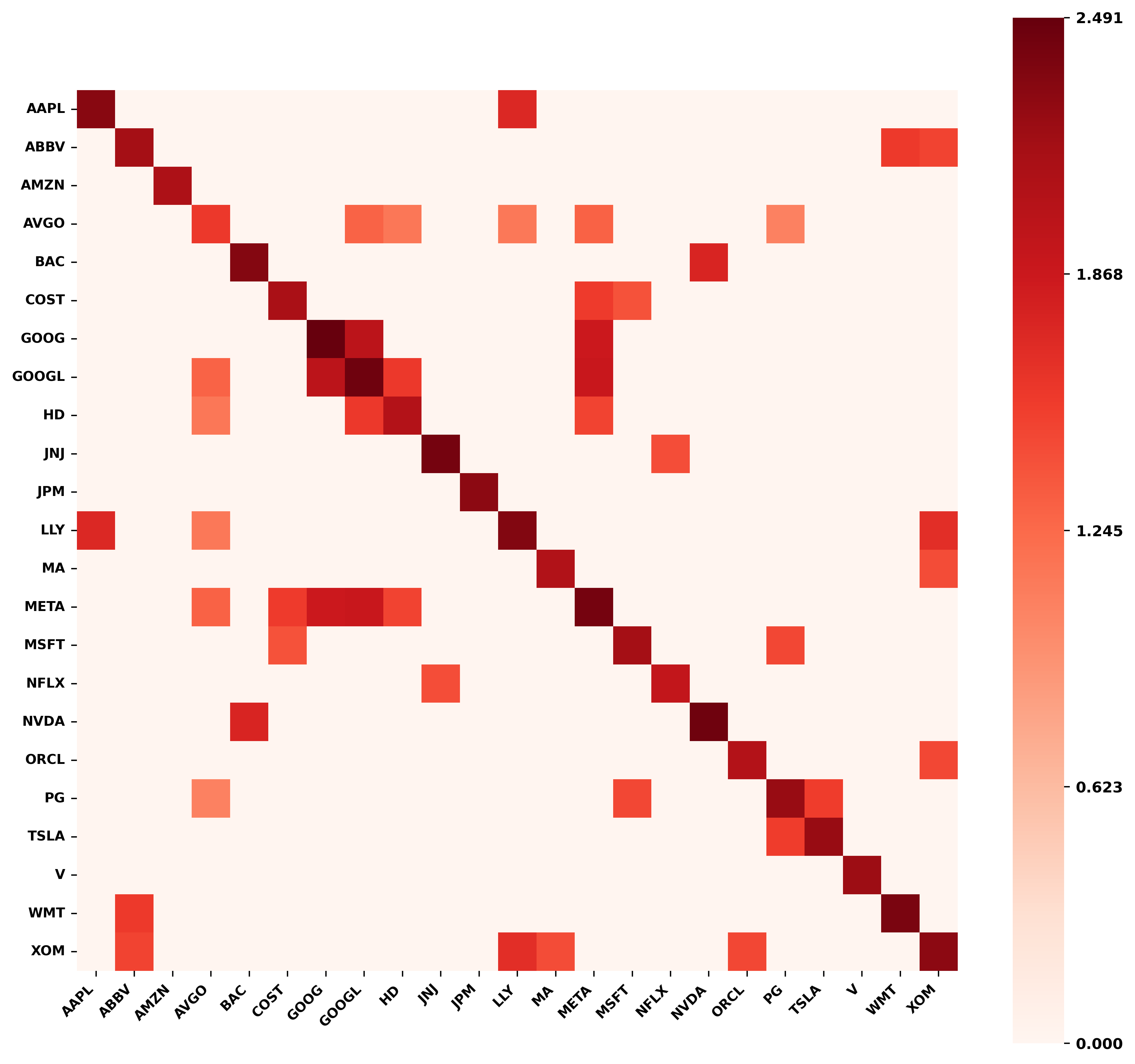}
        \caption{Crash (Cond.\ P-threshold MI)}  \label{5}
    \end{subfigure}
    \hfill
    \begin{subfigure}[b]{0.32\textwidth}
        \centering
        \includegraphics[width=\textwidth]{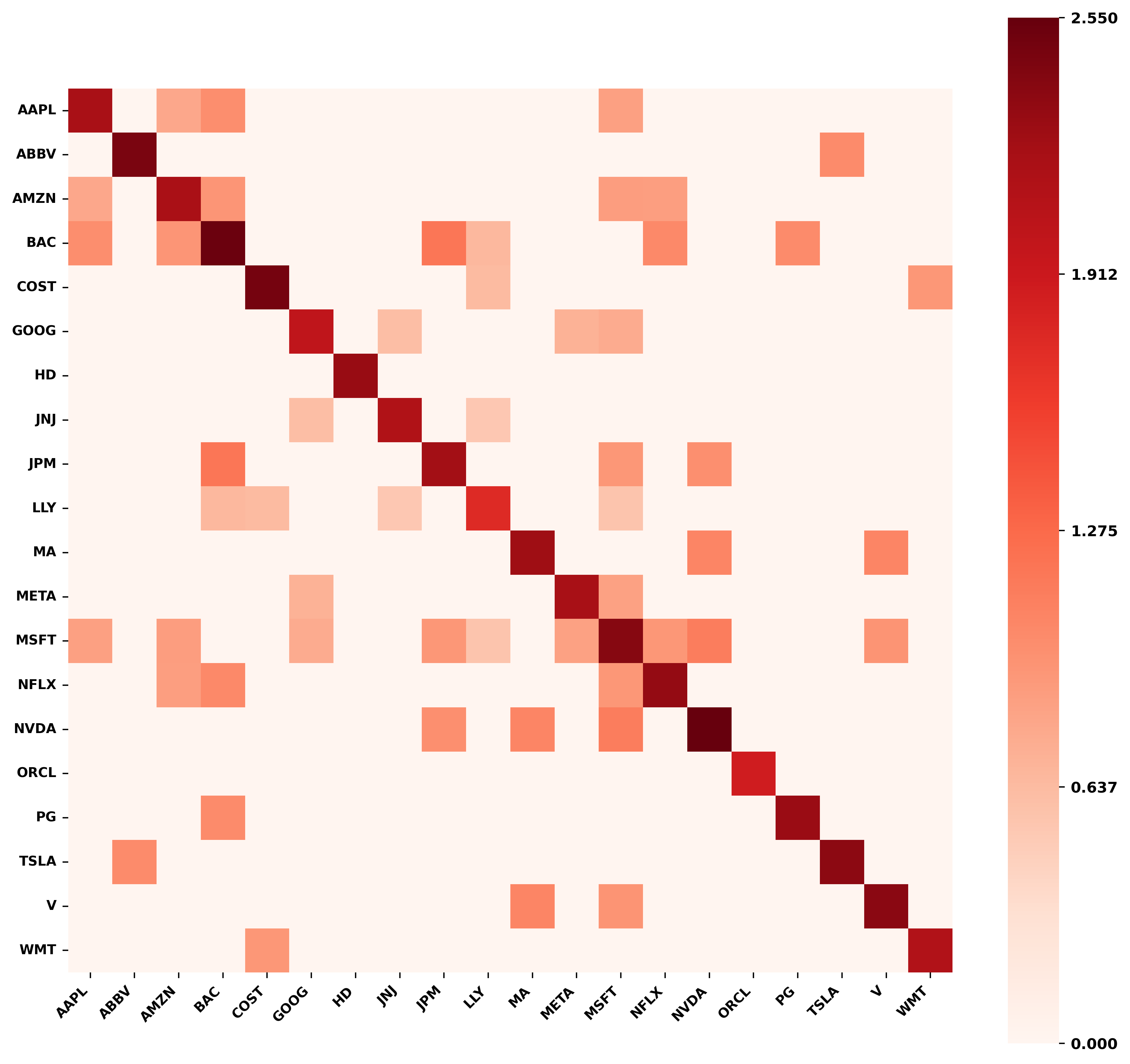}
        \caption{Post-Crash (Cond.\ P-threshold MI)}  \label{6}
    \end{subfigure}

    \caption{Mutual information heatmaps for the 25 largest U.S. stocks across different market regimes. Plot (a)--(c) shows mutual information heatmaps, which are dominated by common market effects and display dense connectivity, especially during the crash. Plot (d)--(f) presents residual-based, significance-filtered MI heatmaps obtained using the conditional P-threshold method. The conditional P-threshold MI heatmaps show a sparse structure, indicating the true nonlinear dependencies between stocks after removing the market effect, which reveals the underlying direct interactions among stocks.} 
    
    \label{fig:6subplots}
\end{figure}

Fig.~\ref{fig:6subplots} shows the heatmap of the MI and conditional p-threshold MI among the top 25 stocks of the U.S. stock market across the pre-crash, crash, and post-crash periods using the method as described in Section\ref{Conditional P-Thresholded MI Matrix}. In these figures, the color spectrum represents the magnitude of MI. Darker shades of red indicate stronger nonlinear dependence between stock pairs, while lighter shades correspond to weaker dependence. Figs.~\ref{1}, \ref{2}, and \ref{3} present the MI heatmaps calculated directly from stock returns, while Figs.~\ref{4}, \ref{5}, and \ref{6} show the residual-based, significance-filtered MI matrices obtained after removing the market index effect and computing MI, followed by the application of the conditional p-threshold at $\alpha = 0.05$. The significance testing is carried out using a permutation-based procedure with 100 permutations per stock pair to retain only statistically meaningful nonlinear dependencies.

From Figs.~\ref{1}, \ref{2}, and \ref{3}, dense MI connectivity is observed across all three periods. The heatmap corresponding to the crash period, as shown in Fig.~\ref{2}, exhibits the highest intensity compared to the pre-crash and post-crash periods, indicating stronger nonlinear dependence during the crash. This suggests enhanced market-wide synchronization driven by common shocks. However, similar to correlation-based measures, MI is also sensitive to overall market movements, and such dense connectivity may reflect a combination of direct stock--stock interactions and indirect dependencies induced by common market factors. Therefore, the removal of these common factors is necessary to analyze the pure relationships between stocks.

To identify purely dependent stock pairs, it is necessary to examine the effect of removing the influence of the market index~\cite{xu2017topological}. This allows us to uncover direct nonlinear relationships between stocks. In our analysis, we removed the effect of the S\&P~500 market index from each stock of the U.S. After removing the market index effect, the number of significant conditional p-threshold MI links decreased across all groups, as clearly shown in Figs.~\ref{4},~ \ref{5} and ~\ref{6}. This reduction in the number of significant links suggests that many dependencies identified without conditioning were driven by overall market movements rather than by direct stock--stock relationships. During the pre-crash period, the conditional p-threshold MI heatmap shown in Fig.~\ref{3} appears sparse, indicating that many apparent dependencies are driven by shared market influences rather than intrinsic nonlinear interactions. During the crash period, the heatmap shown in Fig.~\ref{5} exhibits a clear increase in intensity and the number of significant MI values, reflecting stronger nonlinear dependence among stocks under extreme market stress. This demonstrates that the conditional p-threshold approach captures the amplification of genuine nonlinear interdependencies during these periods. In the post-crash period, the heatmap shown in Fig.~\ref{6} shows a reduction in overall intensity relative to the crash period, while remaining higher than in the pre-crash period. This ordering,
\begin{equation}
    \text{pre-crash} < \text{post-crash} < \text{crash},
\label{compare_pre_crash_post_eq}
\end{equation}
suggests that although market-wide synchronization weakens after the crash, the system does not immediately return to its pre-crisis state. These results indicate that the aftershock effects and memory effects of the main crash.

\begin{figure}[!t]
    \centering

    \begin{subfigure}[b]{0.32\textwidth}
        \centering
        \includegraphics[width=\textwidth]{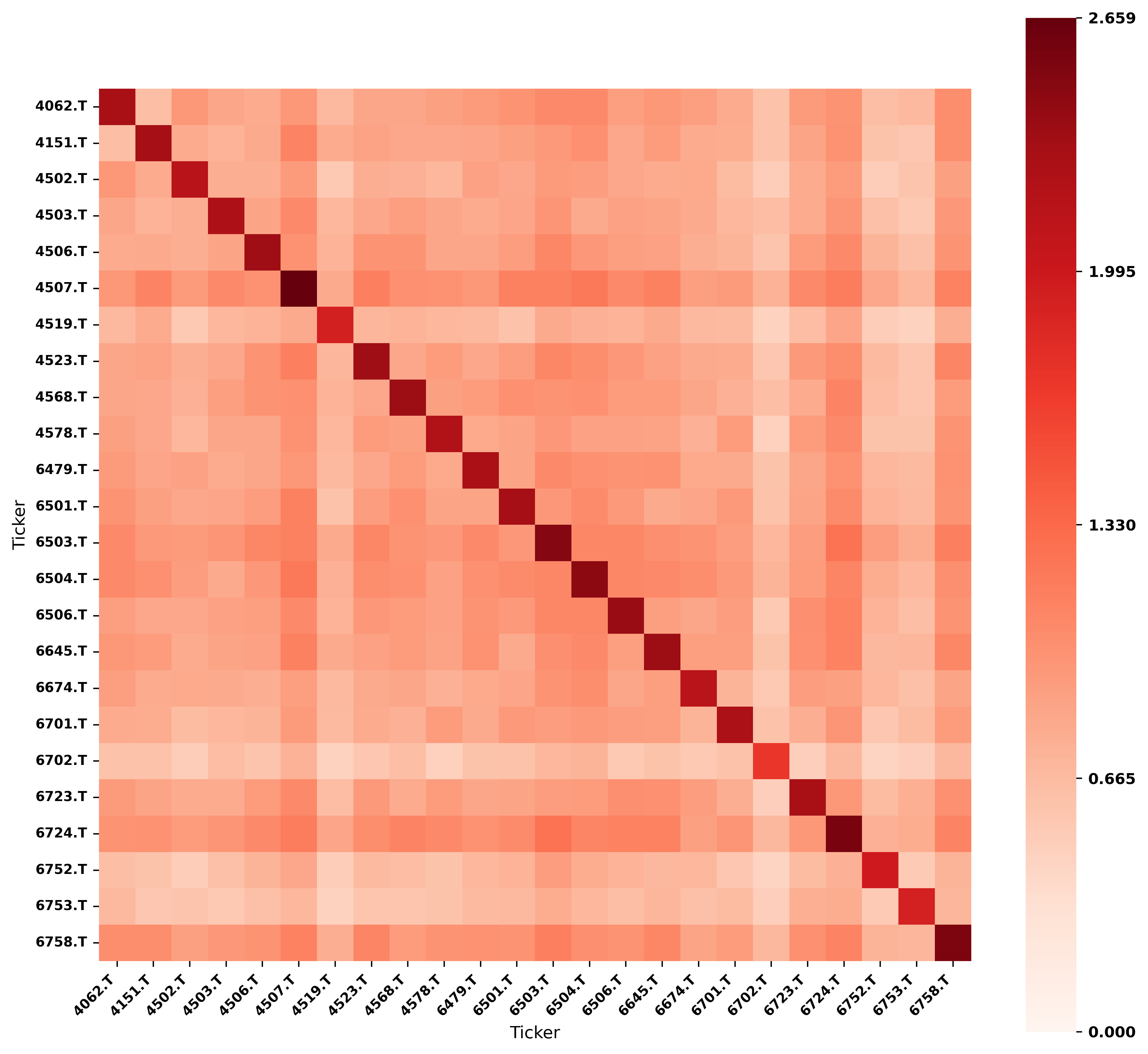}
        \caption{Pre-Crash (MI)}
    \end{subfigure}
    \hfill
    \begin{subfigure}[b]{0.32\textwidth}
        \centering
        \includegraphics[width=\textwidth]{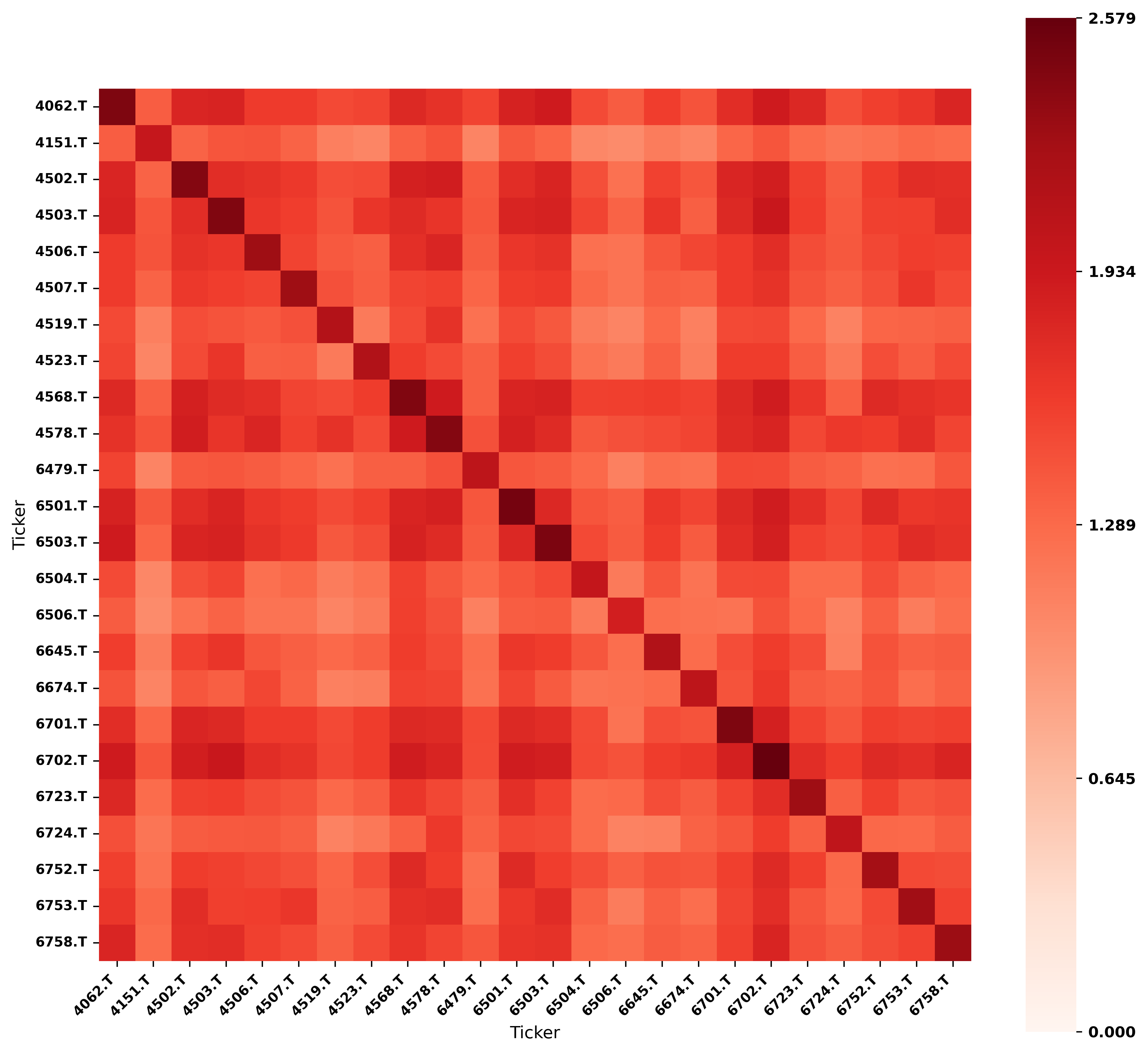}
        \caption{Crash (MI)}
    \end{subfigure}
    \hfill
    \begin{subfigure}[b]{0.32\textwidth}
        \centering
        \includegraphics[width=\textwidth]{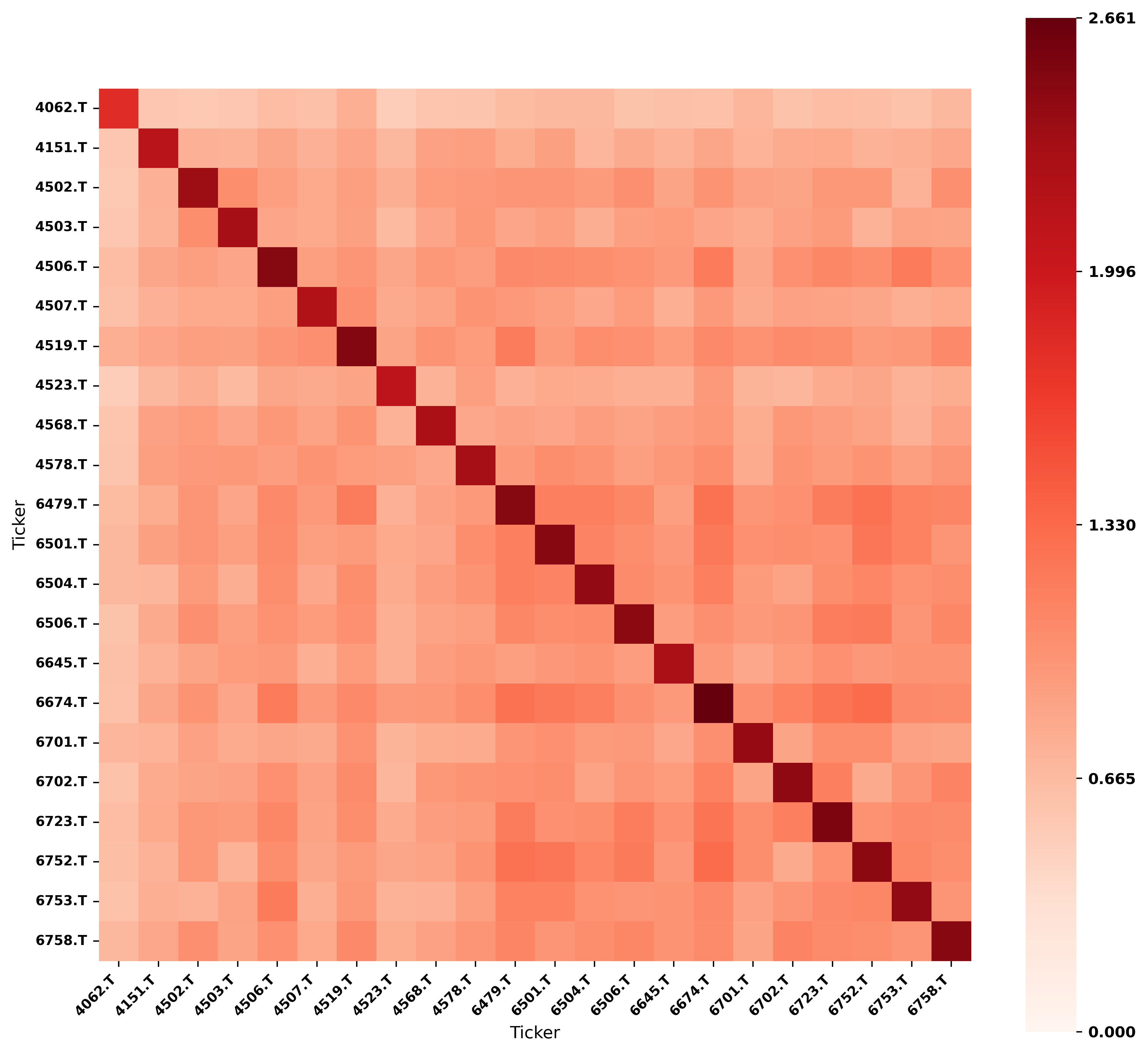}
        \caption{Post-Crash (MI)}
    \end{subfigure}

    \vspace{0.3cm}

    \begin{subfigure}[b]{0.32\textwidth}
        \centering
        \includegraphics[width=\textwidth]{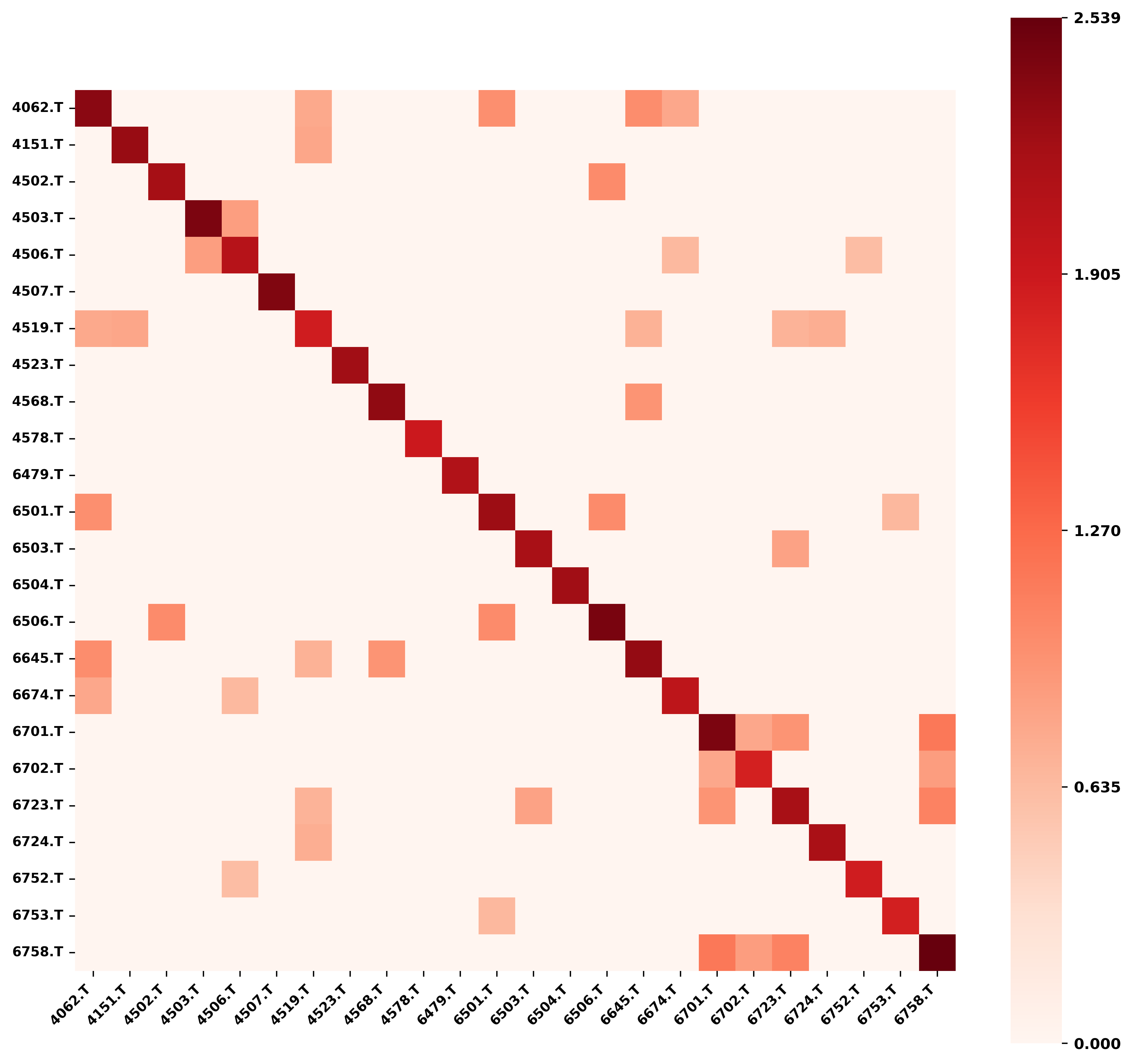}
        \caption{Pre-Crash (Cond.\ P-threshold MI)}
    \end{subfigure}
    \hfill
    \begin{subfigure}[b]{0.32\textwidth}
        \centering
        \includegraphics[width=\textwidth]{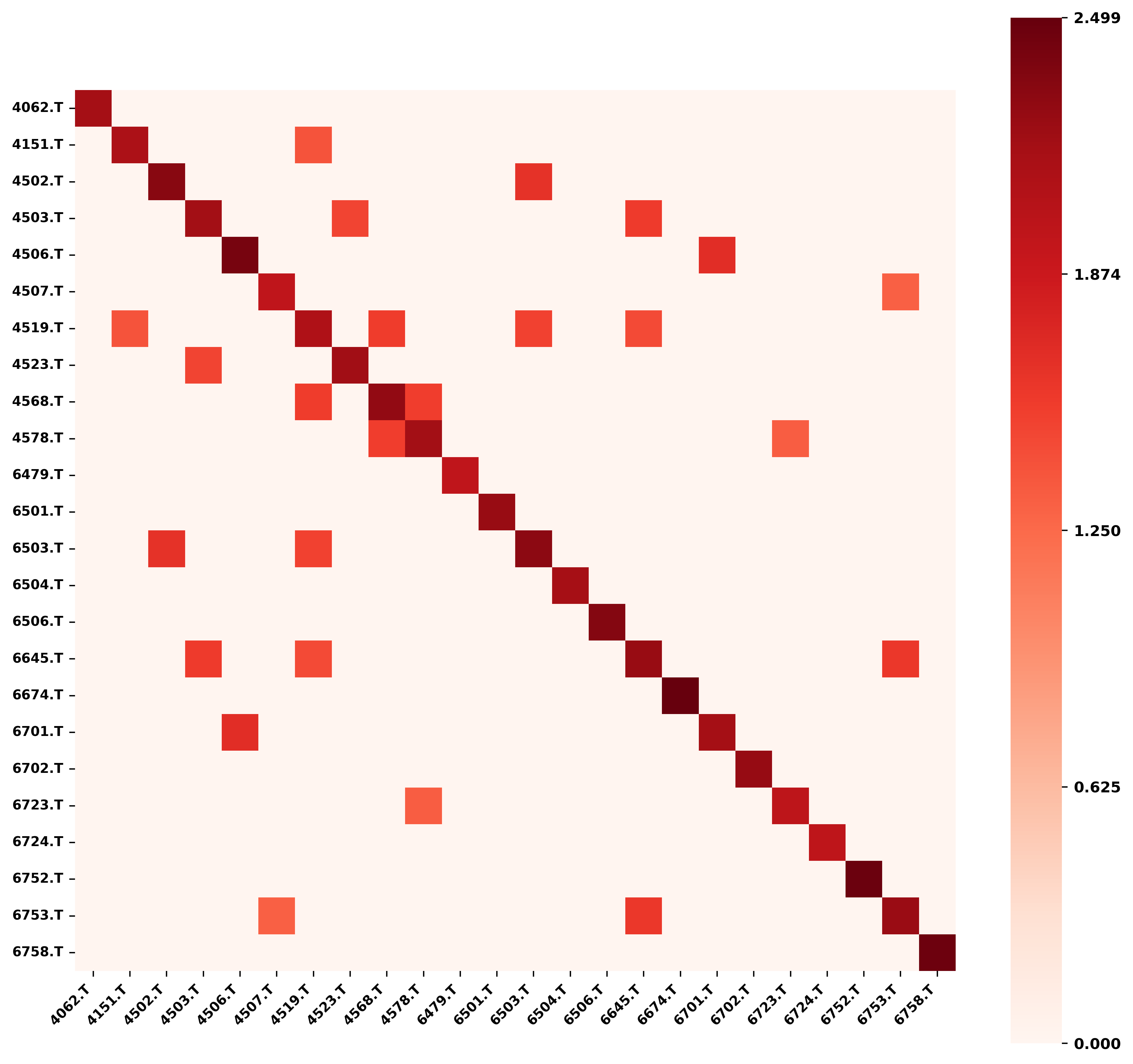}
        \caption{Crash (Cond.\ P-threshold MI)}
    \end{subfigure}
    \hfill
    \begin{subfigure}[b]{0.32\textwidth}
        \centering
        \includegraphics[width=\textwidth]{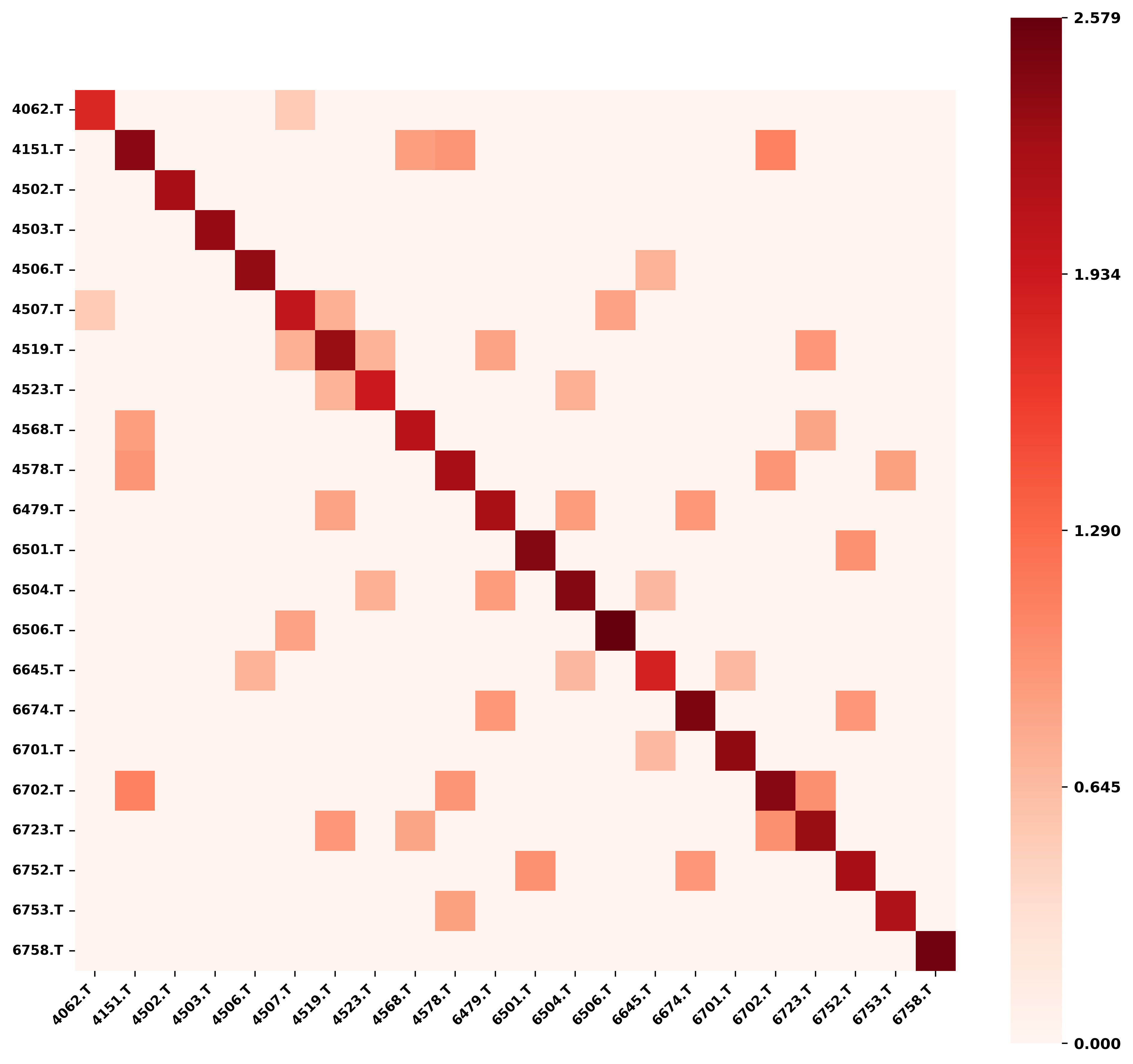}
        \caption{Post-Crash (Cond.\ P-threshold MI)}
    \end{subfigure}

    \caption{Mutual information heatmaps for the 25 largest Japanese stocks across different market periods. Plot (a)--(c) shows raw mutual information (MI) heatmaps, which are dominated by common market effects and exhibit dense connectivity, particularly during the crash period. Plot (d)--(f) display residual-based, significance-filtered MI heatmaps obtained using the conditional P-threshold MI method, highlighting statistically significant and direct nonlinear dependencies. The conditional P-threshold MI heatmaps show a sparse structure, indicating the true nonlinear dependencies between stocks after removing the market effect, which reveals the underlying direct interactions among stocks.}
    
    \label{fig:japan_mi}
\end{figure}

A similar pattern is observed for the Japanese stock market, as shown in Fig.~\ref{fig:japan_mi}. The MI heatmaps display dense connectivity across all periods, with a pronounced strengthening during the crash period. After applying the conditional P-threshold MI filtering, the heatmaps become sparse across all periods, with the highest intensity observed during the crash period compared to the pre-crash and post-crash periods. We also observed a consistent behavior for the Australian and Indian stock markets. The corresponding heatmaps for the Australian and Indian stocks are provided in Appendix~\ref{app:australia_mi} and Appendix~\ref{app:india_mi}, respectively. The same ordering as in Eq.~\eqref{compare_pre_crash_post_eq}, previously observed for the U.S. market in the post-crash period, is preserved for all markets. These indicate the presence of aftershock effects and delayed recovery dynamics.

Overall, these results demonstrate that the conditional p-threshold MI framework robustly isolates statistically significant and direct nonlinear dependencies across different market periods. Unlike MI, which is highly sensitive to market-wide factors, the conditional p-threshold approach provides a refined and economically meaningful representation of market structure. The consistent observation of crash amplification and post-crash aftershock effects across multiple equity markets highlights the suitability of this method for subsequent network construction and for identifying universal features of systemic financial stress.

\subsection{Conditional P-Threshold Network Dynamics}
\label{Conditional P-Threshold Network Dynamics}

We first identify statistically significant dependencies among stocks using the conditional p-threshold MI framework. In this approach, the effect of the market index is removed from individual stock returns, and MI is computed between the resulting residuals to capture direct dependencies. The conditional p-threshold plays a central role by retaining only those MI values that are statistically significant, thereby suppressing spurious connections arising from common market-wide movements. Statistical significance is assessed using a permutation-based hypothesis testing procedure, where the null hypothesis corresponds to the absence of dependence between a given stock pair. Only MI values with associated p-values below a prescribed significance level $\alpha$ are considered meaningful.

The filtered MI matrix obtained from this procedure defines a weighted dependency structure. To construct a sparse and comparable network topology across different market periods, we transform this weighted structure into an MST using a conditional p-threshold MI-based distance measure. The MST extracts the backbone of the conditional P-threshold MI network by retaining the most informative connections while ensuring an identical number of links across periods. All node-level and global topological measures analyzed in this study are computed from these MST networks.

In the primary analysis, we adopt a significance level of $\alpha = 0.05$ for constructing the conditional P-threshold MI networks. To examine robustness, we repeat the analysis using alternative significance levels of $\alpha = 0.01$ and $\alpha = 0.10$. As expected, $\alpha = 0.01$  thresholds yield fewer statistically significant dependencies, and when the thresholds increase to $\alpha = 0.10$, overall connectivity also increases. The network structure and the relative ordering of topological measures across the pre-crash, crash, and post-crash periods remain consistent across different significance levels. Therefore, $\alpha = 0.05$ is selected as an optimal balance between statistical reliability and network interpretability for subsequent analysis.

\subsubsection{Rank-Ordered Distributions of Network Metrics}

\begin{figure}[!t]
\centering

\begin{subfigure}[b]{0.45\textwidth}
\centering
\includegraphics[width=\textwidth]{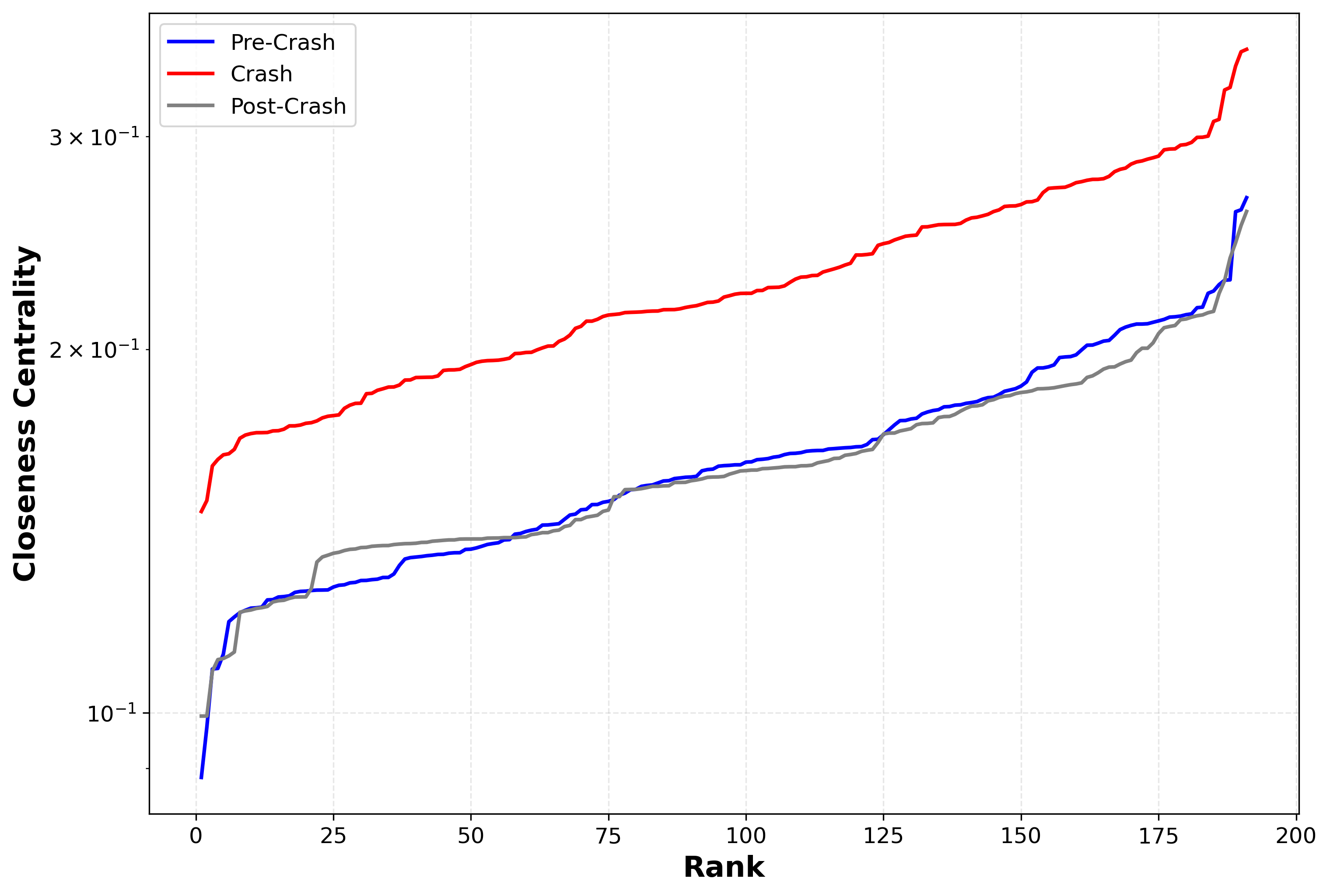}
\caption{Closeness}
\label{MI_MST_closeness_USA_log}
\end{subfigure}
\hfill
\begin{subfigure}[b]{0.45\textwidth}
\centering
\includegraphics[width=\textwidth]{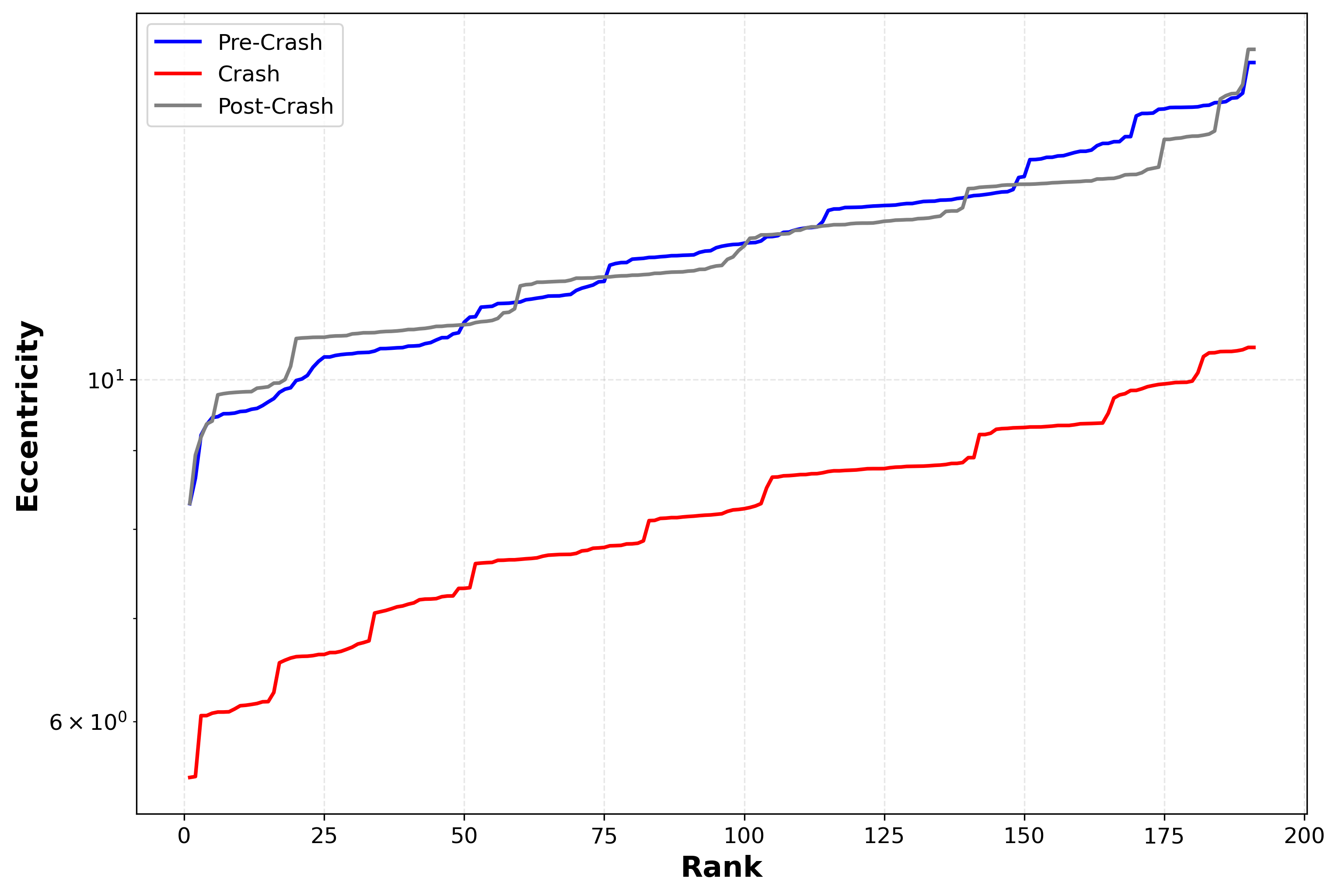}
\caption{Eccentricity}
\label{MI_MST_eccentricity_USA_lo}
\end{subfigure}

\vspace{0.3cm}

\begin{subfigure}[b]{0.45\textwidth}
\centering
\includegraphics[width=\textwidth]{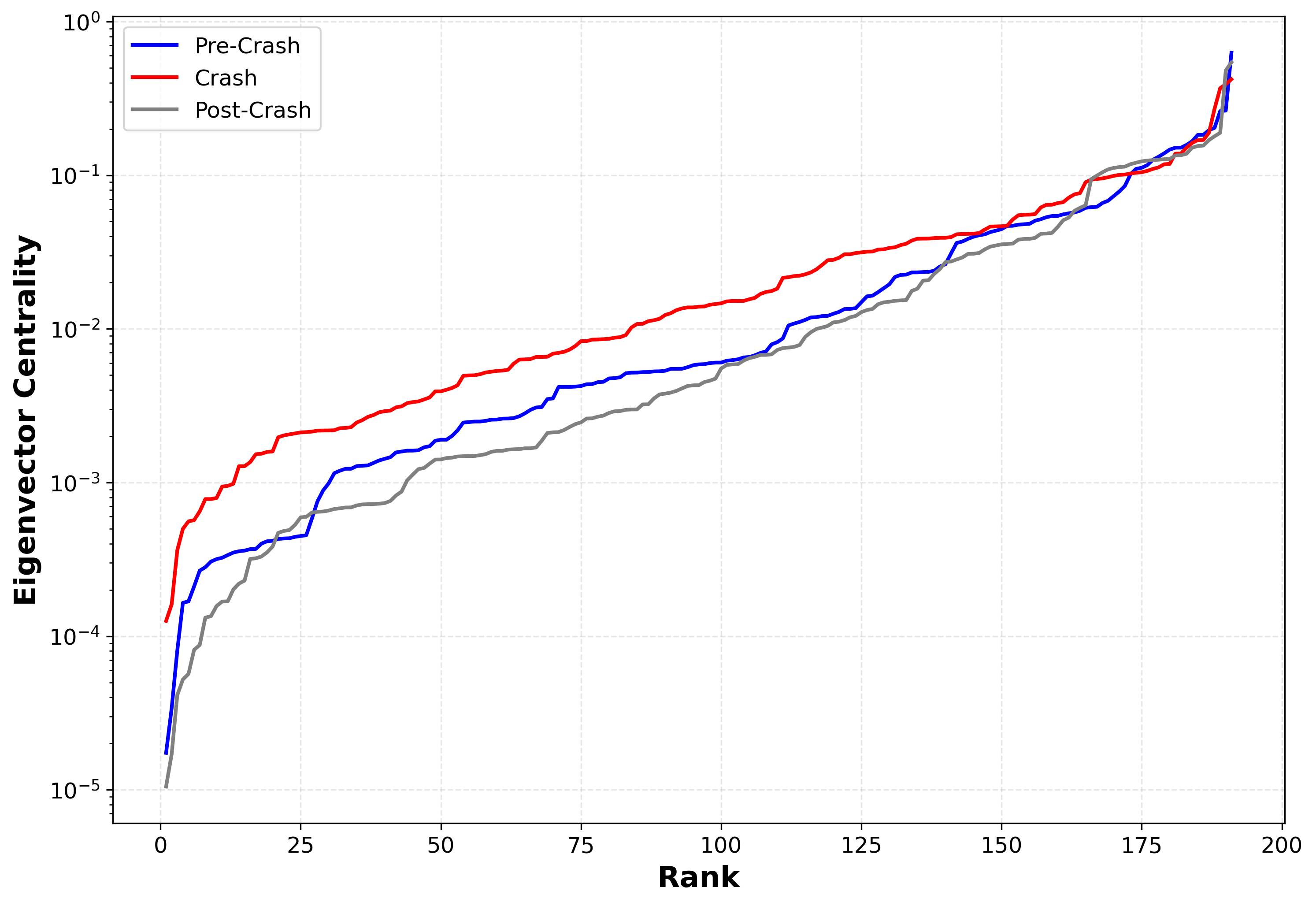}
\caption{Eigenvector Centrality}
\label{MI_MST_eigenvector_USA_log}
\end{subfigure}
\hfill
\begin{subfigure}[b]{0.45\textwidth}
\centering
\includegraphics[width=\textwidth]{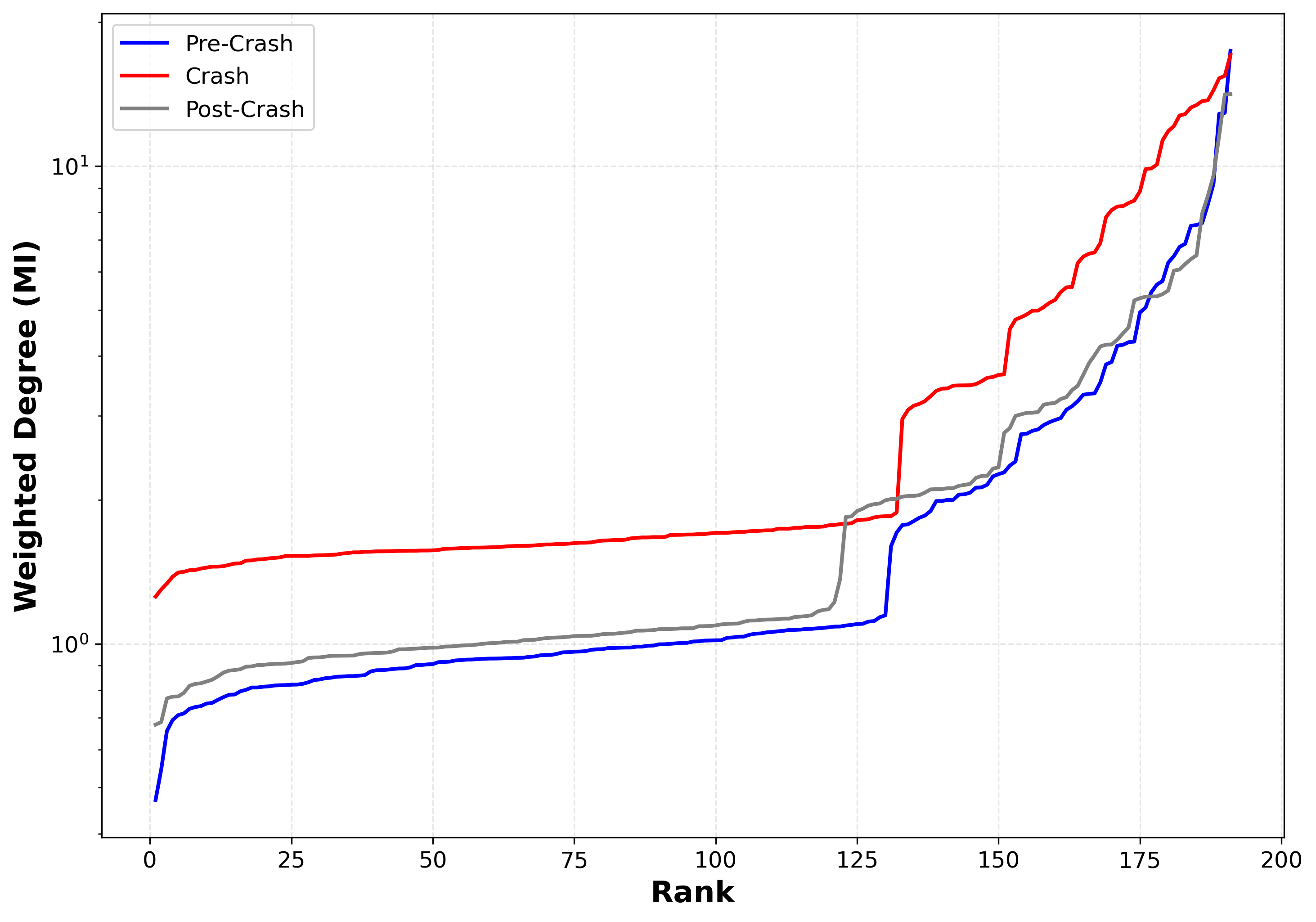}
\caption{Weighted Degree}
\label{MI_MST_weighted_degree_USA_log}
\end{subfigure}

\caption{
    Rank-ordered distributions of MST-based network metrics for the U.S. stock market across pre-crash, crash, and post-crash periods. Plot (a) represents closeness, Plot (b) eccentricity, Plot (c) eigenvector centrality and Plot (d) weighted degree. Stocks are ranked in ascending order for each metric. The curves correspond to the three market periods, highlighting structural reorganization and centralization during the crash period.
}
\label{fig:usa_rank_metrics}
\end{figure}

\begin{figure}[!t]
\centering

\begin{subfigure}[b]{0.45\textwidth}
\centering
\includegraphics[width=\textwidth]{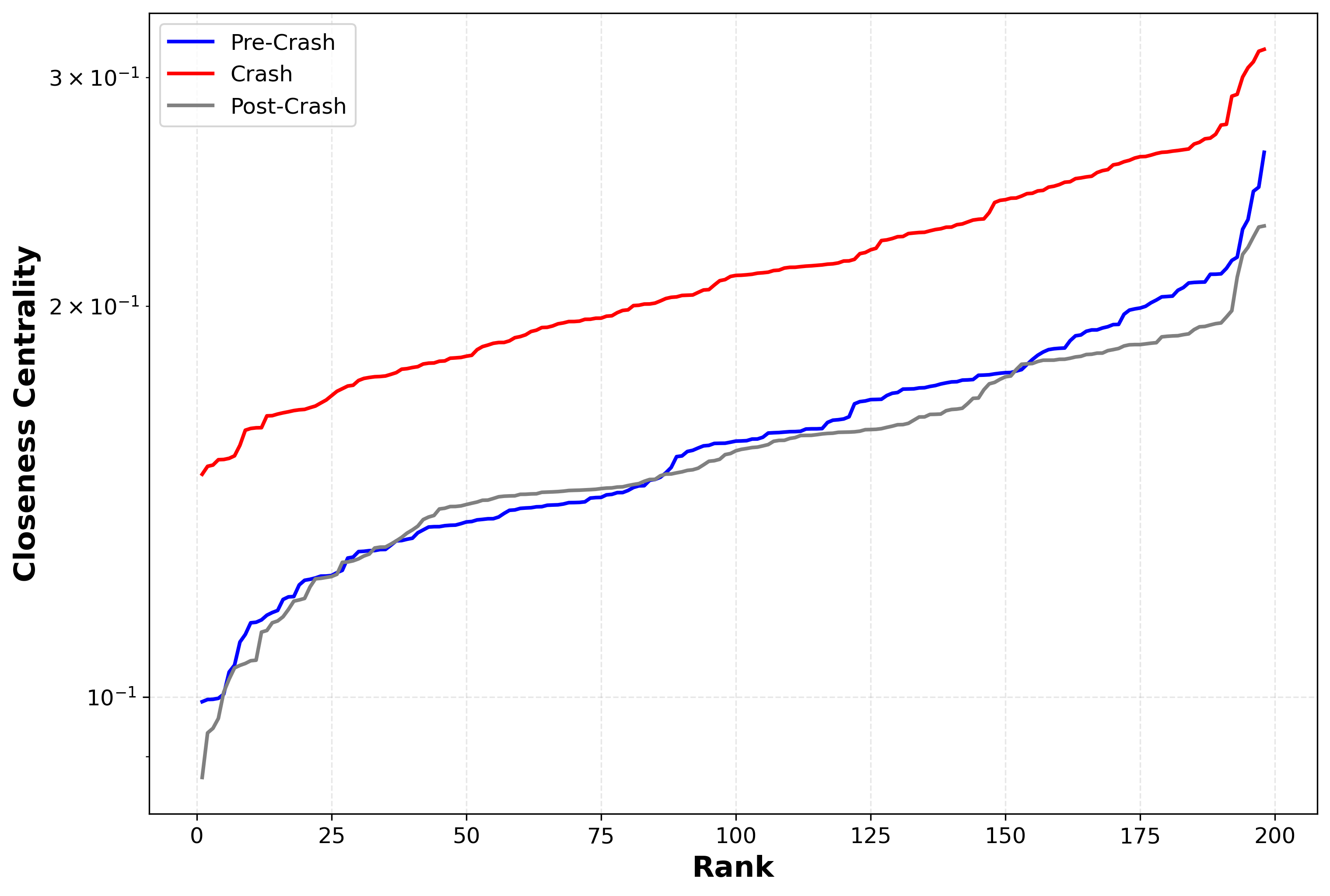}
\caption{Closeness}
\end{subfigure}
\hfill
\begin{subfigure}[b]{0.45\textwidth}
\centering
\includegraphics[width=\textwidth]{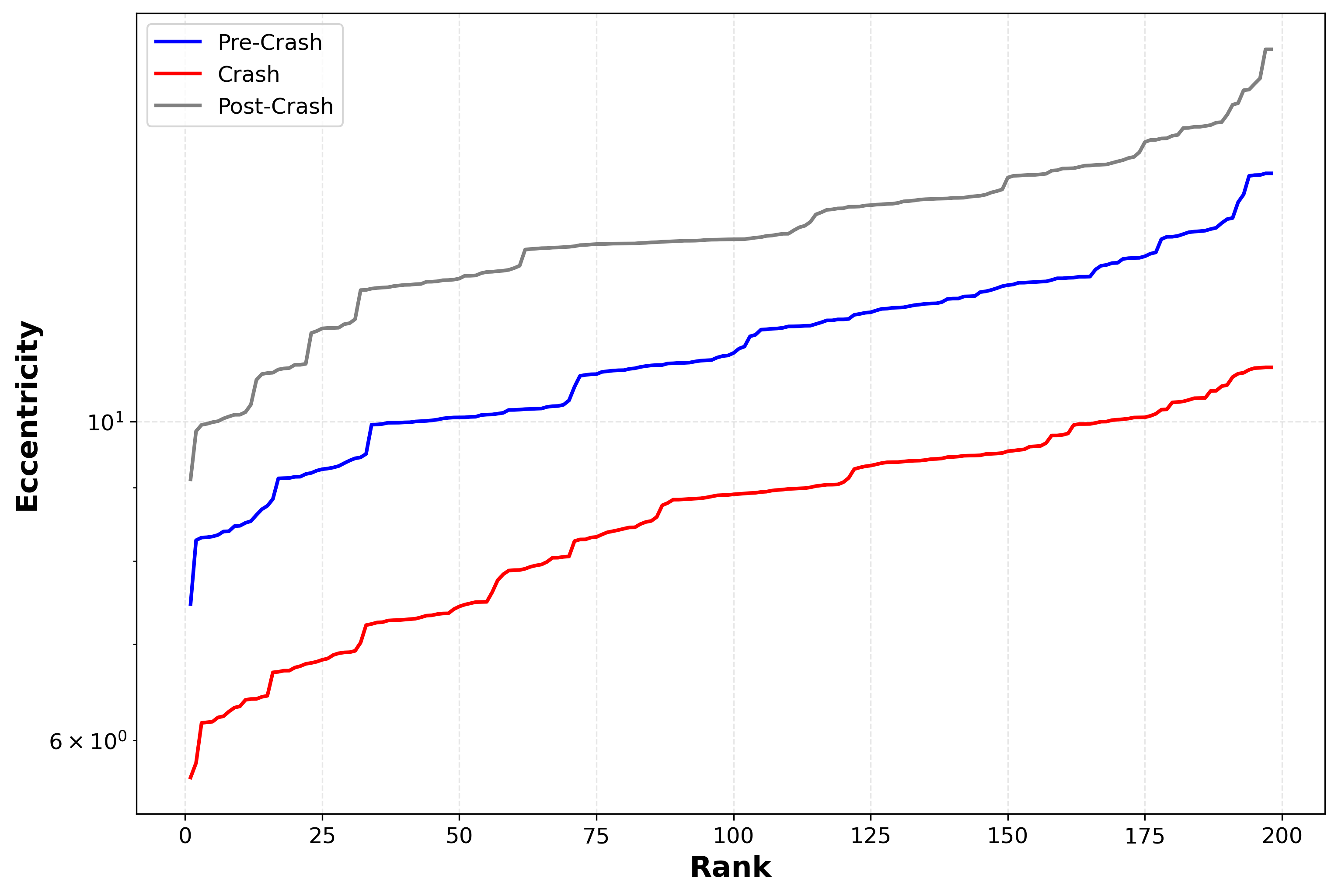}
\caption{Eccentricity}
\end{subfigure}

\vspace{0.3cm}

\begin{subfigure}[b]{0.45\textwidth}
\centering
\includegraphics[width=\textwidth]{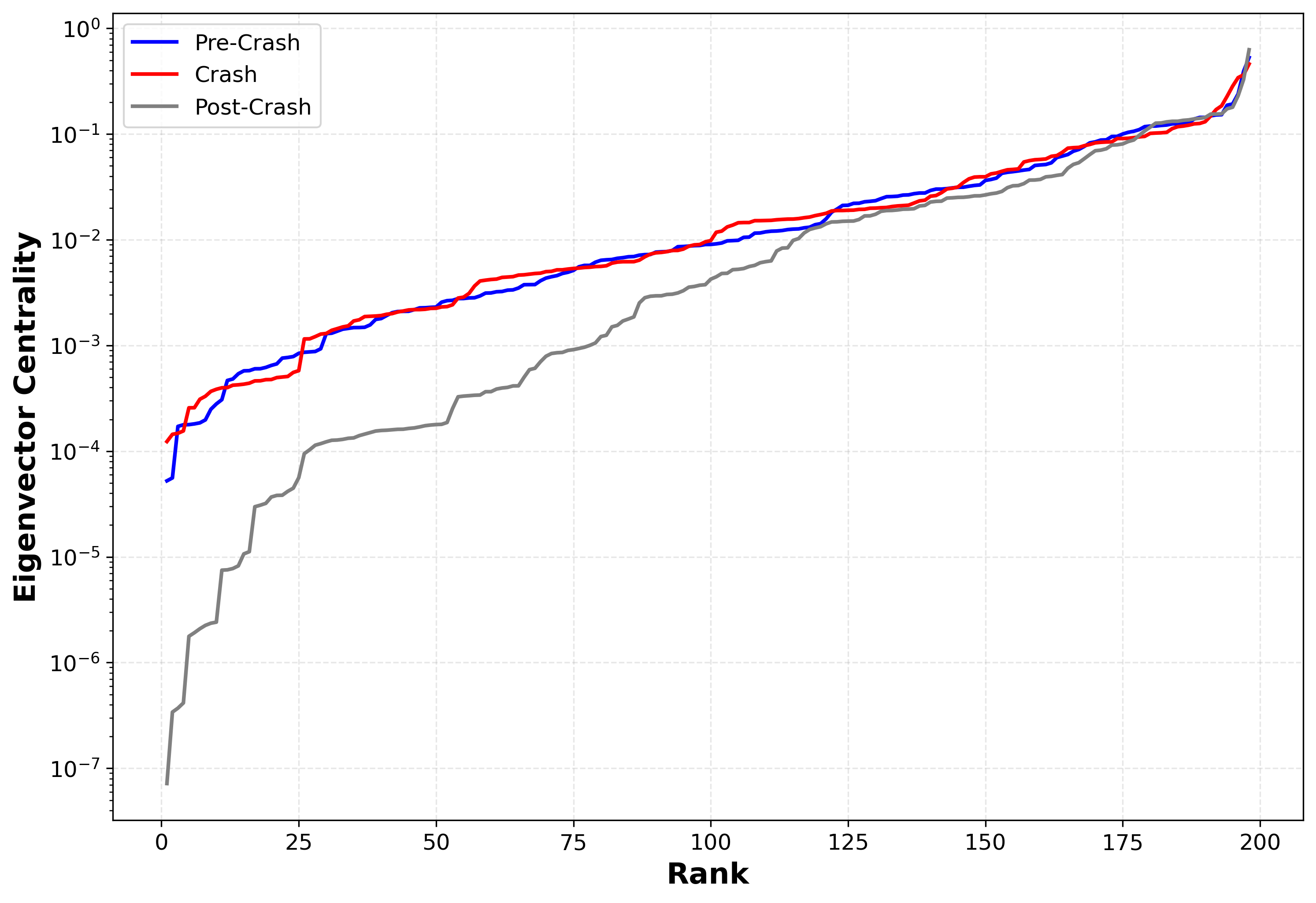}
\caption{Eigenvector Centrality}
\end{subfigure}
\hfill
\begin{subfigure}[b]{0.45\textwidth}
\centering
\includegraphics[width=\textwidth]{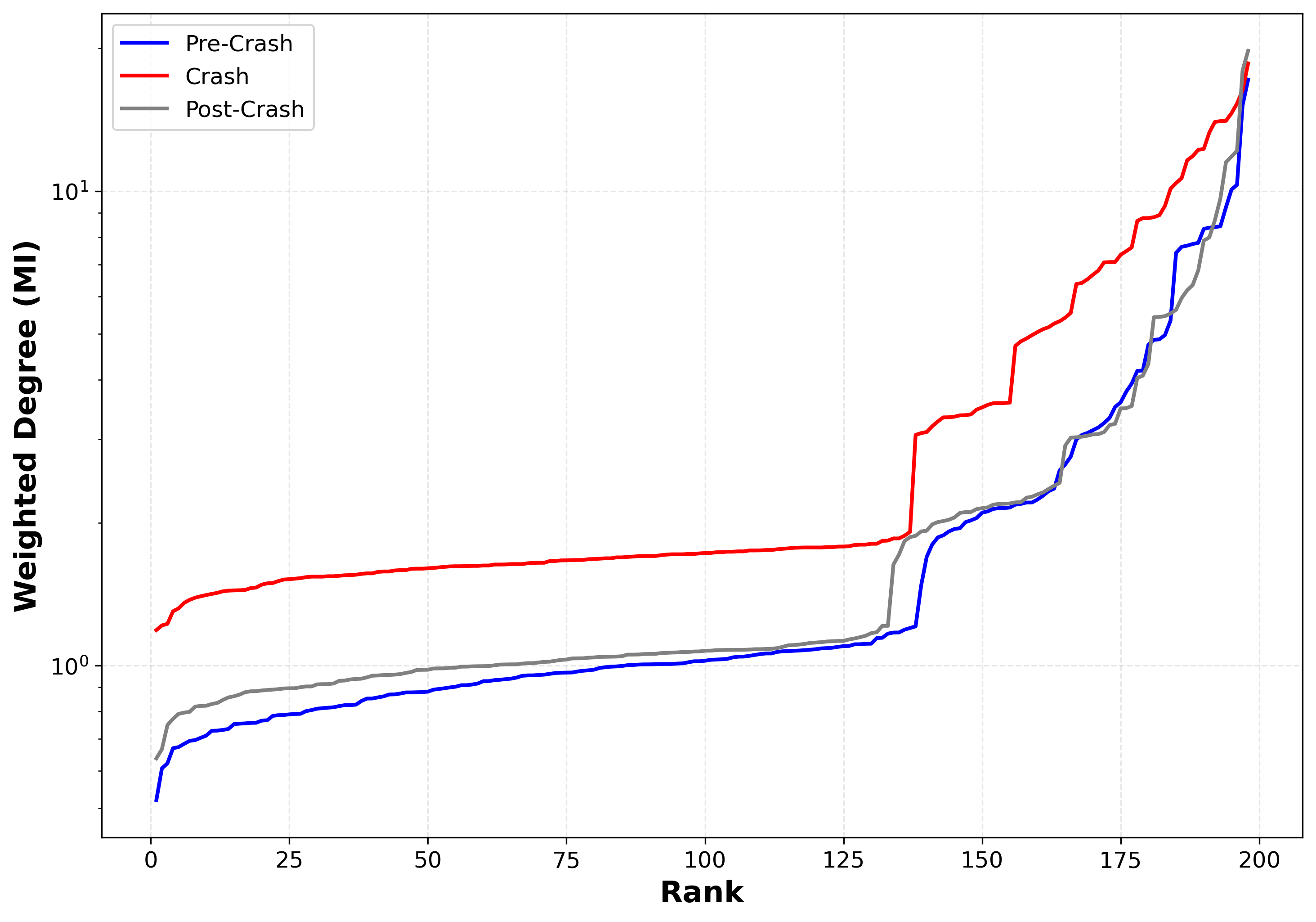}
\caption{Weighted Degree}
\end{subfigure}

\caption{
    Rank-ordered distributions of MST-based network metrics for the Japan stock market across pre-crash, crash, and post-crash periods. Plot (a) represents closeness, Plot (b) eccentricity, Plot (c) eigenvector centrality and Plot (d) weighted degree. Stocks are ranked in ascending order for each metric. The curves correspond to the three market periods, highlighting structural reorganization and centralization during the crash period.
}
\label{fig:japan_rank_metrics}
\end{figure}

\begin{figure}[!t]
\centering

\begin{subfigure}[b]{0.45\textwidth}
\centering
\includegraphics[width=\textwidth]{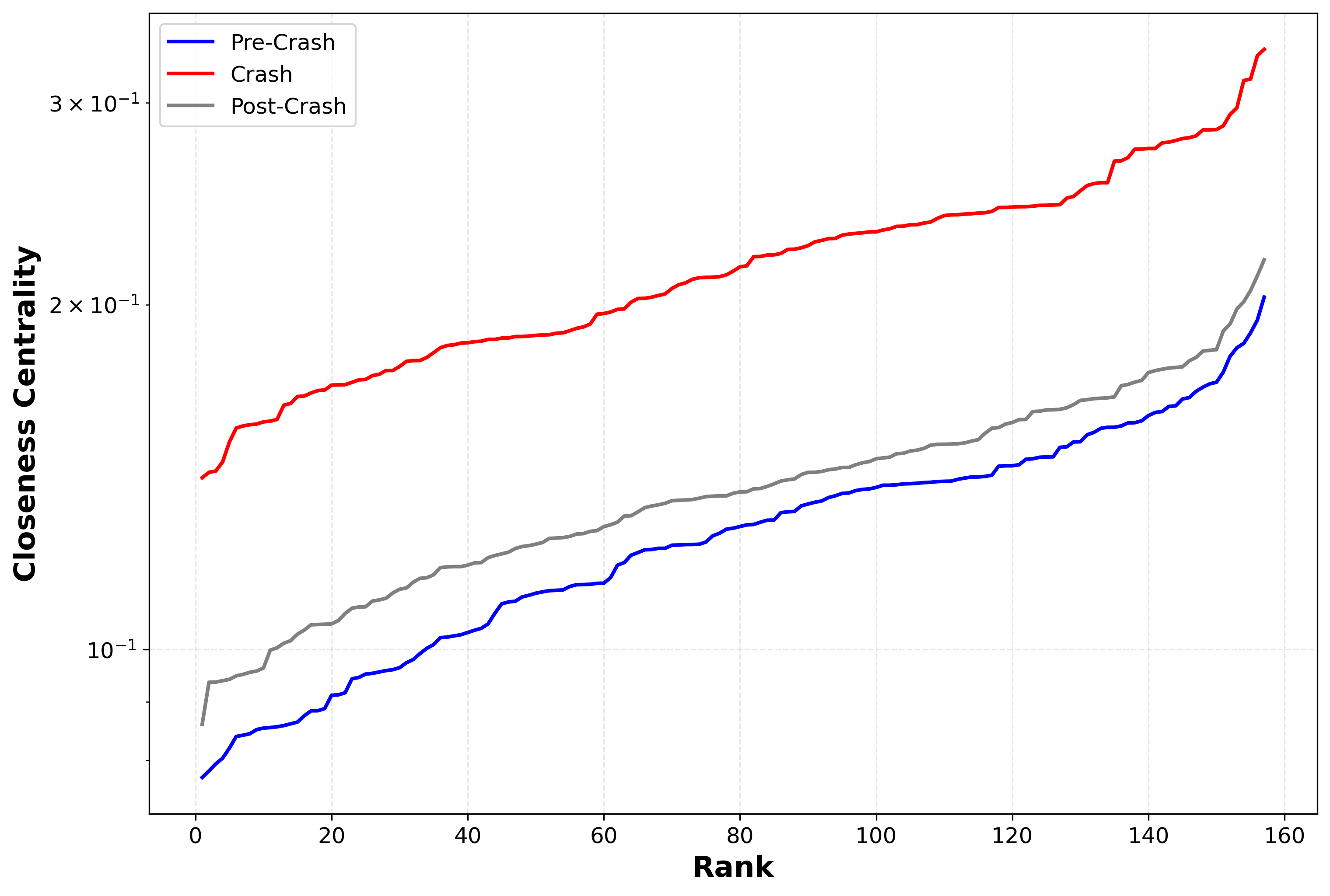}
\caption{Closeness}
\end{subfigure}
\hfill
\begin{subfigure}[b]{0.45\textwidth}
\centering
\includegraphics[width=\textwidth]{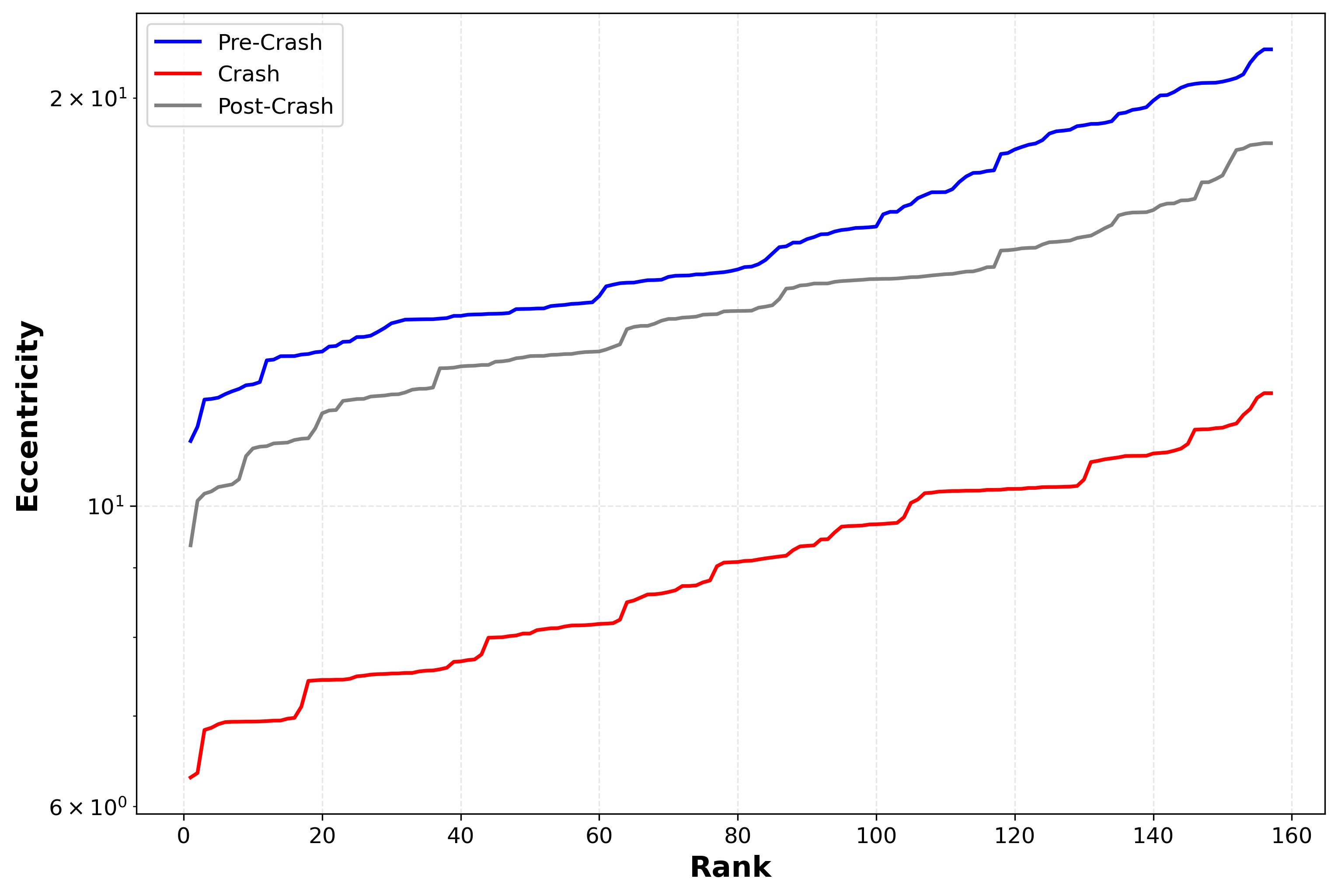}
\caption{Eccentricity}
\end{subfigure}

\vspace{0.3cm}

\begin{subfigure}[b]{0.45\textwidth}
\centering
\includegraphics[width=\textwidth]{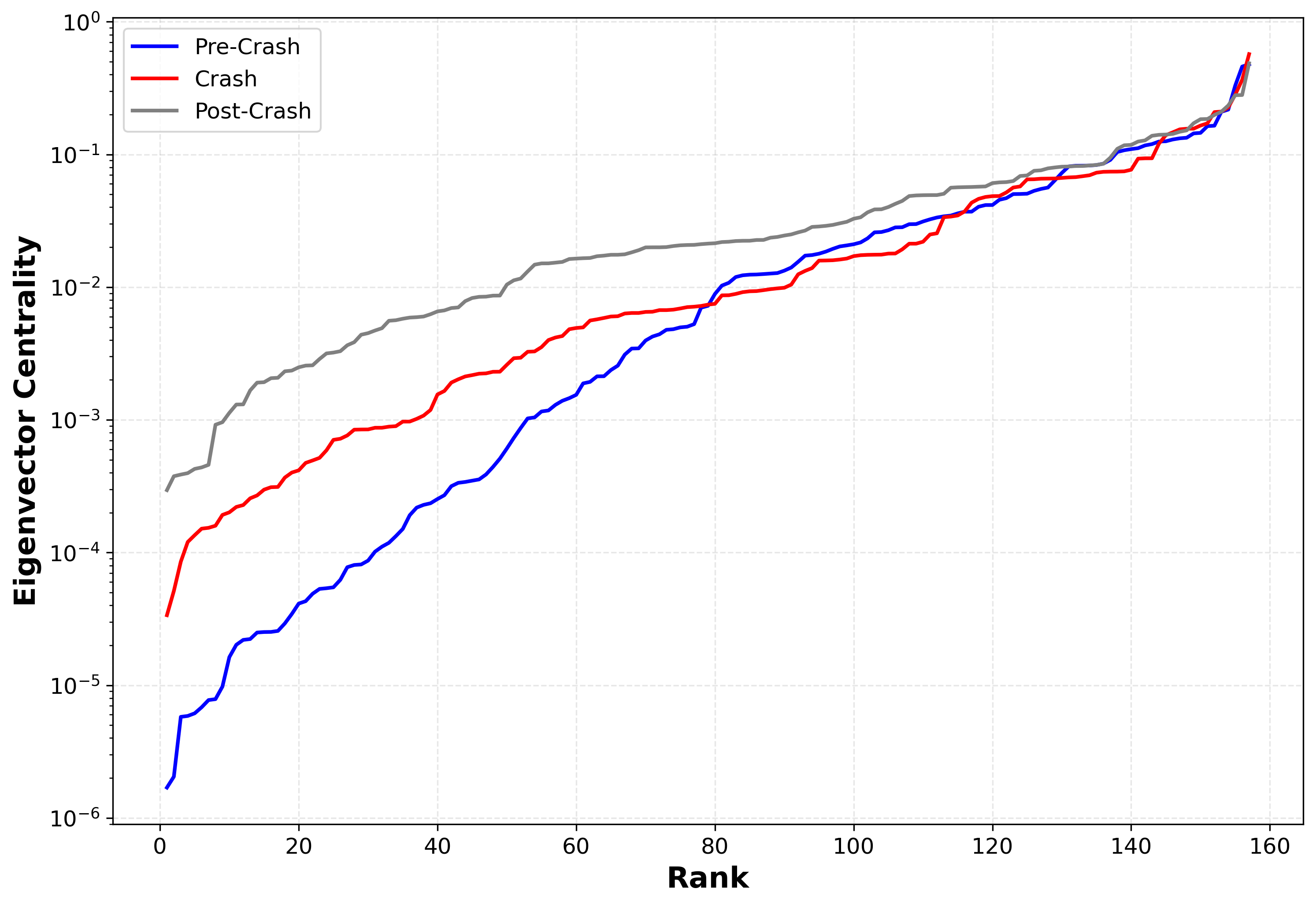}
\caption{Eigenvector Centrality}
\end{subfigure}
\hfill
\begin{subfigure}[b]{0.45\textwidth}
\centering
\includegraphics[width=\textwidth]{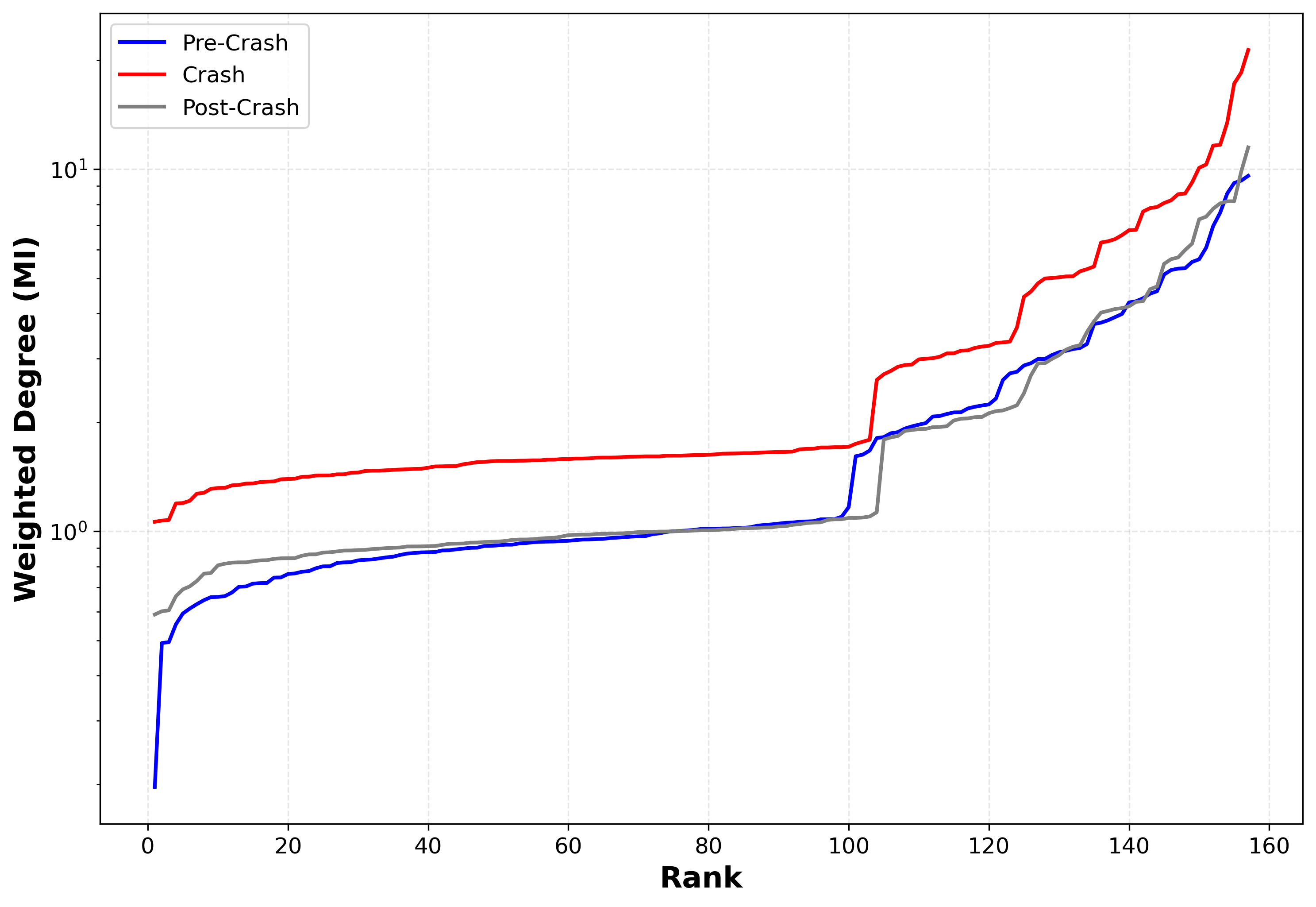}
\caption{Weighted Degree}
\end{subfigure}

\caption{
    Rank-ordered distributions of MST-based network metrics for the Australian stock market across pre-crash, crash and post-crash periods. Plot (a) represents closeness, Plot (b) eccentricity, Plot (c) eigenvector centrality and Plot (d) weighted degree. Stocks are ranked in ascending order for each metric. The curves correspond to the three market periods, highlighting structural reorganization and centralization during the crash period.
}
\label{fig:australia_rank_metrics}
\end{figure}

\begin{figure}[!t]
\centering

\begin{subfigure}[b]{0.45\textwidth}
\centering
\includegraphics[width=\textwidth]{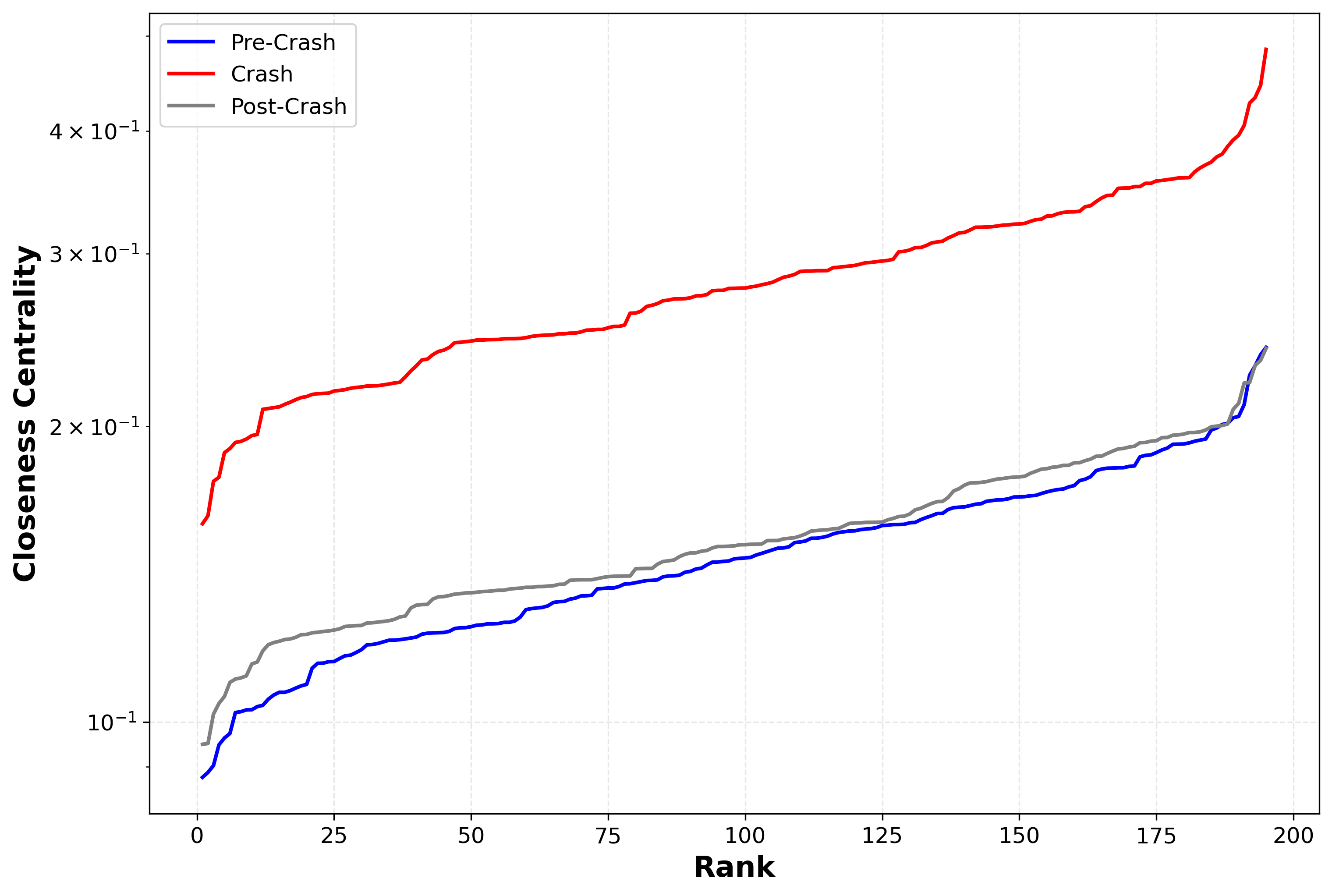}
\caption{Closeness}
\end{subfigure}
\hfill
\begin{subfigure}[b]{0.45\textwidth}
\centering
\includegraphics[width=\textwidth]{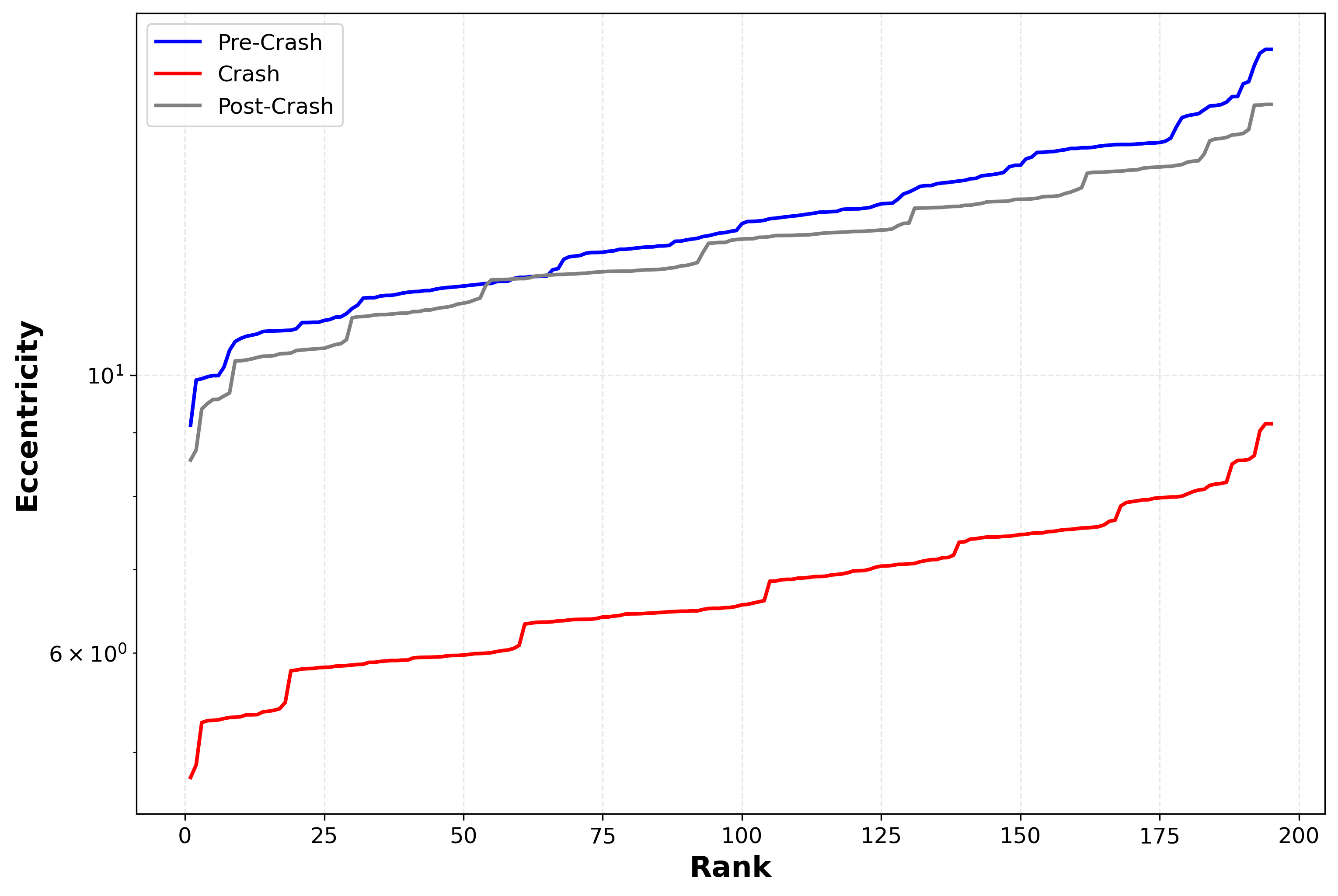}
\caption{Eccentricity}
\end{subfigure}

\vspace{0.3cm}

\begin{subfigure}[b]{0.45\textwidth}
\centering
\includegraphics[width=\textwidth]{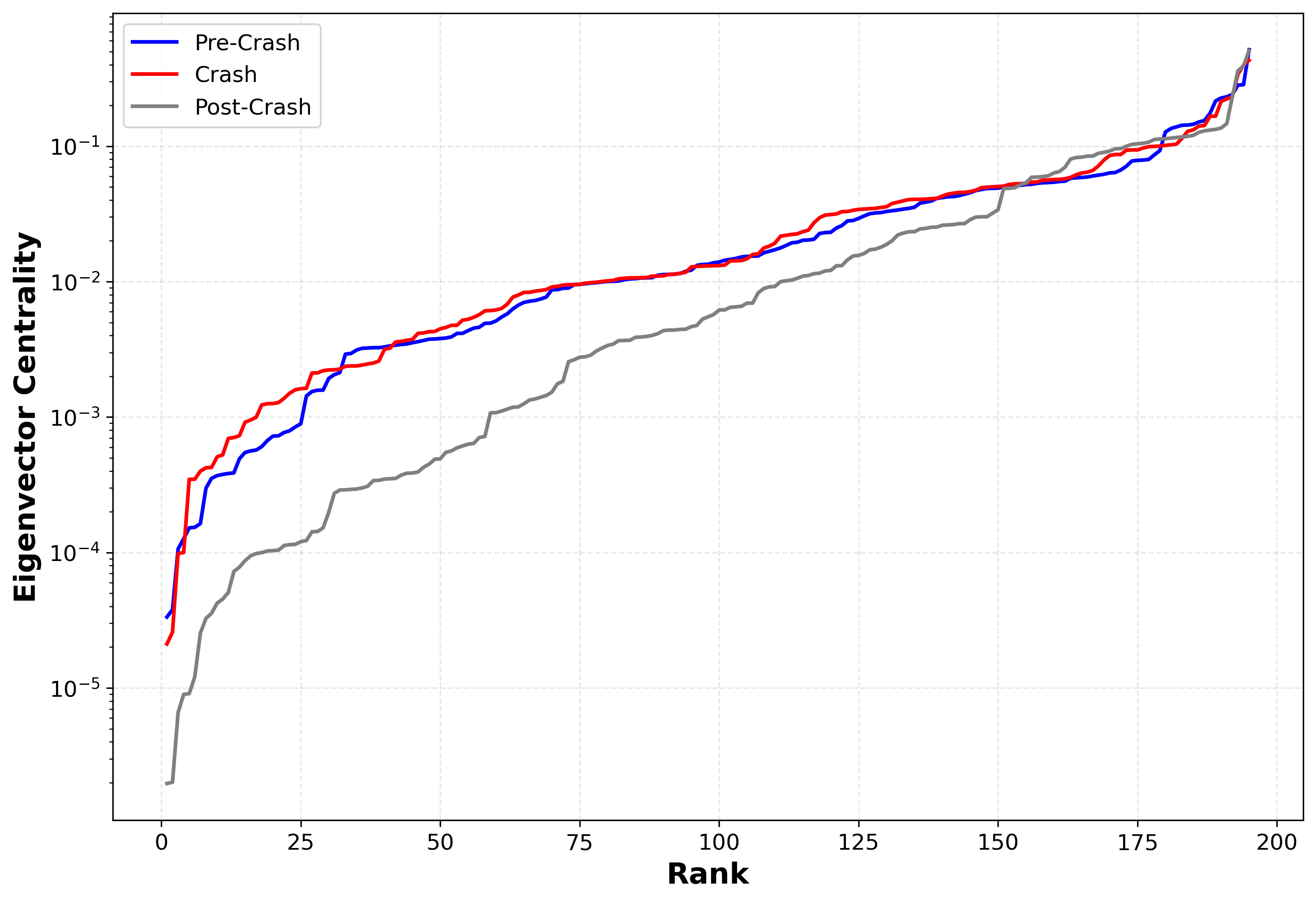}
\caption{Eigenvector Centrality}
\end{subfigure}
\hfill
\begin{subfigure}[b]{0.45\textwidth}
\centering
\includegraphics[width=\textwidth]{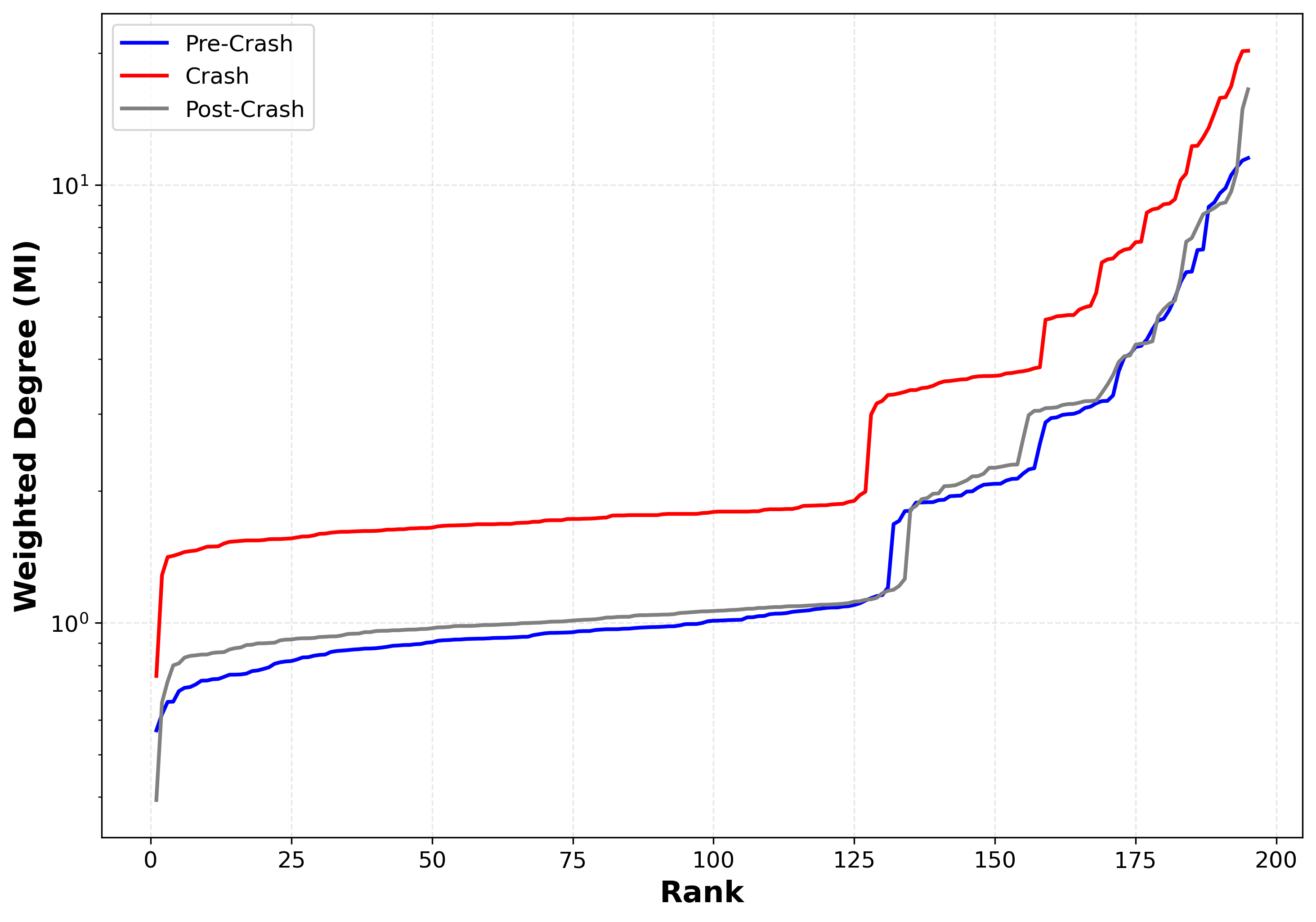}
\caption{Weighted Degree}
\end{subfigure}

\caption{
    Rank-ordered distributions of MST-based network metrics for the Indian stock market across pre-crash, crash and post-crash periods. Plot (a) represents closeness, Plot (b) eccentricity, Plot (c) eigenvector centrality and Plot (d) weighted degree. Stocks are ranked in ascending order for each metric. The curves correspond to the three market periods, highlighting structural reorganization and centralization during the crash period.
}
\label{fig:india_rank_metrics}
\end{figure}

Figs~\ref{fig:usa_rank_metrics},~\ref{fig:japan_rank_metrics},~\ref{fig:australia_rank_metrics} and~\ref{fig:india_rank_metrics} present the rank-ordered distributions of MST-based network metrics: closeness, eccentricity, eigenvector centrality, and weighted degree, derived from the conditional p-threshold MI-MST framework. These network metrics are computed from the sparse, statistically validated adjacency matrices that capture only direct stock-stock dependencies after removing market-wide co-movement. This methodological framework allows us to isolate genuine structural reconfiguration during crises, separating it from the common-factor-driven synchronization that dominates raw correlation or mutual information measures. Across all markets, the rank-ordered plots clearly distinguish the three market periods: pre-crash, crash and post-crash. The crash period consistently exhibits the most pronounced deviation from both pre- and post-crash periods, indicating a fundamental and systematic reorganization of the underlying network structure during the crash. This clear separation across periods, observable in all four QUAD markets, validates the sensitivity and robustness of our conditional p-threshold MI approach in detecting crisis-induced structural changes.

For the U.S. stock market, Fig.~\ref{MI_MST_closeness_USA_log} shows that closeness centrality, which measures how rapidly shocks can propagate through the network, increases significantly during the crash period. This increase, revealed by our filtered network, reflects a contraction of effective network distances and implies faster potential transmission of shocks during a market crash. Similarly, eigenvector centrality and weighted degree, as shown in Figs.~\ref{MI_MST_eigenvector_USA_log} and~\ref{MI_MST_weighted_degree_USA_log}, exhibit elevated values during the crash, indicating that dependence becomes more concentrated around a subset of influential stocks. Eccentricity, defined as the maximum shortest-path distance from a stock to any other stock in the network, decreases during the crash, reflecting network compression and increased synchronization. Importantly, these patterns emerge after removing market effects, suggesting they represent genuine changes in the fabric of stock interdependencies rather than mere reflections of common factor exposure. Following the crash, the post-crash distributions shift back toward pre-crash patterns but do not fully overlap, indicating that the market does not immediately return to its original structural state. This difference reveals persistent aftershock effects, whereby the impact of the crash continues to influence network connectivity beyond the immediate crash period. A similar systematic pattern is observed in the Japanese, Indian, and Australian markets, characterized by a crash deviation followed by partial post-crash recovery, demonstrating the universal applicability of our method.

The consistency of these findings across geographically and economically distinct markets underscores the robustness of the conditional p-threshold MI framework. By filtering out common factors and emphasizing statistically validated nonlinear relationships, our approach provides a clearer and more reliable representation of how stock-specific interactions reorganize during periods of financial crash. The universal detection of network compression, increased centrality concentration, and persistent aftershocks across all QUAD markets confirms that our method captures fundamental aspects of market reconfiguration that are obscured by traditional correlation-based or unfiltered MI measures. In Section~\ref{After_shock}, we further investigate these post-crash aftershock effects using the GR law, providing additional quantitative support for the observed persistence in network reconfiguration.

\subsubsection{Core--Periphery Structure of Stock Market Networks}

\begin{table}[!t]
\centering
\caption{Table represents the core concentration and periphery fragility indices of conditional p-threshold MI--MST networks for QUAD stock markets across pre-crash, crash, and post-crash periods. Core concentration quantifies the dominance of influential stocks, while periphery fragility captures the structural vulnerability of peripheral nodes under market stress.}
\label{tab:core_periphery_indices}
\renewcommand{\arraystretch}{1.15}

\begin{tabular}{llcc}
\toprule
\textbf{Period} & \textbf{Country} &
\textbf{Core Concentration} &
\textbf{Periphery Fragility} \\
\midrule
\multirow{4}{*}{Pre-Crash}
 & Australia & 0.293 & 87.3 \\
 & India     & 0.247 & 96.4 \\
 & Japan     & 0.215 & 78.0 \\
 & USA       & 0.194 & 64.9 \\
\midrule
\multirow{4}{*}{Crash}
 & Australia & 0.168 & 167.5 \\
 & India     & 0.135 & 75.3 \\
 & Japan     & 0.162 & 281.3 \\
 & USA       & 0.164 & 226.5 \\
\midrule
\multirow{4}{*}{Post-Crash}
 & Australia & 0.331 & 102.4 \\
 & India     & 0.210 & 101.9 \\
 & Japan     & 0.192 & 100.3 \\
 & USA       & 0.189 & 53.9 \\
\bottomrule
\end{tabular}
\end{table}

We further investigate the evolution of market structure during the COVID‑19 crash through the lens of core–periphery organization using the conditional p‑threshold MI–MST framework. The core–periphery structure is quantified by two complementary measures: core concentration, which measures the dominance of a cohesive central set of influential stocks, and periphery fragility, which captures the vulnerability of weakly connected peripheral stocks to shock transmission. Table~\ref{tab:core_periphery_indices} reports the evolution of both indices across the three periods. A consistent pattern emerges during the crash: core concentration declines sharply in all four markets, indicating a fragmentation of the influential core and a dispersion of network influence. Simultaneously, periphery fragility increases dramatically in the U.S., Japanese, and Australian markets, revealing that peripheral stocks become structurally more exposed and vulnerable to spillovers. The conditional p‑threshold MI framework uniquely captures this dual shift,  a weakening core and an increasingly fragile periphery, as a key feature of market topology during a crash. The Indian stock presents a partial exception, with periphery fragility decreasing during the crash, suggesting country‑specific differences in market composition.

The topological factors underlying this reconfiguration are summarized in Table~\ref{tab:quad_topology_focused}. During the crash, all markets show a strong decline in average path length (APL), indicating a more compact and highly integrated network. At the same time, the weighted degree increases, reflecting a rise in statistically significant dependencies among stocks. However, algebraic connectivity ($\lambda_2$), a measure of network robustness, decreases, suggesting that higher integration is accompanied by greater structural fragility. This contradiction is explained by the strongly negative assortativity observed during the crash. Negative assortativity indicates that core stocks tend to connect with peripheral stocks, making the periphery a key but fragile channel for shock transmission. By removing common market effects, this disassortative structure is clearly revealed and directly accounts for the high periphery fragility observed during crisis periods. In the post-crash period, the network shows partial recovery. Core concentration increases, periphery fragility decreases, and both average path length and weighted degree move back toward their pre-crash levels. However, this recovery is neither complete nor uniform across markets. For Australia, post-crash core concentration exceeds its pre-crash value, while periphery fragility remains high, indicating a persistent change in network structure. In contrast, the U.S. and Japan exhibit lower post-crash algebraic connectivity than before the crash, suggesting continued structural vulnerability. This heterogeneous aftershock pattern, observed only in the residual, statistically validated network, demonstrates that the conditional p-threshold MI framework captures not only the crash period but also lasting structural effects that differ across markets.

\begin{table}[!t]
\centering
\caption{Table represents the average network topological measures of conditional p-threshold MI--MST networks for QUAD stock markets across pre-crash, crash, and post-crash periods. W.Deg denotes weighted degree, Ecc denotes eccentricity, Eff denotes global efficiency, APL denotes average path length, $\lambda_2$ denotes algebraic connectivity, Tree Len denotes total tree length, and Assort denotes degree assortativity. These metrics collectively characterize changes in network integration, connectivity, and core--periphery structure across market periods.}

\label{tab:quad_topology_focused}
\renewcommand{\arraystretch}{1.15}

\resizebox{\textwidth}{!}{
\begin{tabular}{llccccccc}
\toprule
\textbf{Period} & \textbf{Country} &
\textbf{W.Deg} &
\textbf{Ecc} &
\textbf{Eff} &
\textbf{APL} &
$\boldsymbol{\lambda_2}$ &
\textbf{Tree Len} &
\textbf{Assort} \\
\midrule

\multirow{4}{*}{Pre-Crash}
 & USA        & 1.959 & 12.22 & 0.181 & 6.39 & 0.0059 & 197.41 & -0.383 \\
 & Japan      & 1.979 & 11.25 & 0.176 & 6.54 & 0.0051 & 203.64 & -0.398 \\
 & India      & 1.949 & 13.23 & 0.167 & 7.02 & 0.0044 & 202.17 & -0.424 \\
 & Australia  & 1.910 & 15.83 & 0.157 & 8.31 & 0.0041 & 170.97 & -0.358 \\
\midrule

\multirow{4}{*}{Crash}
 & USA        & 3.333 & 8.21  & 0.161 & 4.55 & 0.0020 & 114.26 & -0.453 \\
 & Japan      & 3.381 & 8.62  & 0.154 & 4.84 & 0.0015 & 117.16 & -0.422 \\
 & India      & 3.479 & 6.74  & 0.178 & 3.71 & 0.0047 & 112.20 & -0.354 \\
 & Australia  & 3.156 & 9.05  & 0.167 & 4.79 & 0.0024 & 99.45  & -0.402 \\
\midrule

\multirow{4}{*}{Post-Crash}
 & USA        & 2.073 & 12.11 & 0.174 & 6.42 & 0.0046 & 184.63 & -0.248 \\
 & Japan      & 2.078 & 13.56 & 0.177 & 6.66 & 0.0029 & 191.35 & -0.291 \\
 & India      & 2.077 & 12.64 & 0.172 & 6.68 & 0.0032 & 189.15 & -0.326 \\
 & Australia  & 1.953 & 13.97 & 0.167 & 7.49 & 0.0045 & 161.56 & -0.461 \\
\bottomrule
\end{tabular}}
\end{table}

Overall, the conditional p‑threshold MI–MST framework provides a clear view of how crisis‑induced reconfiguration operates through the core–periphery structure. The method filters out the diffuse co‑movement driven by common factors, revealing a distinct structural change: a weakened core, a weakened core, an increasingly fragile and actively connected periphery, and disassortative mixing that amplifies vulnerability. These patterns are consistent across geographically diverse markets, confirming that the approach captures universal features of crisis topology that are masked by traditional, non‑filtered network measures.

\section{Community Structure of Stock Market Networks}

\begin{figure}[!t]
    \centering
    \includegraphics[width=0.8\textwidth]{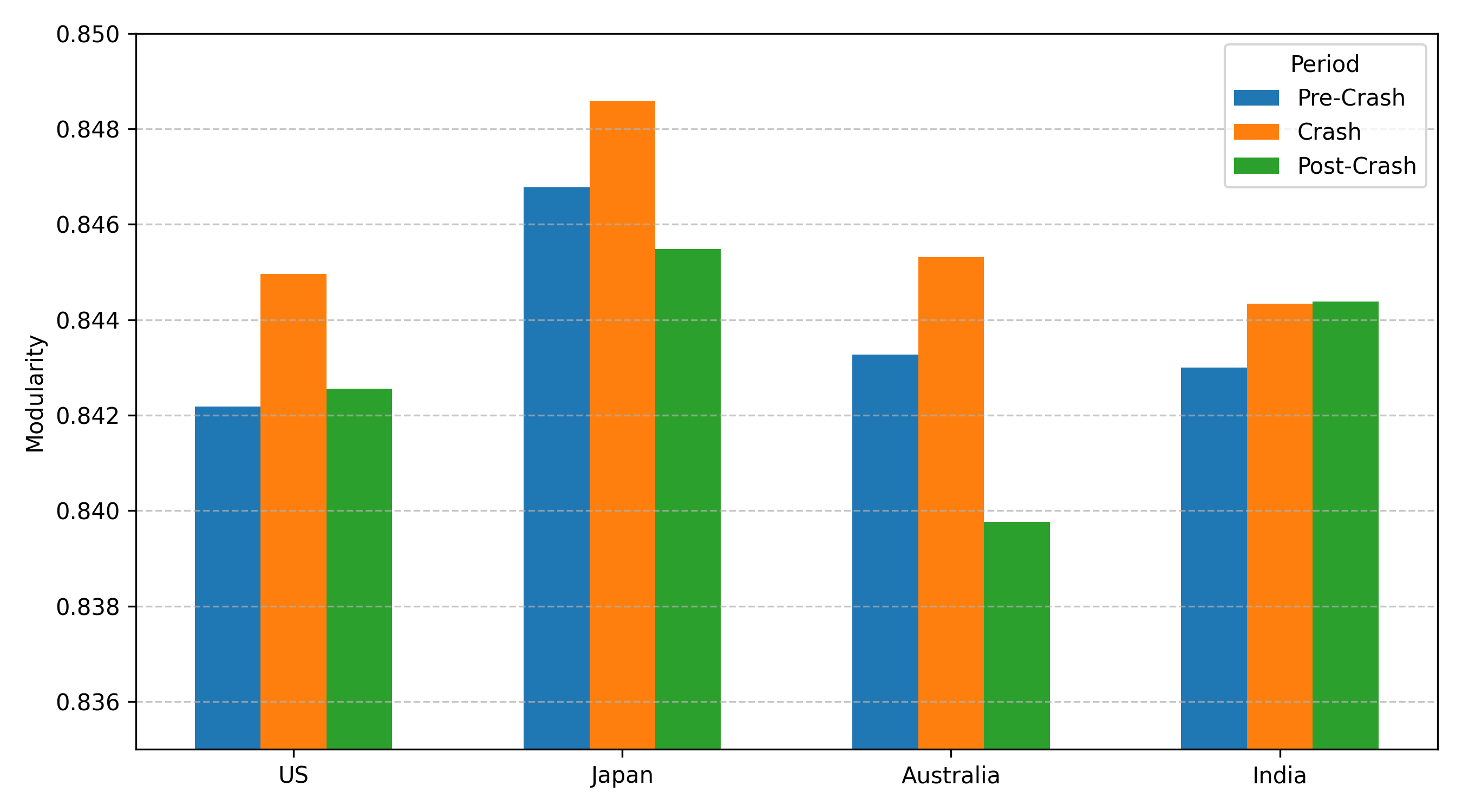} 
        \caption{Modularity of conditional p-threshold MI--MST networks for QUAD stock markets(US, Japan, Australia and India) across pre-crash, crash, and post-crash periods. Modularity values were computed using Clauset-Newman-Moore greedy modularity maximization applied to the conditional p-threshold MI--MST network.}

    \label{fig:modularity_comparison}
\end{figure}

To analyze the community structure of stock markets during periods of financial crash, we computed network modularity using the Clauset-Newman-Moore greedy modularity maximization~\cite{clauset2004finding} algorithm applied to conditional p-threshold MI--MST networks. The greedy modularity maximization method starts by assigning each node to its own community and iteratively merging pairs of communities that yield the largest increase in modularity, until no further improvement is possible and a (local) maximum of the modularity function is reached. Modularity $Q$~\cite{clauset2004finding} quantifies the strength of community structure by comparing the observed fraction of intra-community edges to that expected under a degree-preserving null model and is given by
\begin{equation}
Q = \sum_{c} \left( \frac{E_{c}}{E} - \left( \frac{k_c}{2E} \right)^2 \right),
\end{equation}
where $E_{c}$ is the number of edges within community $c$, $E$ is the total number of edges and $k_c$ is the sum of degrees of nodes in community $c$.

Fig.~\ref{fig:modularity_comparison} shows that, across all four markets: the US, Japan, Australia, and India, the modularity increases systematically during the crash period compared to the pre- and post-crash period. This indicates a strengthening of community structures under market stress. During crash episodes, assets tend to exhibit heightened synchronization within economically or sectorally related groups, while inter-community interactions weaken. Such correlation clustering enhances intra-community connectivity relative to null models, leading to higher modularity values.

To evaluate the statistical significance of community structures in the financial networks across all four markets during the pre-crash, crash, and post-crash periods, we compared the observed modularity ($Q_{\rm obs}$) with modularity values from randomized networks ($Q_{\rm rand}$). The hypotheses for each market-period pair were formulated as follows:
\[
\begin{aligned}
H_0 &: Q_{\rm obs} \le Q_{\rm rand} \quad \text{(no significant community structure)} \\
H_1 &: Q_{\rm obs} > Q_{\rm rand} \quad \text{(significant community structure)}
\end{aligned}
\]
For the testing procedure, $Q_{\rm obs}$ was calculated using the greedy modularity maximization method. An ensemble of $N=1000$ degree-preserving randomized networks was generated using double-edge swaps, and modularity $Q_{\rm rand}$ was calculated for each randomized network. The empirical p-value was determined as:
\[
p = \frac{1}{N} \sum_{i=1}^{N} \mathbb{I}\left(Q_{\rm rand}^{(i)} \ge Q_{\rm obs}\right),
\]
where $\mathbb{I}(\cdot)$ is the indicator function. The null hypothesis $H_0$ was rejected if $p < 0.05$.
Across all markets and periods, we observed that the modularity values were consistently high and empirical p-values were significantly small with $p < 0.01$, indicating that the detected community structures are highly unlikely to occur by chance. These results confirm that the financial networks possess robust and statistically significant community structures that persist across different market conditions.

\subsection{Aftershock in the Stock Market}
\label{After_shock}

\begin{figure}[!t]
\centering

\begin{subfigure}[b]{0.45\textwidth}
\centering
\includegraphics[width=\textwidth]{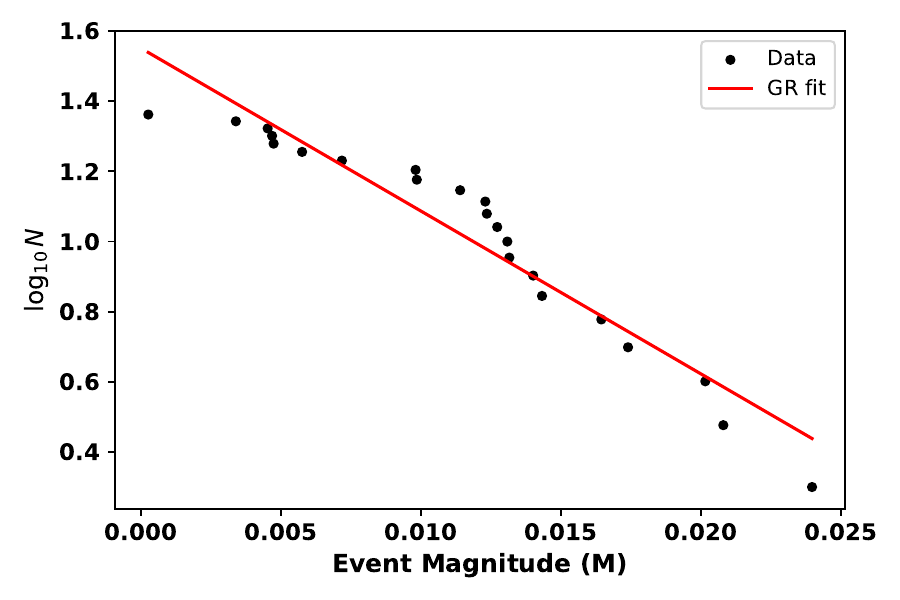}
\caption{S\&P 500}
\end{subfigure}
\hfill
\begin{subfigure}[b]{0.45\textwidth}
\centering
\includegraphics[width=\textwidth]{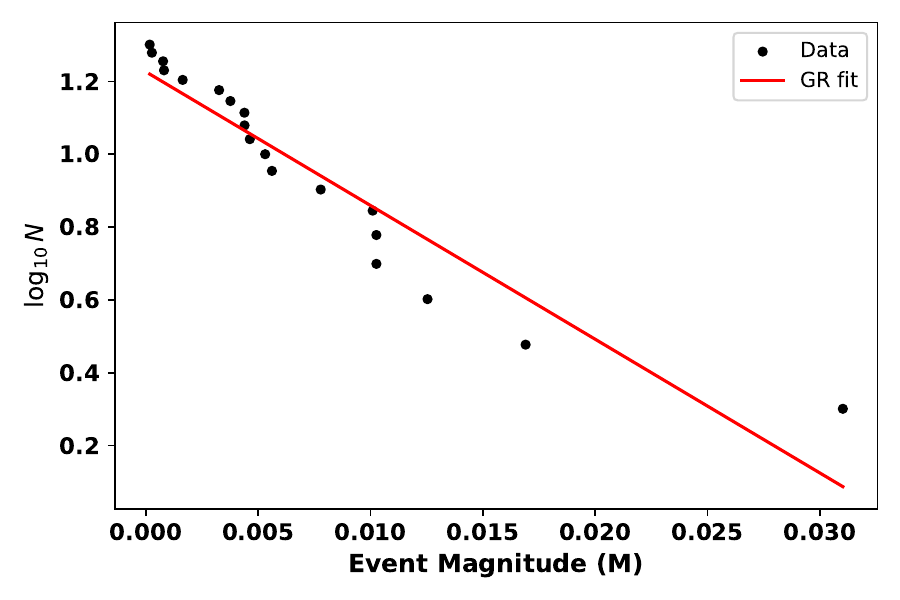}
\caption{Nikkei 225}
\end{subfigure}

\vspace{0.3cm}

\begin{subfigure}[b]{0.45\textwidth}
\centering
\includegraphics[width=\textwidth]{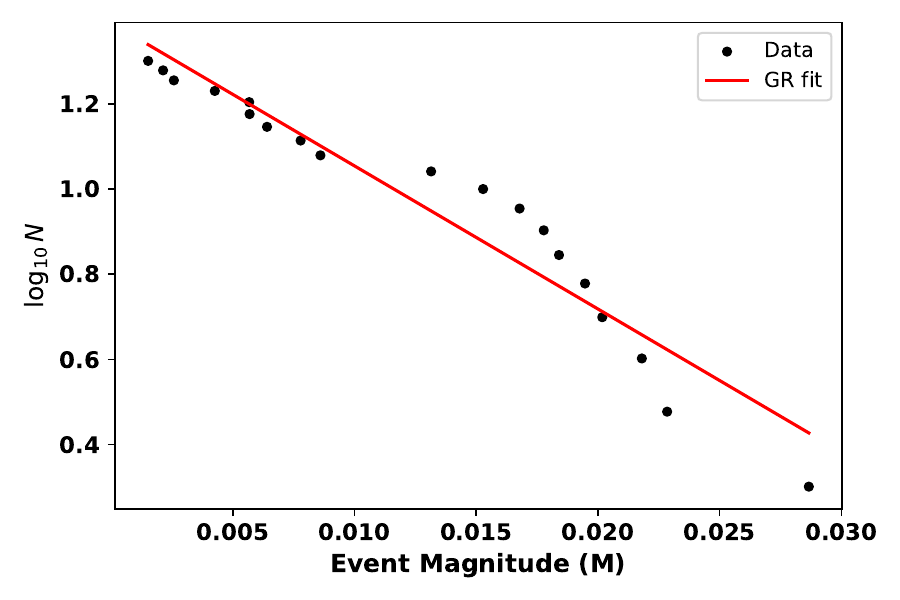}
\caption{S\&P/ASX 200}
\end{subfigure}
\hfill
\begin{subfigure}[b]{0.45\textwidth}
\centering
\includegraphics[width=\textwidth]{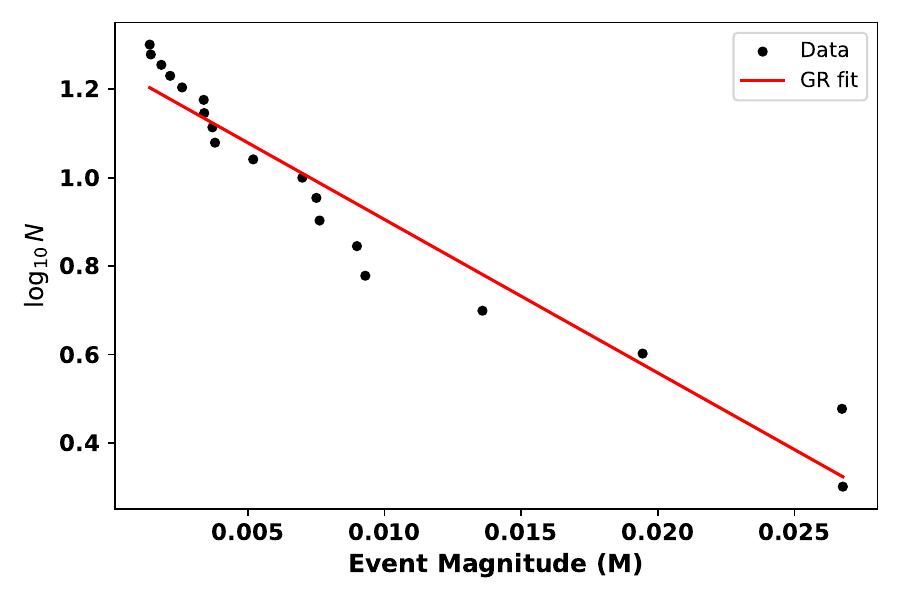}
\caption{Nifty50}
\end{subfigure}

\caption{Gutenberg--Richter (GR) plots for QUAD stock market indices. Figures (a), (b), (c), and (d) represent the GR plot for the S\&P 500 (US), the Nikkei 225 (Japan), the S\&P/ASX 200 (Australia), and the Nifty50 (India). respectively. Each figure shows the cumulative log--rank distribution of peak--trough log-price fluctuations during the post-crash period, along with the fitted GR law.}
\label{fig:GR}
\end{figure}

Fig.~\ref{fig:GR} shows the log--log relationship between the cumulative number of post-crash volatility events $N(M)$ following the COVID-19 stock market crash and their corresponding magnitudes $M$ for major QUAD stock market indices. We fitted the data using Eq.~\ref{Gr_eq}. The plots of $\log_{10} N(M)$ versus $M$ show good agreement with the post-crash data, indicating that peak--trough log-price fluctuations follow the GR power-law behavior. The straight line in Fig.~\ref{fig:GR} represents the best linear fit to the data. For all the QUAD stock market indices, the cumulative log–rank distributions of peak–trough log-price fluctuations exhibit an approximately linear scaling over a broad range of magnitudes, indicating that post-crash volatility events follow a GR-type power-law relationship. Table~\ref{tab:gr_tab} presents the comparison of GR scaling parameters between pre-crash and post-crash periods. We observed that across all the indices and stocks, the estimated $b$-values decrease significantly in the post-crash period compared to the pre-crash period. The systematic reduction in $b$-values indicates that post-crash market dynamics are characterized by a higher relative frequency of large volatility events, consistent with the persistence of market stress rather than a return to pre-crash stability. This aligns with our network topology findings, where post-crash periods show continued structural changes. Lower $b$-values correspond to aftershock-like behavior where large volatility events remain more probable, reflecting ongoing market fragility despite apparent recovery. Since lower $b$-values correspond to a higher relative occurrence of large volatility events, this result indicates increased persistence of large fluctuations following the crash. Together with statistically valid GR fits, these findings suggest that post-crash market dynamics are characterized by aftershock-like volatility rather than an immediate return to pre-crash conditions. This evidence supports and complements the earlier network topology analysis.

\begin{table}[!t]
\centering
\caption{Table represents a comparison of Gutenberg--Richter (GR) scaling parameters between pre-crash and post-crash periods for QUAD market indices and representative stocks. A consistent decrease in $b$-values ($\Delta b < 0$) is observed across all assets, indicating an increased occurrence of large volatility events and aftershock-like behavior in the post-crash period.}
\label{tab:gr_tab}
\renewcommand{\arraystretch}{1.15}

\resizebox{\textwidth}{!}{
\begin{tabular}{llccccc}
\toprule
\textbf{Category} & \textbf{Asset} &
$\boldsymbol{b_{\text{pre}}}$ &
$\boldsymbol{b_{\text{post}}}$ &
$\boldsymbol{\Delta b}$ &
$\boldsymbol{R^2_{\text{post}}}$ &
\textbf{KS $p_{\text{post}}$} \\
\midrule

\multirow{4}{*}{Indices}
 & \textasciicircum GSPC & 103.35 & 46.38 & $-56.97$ & 0.921 & 0.872 \\
 & \textasciicircum N225 & 90.56  & 36.72 & $-53.84$ & 0.898 & 0.808 \\
 & \textasciicircum NSEI & 98.98  & 34.69 & $-64.29$ & 0.932 & 0.538 \\
 & \textasciicircum AXJO & 106.88 & 33.60 & $-73.28$ & 0.927 & 0.978 \\
\midrule

\multirow{6}{*}{Stocks}
 & AAPL     & 62.89 & 25.04 & $-37.85$ & 0.953 & 0.990 \\
 & MSFT    & 93.19 & 26.58 & $-66.61$ & 0.920 & 1.000 \\
 & AMZN    & 78.29 & 16.92 & $-61.37$ & 0.918 & 0.921 \\
 & NVDA    & 46.75 & 18.23 & $-28.52$ & 0.953 & 0.983 \\
 & JPM     & 66.76 & 14.77 & $-51.99$ & 0.955 & 0.952 \\
 & INFY.NS & 59.98 & 33.13 & $-26.85$ & 0.986 & 1.000 \\

\bottomrule
\end{tabular}}
\end{table}

\FloatBarrier

\section{Conclusion}
\label{conclusion}

This study provides a comprehensive examination of the evolution of nonlinear dependency structures and network topology in global equity markets during the COVID-19 crisis. Using a conditional $p$-threshold mutual information (MI) framework, we analyzed stock market dynamics across the QUAD economies, namely the United States, Japan, Australia, and India, by systematically comparing pre-crash, crash, and post-crash periods. The COVID-19 market crash was identified using a robust two-stage approach. First, a rolling Hellinger Distance (HD) detected statistically significant structural breaks in cross-sectional return distributions through synchronized threshold exceedances ($H_D > \mu_H + 2\sigma_H$) across all four markets. Second, the Hilbert Spectrum (HS) characterized the internal temporal dynamics of the identified crash period, revealing a sustained concentration of high instantaneous energy during February–March 2020. Together, these methods confirm that the detected event corresponds to a prolonged high-volatility market period.

A comparison between MI and conditional $p$-threshold MI highlights clear differences in the resulting dependency structures. The MI matrices were found to be dense and dominated by common market-wide effects, particularly during the crash, which hide pure stock-level interactions. In contrast, conditioning on market index returns and applying permutation-based statistical filtering yielded sparse and economically interpretable dependency structures. The conditional $p$-threshold MI framework thus effectively isolated direct and statistically significant nonlinear stock–stock dependencies, providing a reliable basis for network construction. Transforming these filtered dependencies into Minimum Spanning Tree (MST) networks revealed consistent and universal topological reorganization across all QUAD markets during the crash. The networks exhibited clear compression, characterized by a reduction in average path length and eccentricity alongside increases in closeness and weighted degree, implying faster potential propagation of shocks. Despite this increased integration, algebraic connectivity declined, indicating that the crash-induced connectivity was accompanied by greater structural fragility. The crisis period was also marked by a pronounced reconfiguration of the core–periphery structure. Core concentration declined systematically across all markets, indicating a fragmentation and decentralization of influence away from a stable core. At the same time, periphery fragility increased in most markets, indicating greater vulnerability of weakly connected stocks to shock transmission. This shift was reinforced by increasingly negative assortativity, reflecting preferential connections between influential core stocks and fragile peripheral nodes, thereby amplifying systemic risk. Beyond node-level and global topology, community structure analysis revealed a strengthening of modular organization during the crash. Increased modularity indicates that stocks clustered more tightly into internally cohesive groups, while inter-community connections weakened, reflecting segmentation of market interactions under stress. This clustering further contributes to the uneven transmission of shocks across the network.

In the post-crash period, the network did not fully revert to its pre-crisis configuration. Several topological measures and core–periphery indices showed only partial recovery, indicating persistent structural aftereffects. This delayed normalization was supported by an aftershock analysis based on the Gutenberg–Richter law. A systematic reduction in the scaling parameter ($b$-value) across indices and representative stocks implies a higher relative frequency of large volatility events following the crash, consistent with continued market fragility rather than immediate stabilization.

Overall, the conditional $p$-threshold MI–MST framework offers a refined lens for examining crisis-driven market reorganization. By filtering out common market effects and emphasizing statistically validated nonlinear dependencies, the approach uncovers universal structural signatures of financial crises, including network compression, weakened core dominance, heightened periphery vulnerability, strengthened community structure, and persistent aftershocks. These findings provide meaningful insights into systemic risk propagation and recovery dynamics across geographically distinct markets. In future work, we will extend this framework to other asset classes and crisis periods.

\section*{Acknowledgments}

The authors, Kundan Mukhia and S.R. Luwang, would like to acknowledge the National Institute of Technology Sikkim for providing doctoral research fellowships. Imran Ansari acknowledges support from the Kotak IISc AI–ML Centre (KIAC) at the Indian Institute of Science, Bengaluru, India.

\FloatBarrier

\section*{Appendix}

\begin{figure}[H]
    \centering

    \begin{subfigure}[b]{0.32\textwidth}
        \centering
        \includegraphics[width=\textwidth]{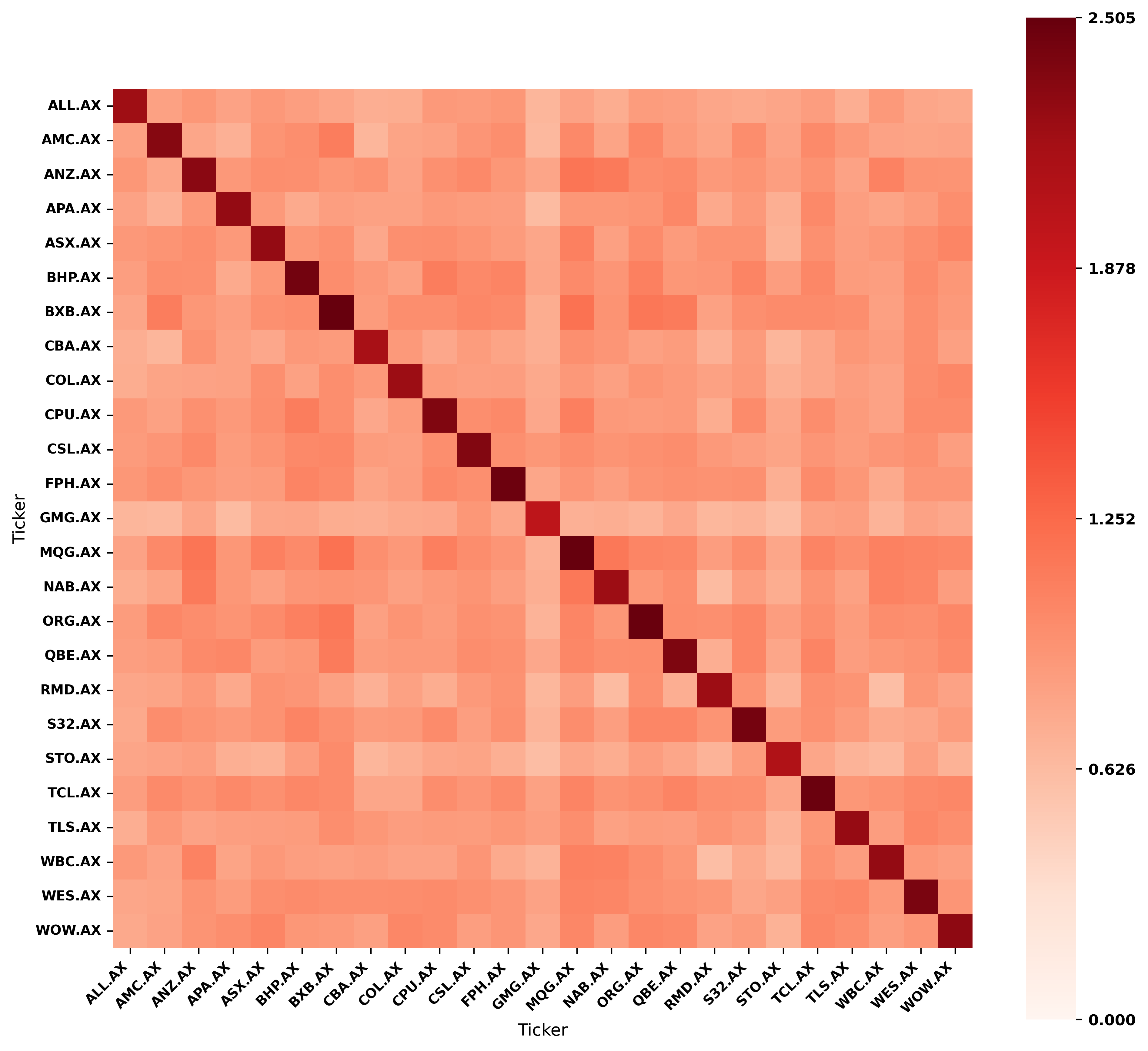}
        \caption{Pre-Crash (Raw MI)}
    \end{subfigure}
    \hfill
    \begin{subfigure}[b]{0.32\textwidth}
        \centering
        \includegraphics[width=\textwidth]{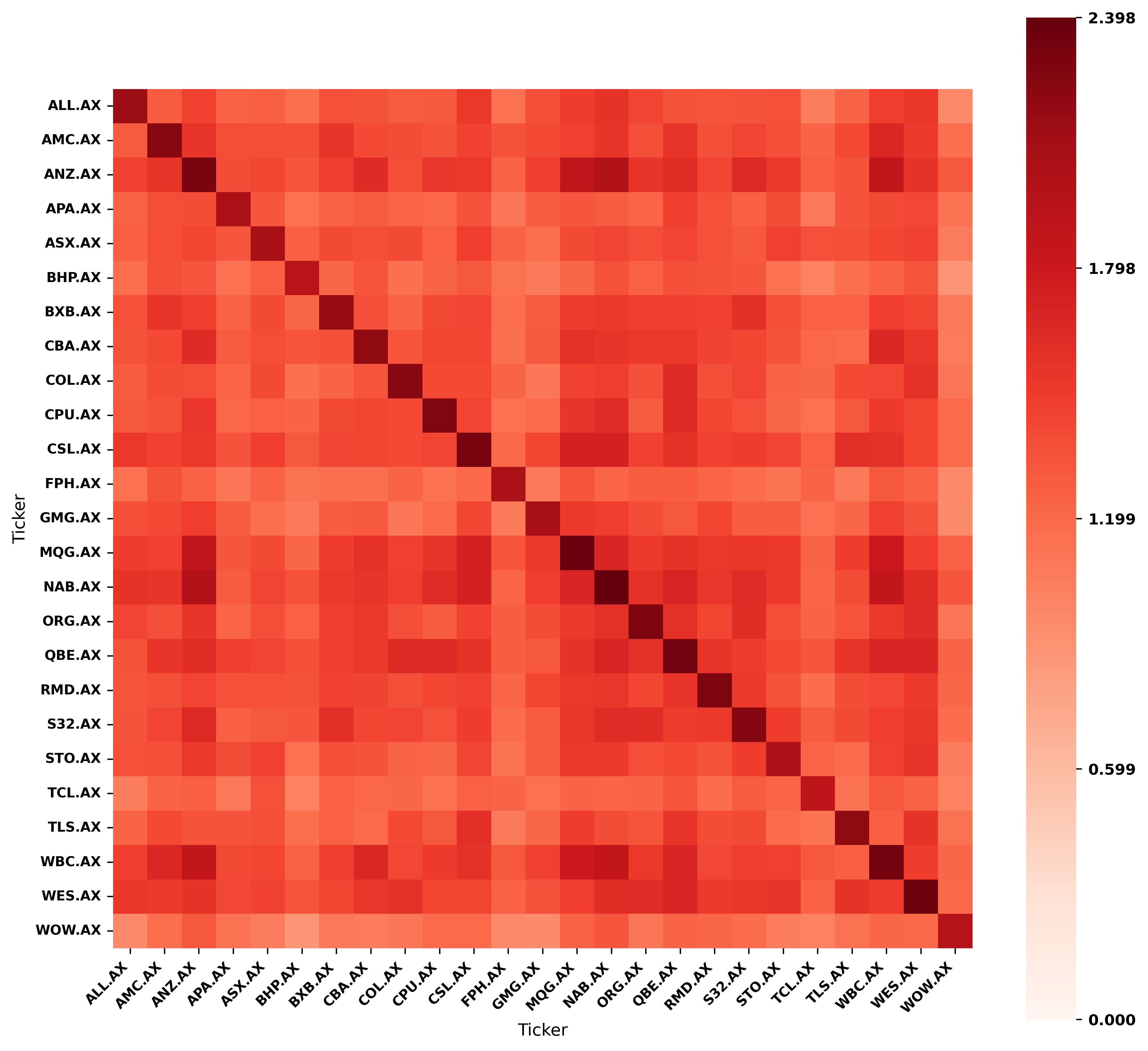}
        \caption{Crash (Raw MI)}
    \end{subfigure}
    \hfill
    \begin{subfigure}[b]{0.32\textwidth}
        \centering
        \includegraphics[width=\textwidth]{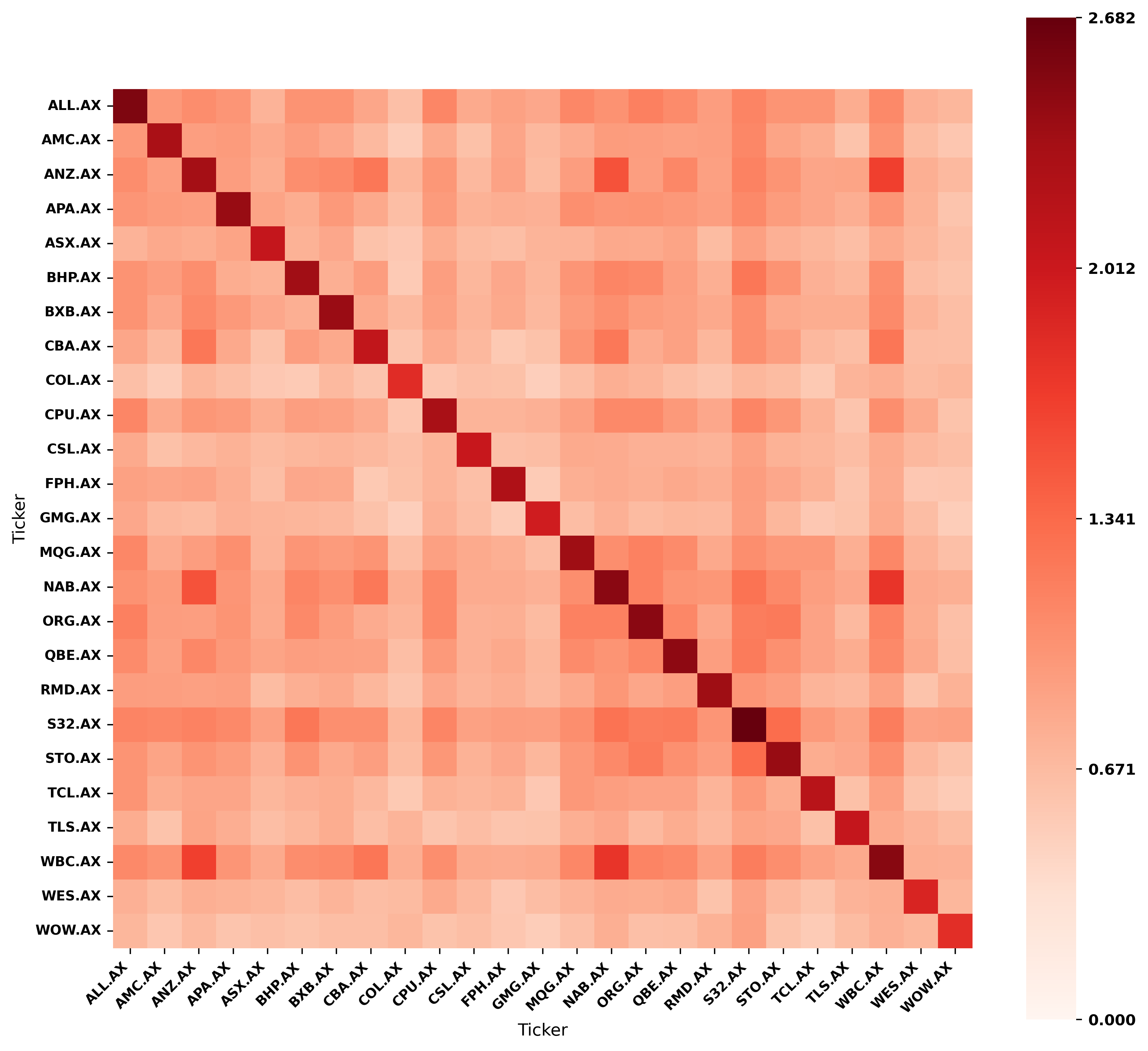}
        \caption{Post-Crash (Raw MI)}
    \end{subfigure}

    \vspace{0.3cm}

    \begin{subfigure}[b]{0.32\textwidth}
        \centering
        \includegraphics[width=\textwidth]{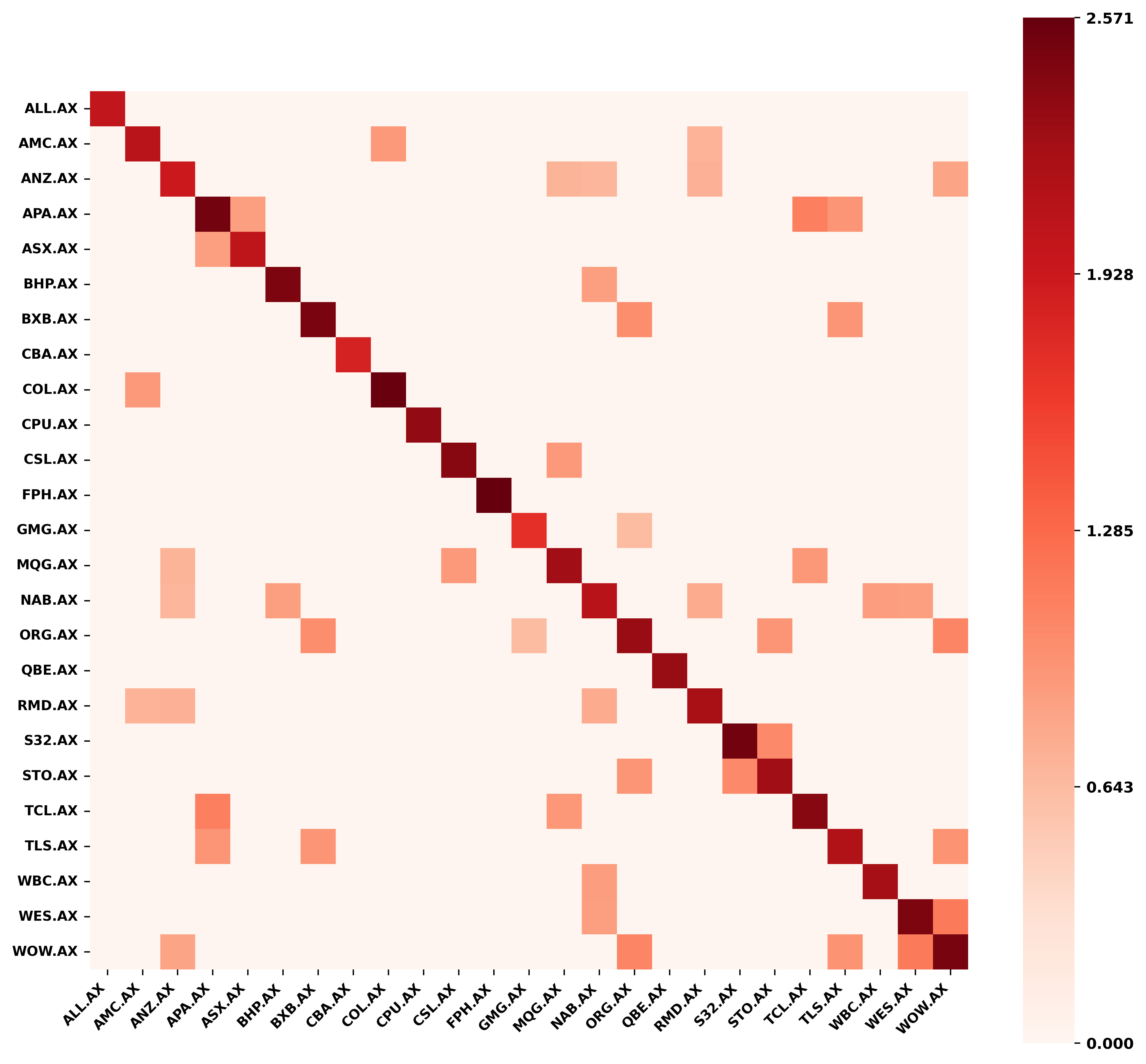}
        \caption{Pre-Crash (Cond.\ P-threshold)}
    \end{subfigure}
    \hfill
    \begin{subfigure}[b]{0.32\textwidth}
        \centering
        \includegraphics[width=\textwidth]{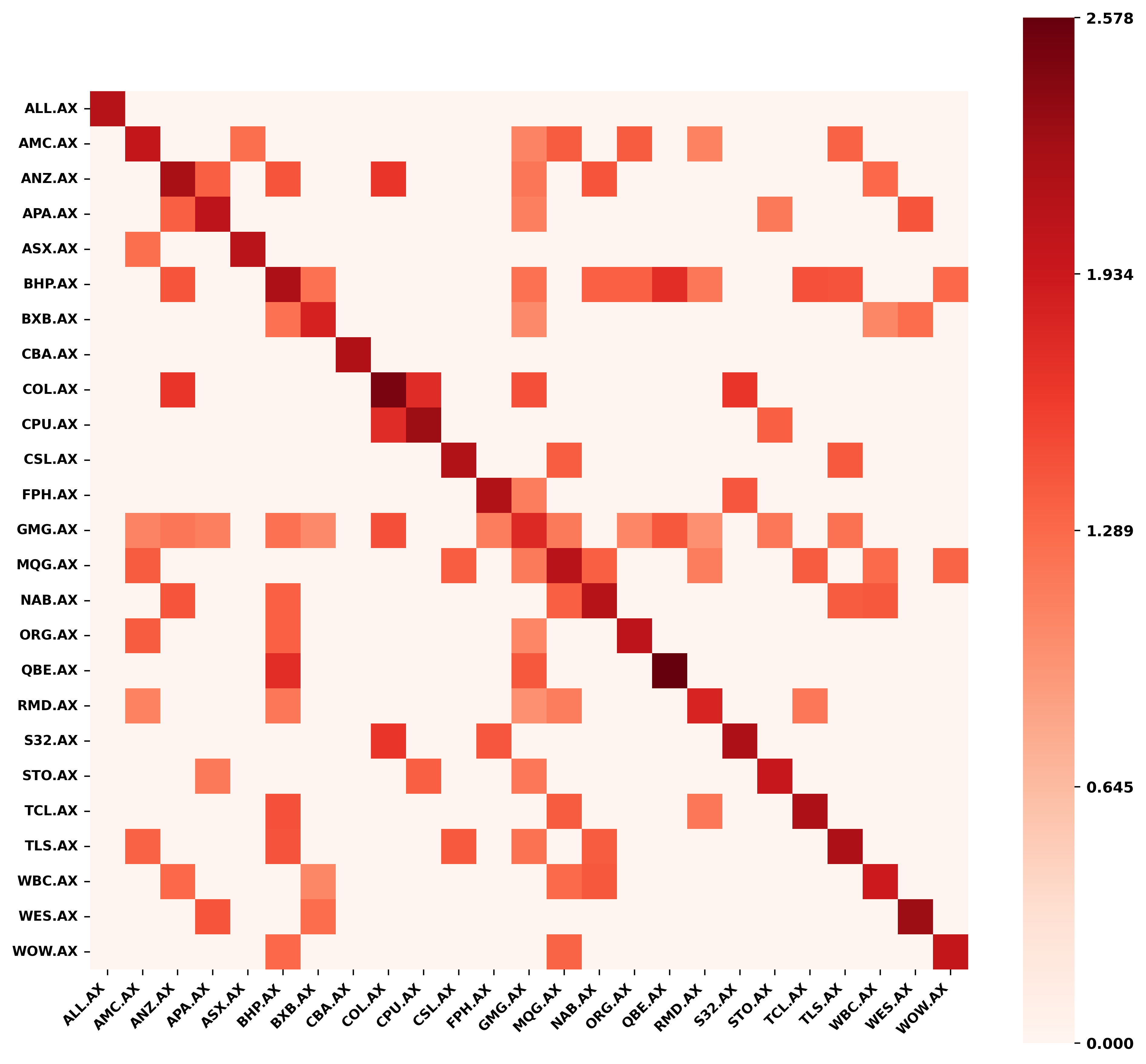}
        \caption{Crash (Cond.\ P-threshold)}
    \end{subfigure}
    \hfill
    \begin{subfigure}[b]{0.32\textwidth}
        \centering
        \includegraphics[width=\textwidth]{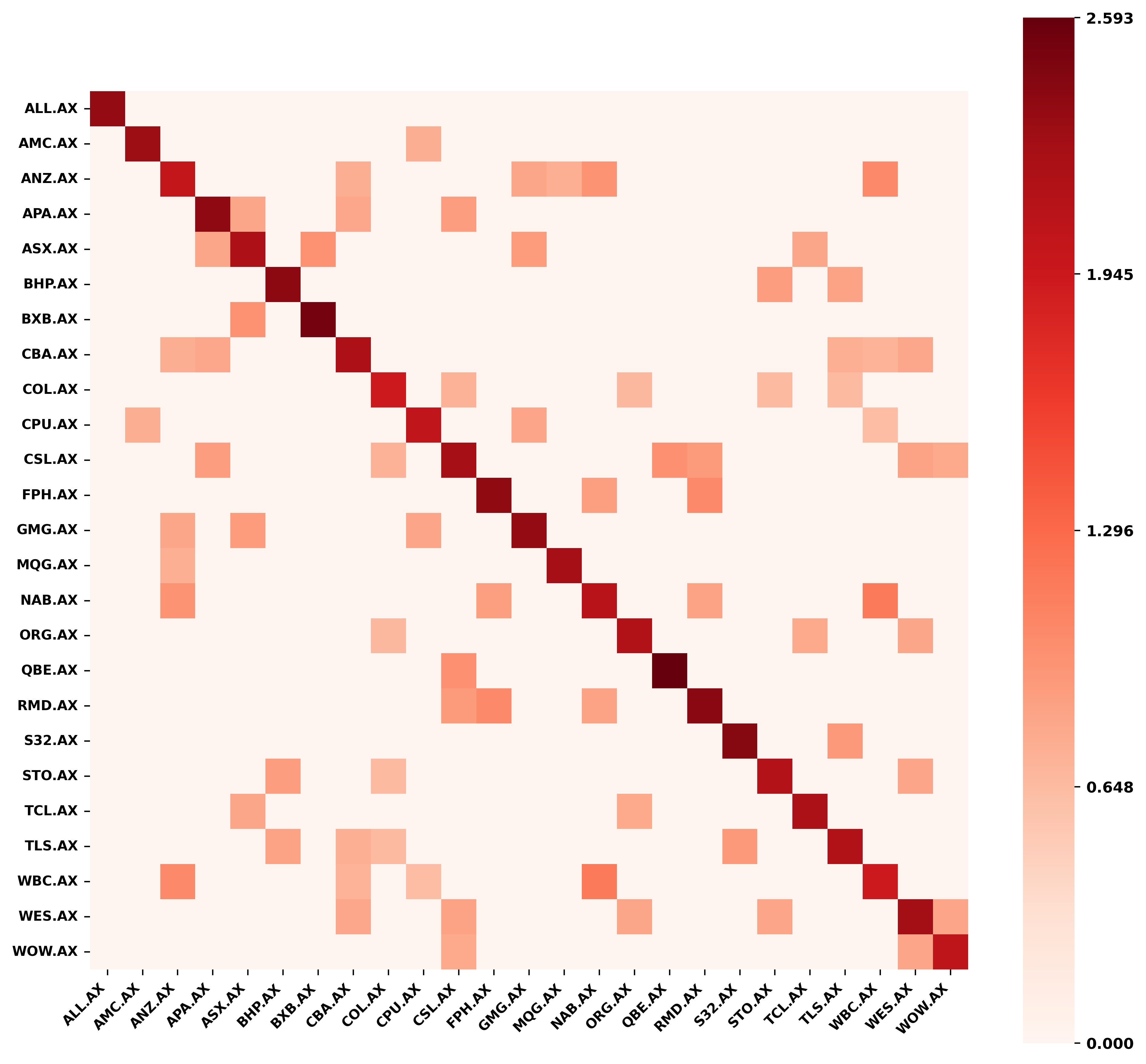}
        \caption{Post-Crash (Cond.\ P-threshold)}
    \end{subfigure}

    \caption{Mutual information heatmaps for the 25 largest Australia stocks across different market periods. Plot (a)--(c) shows raw mutual information (MI) heatmaps, which are dominated by common market effects and exhibit dense connectivity, particularly during the crash period. Plot (d)--(f) display residual-based, significance-filtered MI heatmaps obtained using the conditional P-threshold MI method, highlighting statistically significant and direct nonlinear dependencies. The conditional P-threshold MI heatmaps show a sparse structure, indicating the true nonlinear dependencies between stocks after removing the market effect, which reveals the underlying direct interactions among stocks}
    
    \label{app:australia_mi}
\end{figure}

\begin{figure}[!t]
    \centering

    \begin{subfigure}[b]{0.32\textwidth}
        \centering
        \includegraphics[width=\textwidth]{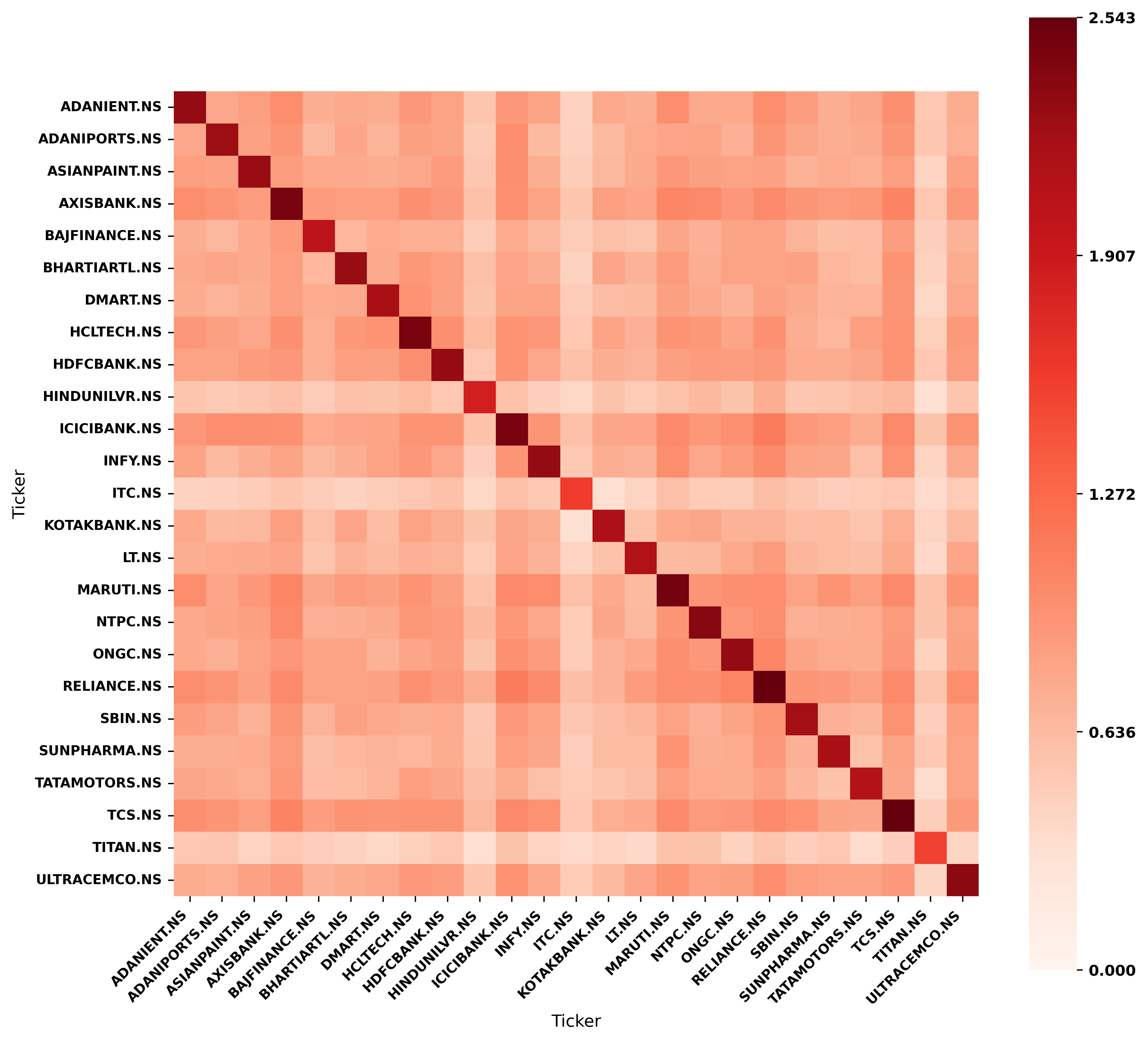}
        \caption{Pre-Crash (Raw MI)}
    \end{subfigure}
    \hfill
    \begin{subfigure}[b]{0.32\textwidth}
        \centering
        \includegraphics[width=\textwidth]{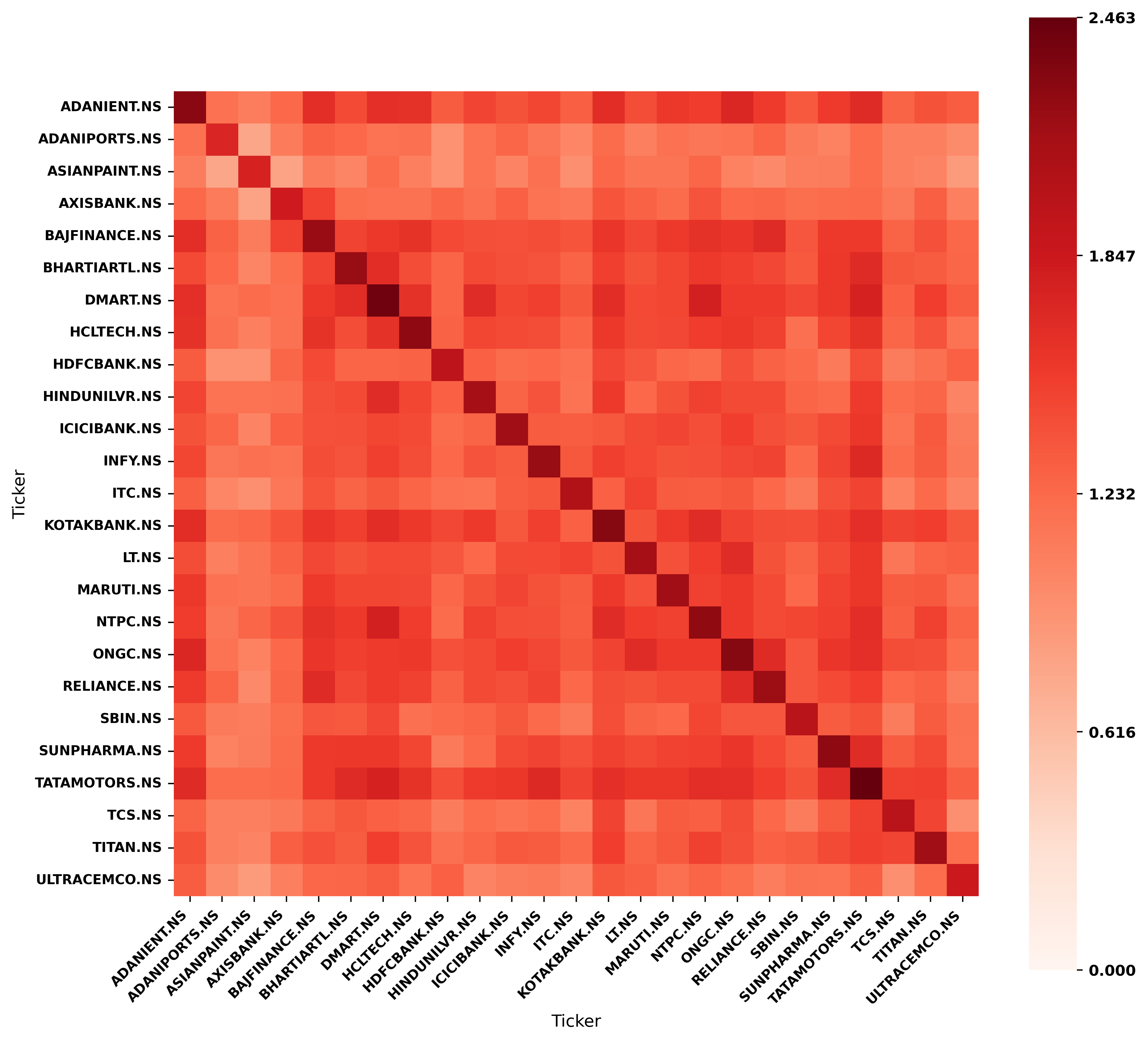}
        \caption{Crash (Raw MI)}
    \end{subfigure}
    \hfill
    \begin{subfigure}[b]{0.32\textwidth}
        \centering
        \includegraphics[width=\textwidth]{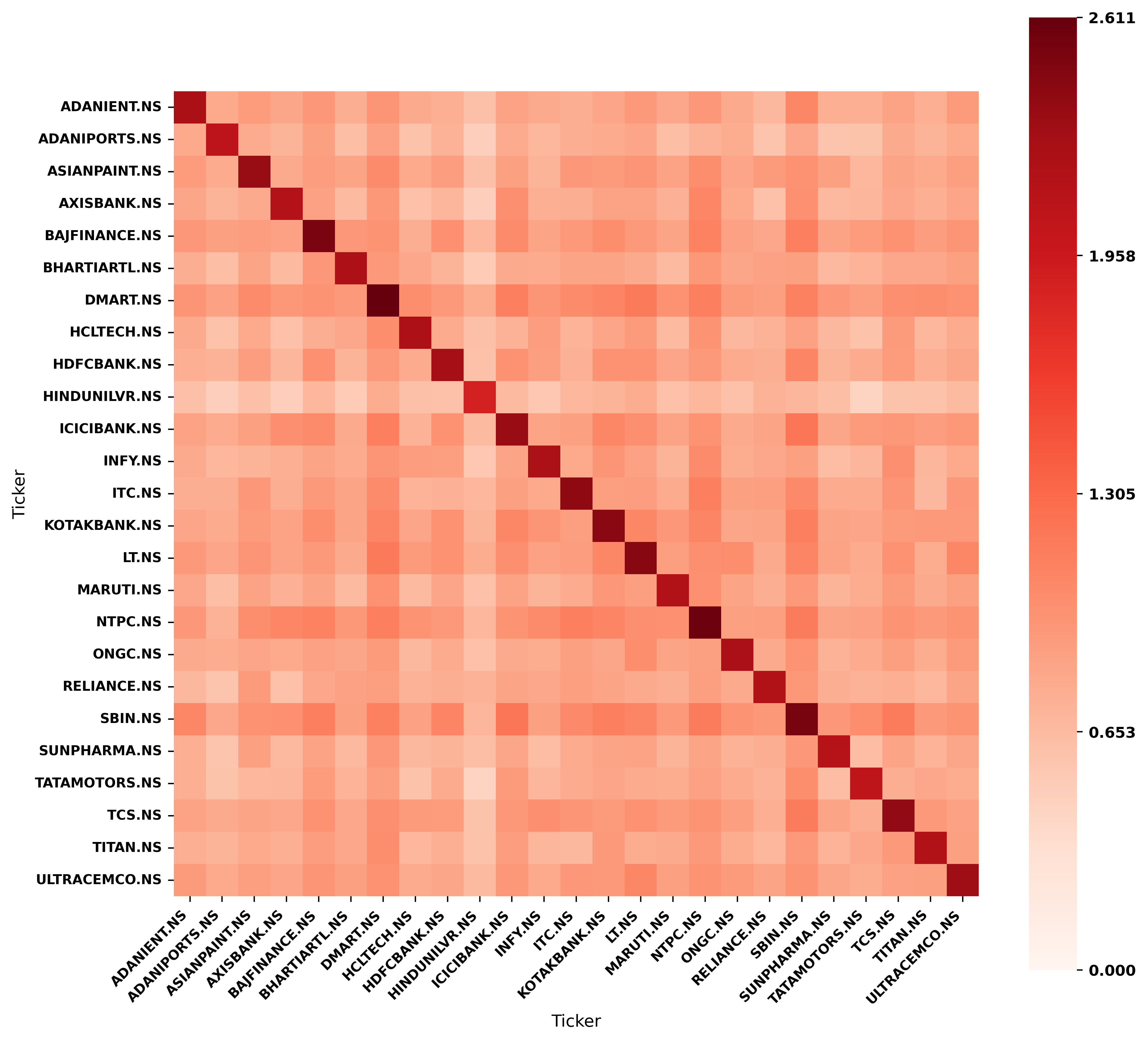}
        \caption{Post-Crash (Raw MI)}
    \end{subfigure}

    \vspace{0.3cm}

    \begin{subfigure}[b]{0.32\textwidth}
        \centering
        \includegraphics[width=\textwidth]{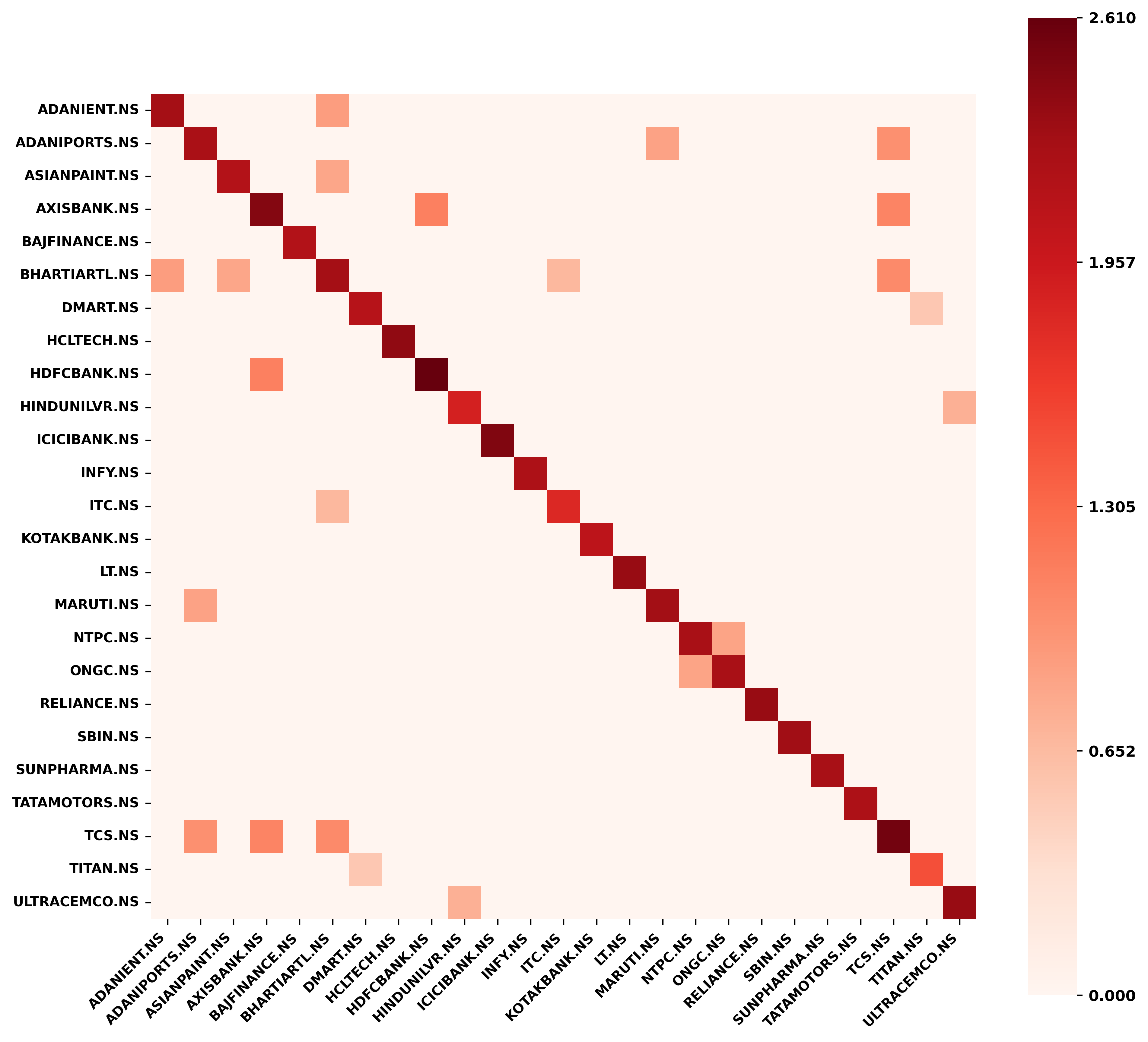}
        \caption{Pre-Crash (Cond.\ P-threshold)}
    \end{subfigure}
    \hfill
    \begin{subfigure}[b]{0.32\textwidth}
        \centering
        \includegraphics[width=\textwidth]{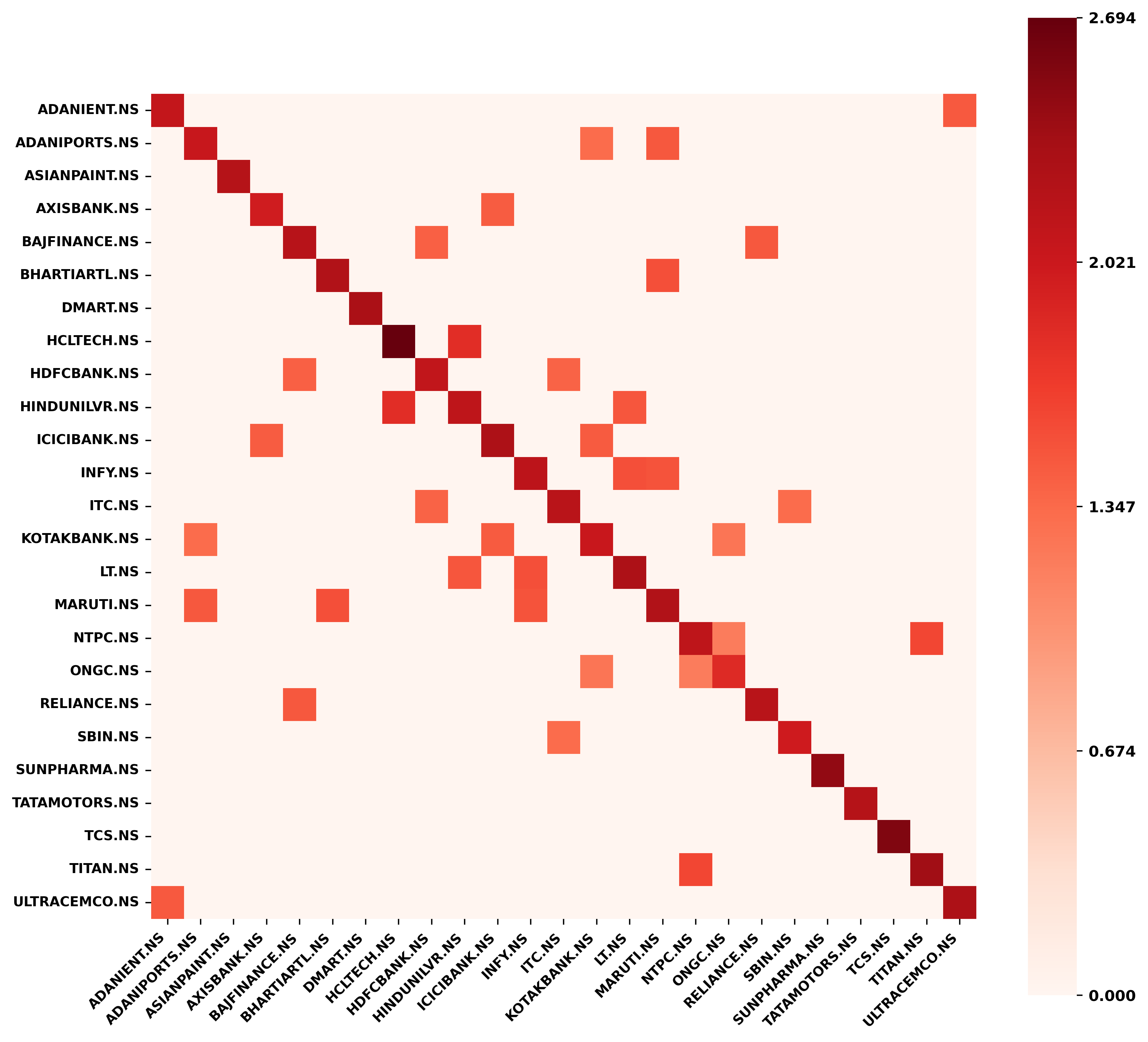}
        \caption{Crash (Cond.\ P-threshold)}
    \end{subfigure}
    \hfill
    \begin{subfigure}[b]{0.32\textwidth}
        \centering
        \includegraphics[width=\textwidth]{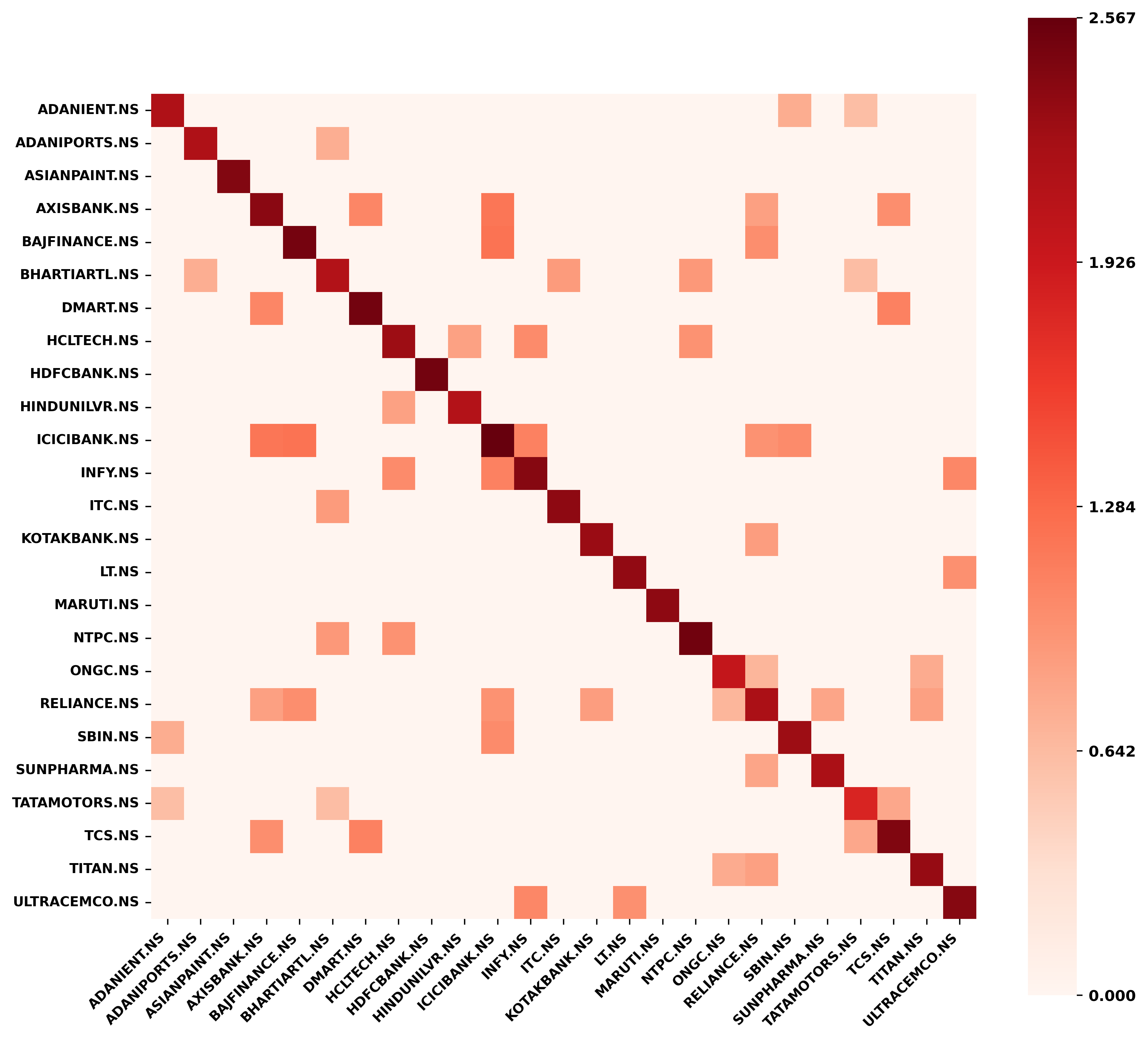}
        \caption{Post-Crash (Cond.\ P-threshold)}
    \end{subfigure}

    \caption{Mutual information heatmaps for the 25 largest Indian stocks across different market periods. Plot (a)--(c) shows raw mutual information (MI) heatmaps, which are dominated by common market effects and exhibit dense connectivity, particularly during the crash period. Plot (d)--(f) display residual-based, significance-filtered MI heatmaps obtained using the conditional P-threshold MI method, highlighting statistically significant and direct nonlinear dependencies. The conditional P-threshold MI heatmaps show a sparse structure, indicating the true nonlinear dependencies between stocks after removing the market effect, which reveals the underlying direct interactions among stocks}
    
    \label{app:india_mi}
\end{figure}

\FloatBarrier

\bibliographystyle{elsarticle-num} 
\bibliography{References}

\end{document}